\documentclass[aip,amsmath,amssymb,reprint]{revtex4-1}

\usepackage{graphicx}

\usepackage[T1]{fontenc}
\usepackage{newtxtext,newtxmath}
\usepackage[cal=boondoxo]{mathalfa}

\usepackage[defaultsans]{lato}
\usepackage[SquareTraceBrackets]{quantum}

\usepackage{bm,mathrsfs}
\usepackage{graphicx}
\usepackage[dvipsnames]{color}

\newcommand{\povm}{{\sc povm}}
\newcommand{\qfi}{{\sc qfi}}
\newcommand{\sld}{{\sc sld}}
\newcommand{\rld}{{\sc rld}}

\newcommand{\mse}{{\sc mse}}
\newcommand{\crb}{{\sc crb}}

\newcommand{\qcrb}{{\sc qcrb}}

\DeclareMathOperator{\cov}{Cov}
\DeclareMathOperator{\var}{Var}
\DeclareMathOperator{\ex}{\mathbb{E}}
\DeclareMathOperator{\vect}{vec}

\newcommand{\caron}[1]{\check{#1}}

\newcommand{\x}{\mathbf{x}}
\newcommand{\y}{\mathbf{y}}
\newcommand{\z}{\mathbf{z}}

\renewcommand{\identity}{\mathbb{I}}
\renewcommand{\Re}{\mathrm{Re}\,}
\renewcommand{\Im}{\mathrm{Im}\,}

\DeclarePairedDelimiterX{\infdivx}[2]{(}{)}{%
  #1\,\delimsize\|\,#2%
}
\newcommand{\infdiv}{\infdivx}

\usepackage{float}								
\usepackage{booktabs}
\usepackage{makecell}							
		
\graphicspath{{./figs/}}

\newcommand*\dif{\mathop{}\!\mathrm{d}}				
\definecolor{RoyalAzure}{rgb}{0.0, 0.22, 0.66}			
\usepackage[a4paper=true,colorlinks=true,citecolor=RoyalAzure,linkcolor=RoyalAzure,urlcolor=RoyalAzure]{hyperref}

\usepackage[caption=false]{subfig}					
\DeclareGraphicsExtensions{.pdf, .jpg, .eps, .svg}

\DeclareMathAlphabet{\mathscrbf}{OMS}{mdugm}{b}{n} 	
\DeclareMathAlphabet\mathscrbf{OMS}{cmsy}{b}{n}		

\begin{document}


\title{A Geometric Perspective on Quantum Parameter Estimation}

\author{Jasminder S. Sidhu}
 \email{jsmdrsidhu@gmail.com}
\author{Pieter Kok}
 \email{p.kok@sheffield.ac.uk}
\affiliation{ 
Department of Physics and Astronomy, The University of Sheffield, Sheffield, S3 7RH, United Kingdom.}

\date{\today}

\begin{abstract}
Quantum metrology holds the promise of an early practical application of quantum technologies, in which measurements of physical quantities can be made with much greater precision than what is achievable with classical technologies. In this review, we collect some of the key theoretical results in quantum parameter estimation by presenting the theory for the quantum estimation of a single parameter, multiple parameters, and optical estimation using Gaussian states. We give an overview of results in areas of current research interest, such as Bayesian quantum estimation, noisy quantum metrology, and distributed quantum sensing. We address the question how minimum measurement errors can be achieved using entanglement as well as more general quantum states. This review is presented from a geometric perspective. This has the advantage that it unifies a wide variety of estimation procedures and strategies, thus providing a more intuitive big picture of quantum parameter estimation.
\end{abstract}

\maketitle

\tableofcontents


\section{Introduction}\label{sec:intro}
\noindent
When we measure a physical quantity, we need to quantify the error in that measurement, for example via the Mean Square Error (\mse). The actual \mse\ is hard to calculate directly, because it depends on the true unknown value of the quantity we want to measure. Remarkably, we can bound the \mse\ such that for any given experiment it cannot be smaller than a value that we \emph{can} calculate. In addition, we can establish general conditions under which this bound can be saturated, i.e., the minimum \mse\ is in fact achieved. This is in broad strokes what parameter estimation theory is about. 

With the classical theory of parameter estimation well-established in terms of the probability distribution of the observed data given the value of the physical quantity, we turn our attention to the space of probability distributions. Intuitively, when probability distributions move a large distance in this space under a change in value of the physical quantity, our measurements can pick up this change more easily than when the probability distribution moves hardly at all. This provides the link between measurement sensitivity (with correspondingly small \mse) and distance functions in the space of probabilities. Our task is to find a distance measure that allows us to make this link quantitative such that the distance measure can be directly related to the \mse. This gives us a metric in the space of probabilities, which is called the Fisher information. We will review key properties of the Fisher information.

The extension to quantum mechanics requires that we consider density operators instead of classical probability distributions. These operators also live in a linear vector space, and we can again construct distance measures in the space that connects the density operator to the minimal \mse\ in a physical experiment. However, since the density operators have a richer structure than the classical probability distributions (e.g., allowing non-commutative operators), the corresponding distance measures are also more complex. Based on the choice of inner product in the space of density operators, different distance measures can be constructed that have subtly different physical interpretations. The direct generalisation of the Fisher information, the \emph{quantum} Fisher information, is a well-studied quantity that provides an  attainable bound for the estimation of a single parameter that is imprinted on the quantum state via a unitary evolution. When multiple parameters come into play, the picture complicates considerably due to the potential non-commutativity of the observables that best estimate the individual parameters. Here, there are close links to generalised uncertainty relations. Alternative metrics can be more appropriate for different uses. For example, the Kubo-Mori information will be shown to have a particular relevance when thermal states are considered, and when we are interested in the relation between parameter estimation and conserved quantities we may turn to the Wigner-Yanase information.

\begin{figure}[t!]
\centering
\includegraphics[width =0.95\columnwidth]{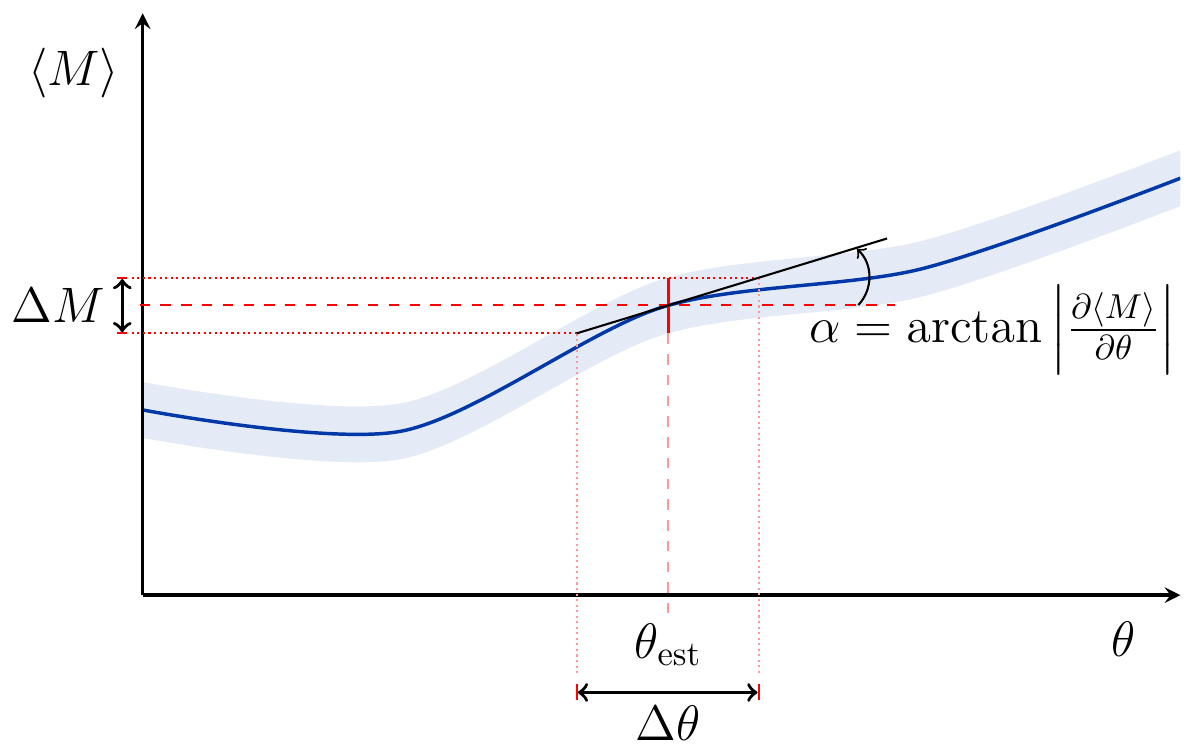}
\caption{We measure an observable $M$ to obtain an estimate $\theta_{\rm est}$ for a parameter of interest $\theta$. We can relate the error $\Delta\theta$ in the estimate to the standard deviation $\Delta M$ by relating the tangent at $\theta_{\rm est}$ with angle $\alpha$ to the ratio between $\Delta M$ (red vertical line) and the error $\Delta\theta$.}
\label{fig:errorformula}
\end{figure}

Once we have established lower bounds for the error in the parameters, we consider how we can attain these bounds. For quantum estimation, entanglement plays an important, albeit subtle role. We review two important special cases in quantum estimation theory, namely estimation procedures using Gaussian states in quantum optics, and the extension of quantum estimation to the case where the evolution does not have a simple phase-like unitary structure. The final two sections are devoted to current areas of research interest, including Bayesian quantum estimation, noisy quantum metrology and some fault tolerant solutions, and distributed quantum sensing. Throughout this review, we aim to emphasise the bigger picture of quantum parameter estimation and how it very naturally arises from a geometrical view of quantum states. To achieve this we have prioritised practical examples and intuition over mathematical rigour. Readers interested in the technical mathematical details are referred to the references, and more mathematically inclined readers are referred to the collection of selected papers on quantum statistical inference by Hayashi~\cite{Hayashi05}.


\section{Classical estimation theory}\label{sec:ce_theory}
\noindent
In this section, we briefly review classical estimation theory. We introduce the expectation value and the variance between the measured and the true value. We formulate a family of generic lower bounds that constrain the variance of parameter estimates, culminating in the well-known Cram\'er-Rao bound. For a general introduction to classical estimation theory, see Kay (1993)~\cite{Kay1993}.

Several features of estimation theory can be understood by considering the following heuristic argument~\cite{Toth2013_PRA}: given a measurement of the observable $M$ whose outcomes depend on a parameter $\theta$, we can associate an estimate and error to $\theta$. The average value $\braket{M}$ over many measurements of $M$ depends on $\theta$, and the error formula can be obtained from Fig.~\ref{fig:errorformula} as
\begin{align}\label{eq:n304riwerofjdnk}
 \delta\theta = \frac{\Delta M}{\abs{\partial_\theta \braket{M}}},
\end{align}
where $\smash{\partial_\theta=\partial/\partial\theta}$, and $\smash{\Delta M = [\braket{M^2} - \braket{M}^2]^{\frac12}}$ is the standard deviation in the measurement outcomes. The denominator $\abs{\partial_\theta \braket{M}}$ can be viewed as a local correction in the units of $\delta\theta$. The steeper the tangent at $\theta_{\rm est}$, the more precise the estimate of $\theta$ at that point, i.e., the smaller $\delta \theta$. On the other hand, the larger the standard deviation $\Delta M$, the lower the precision.

\begin{figure}[t!]
\centering
\includegraphics[width =0.93\columnwidth]{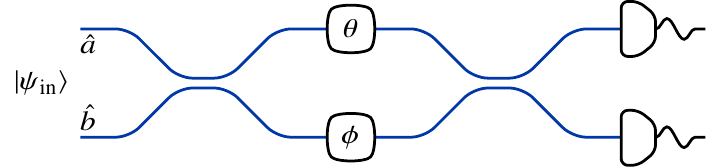}
\caption[Mach-Zehnder interferometer]{The Mach-Zehnder interferometer, drawn with two evanescently coupled 50:50 beam splitters. Here, $\smash{\hat{a}}$ and $\smash{\hat{b}}$ define two input modes. An optical path difference between the internal arms of the interferometer results in a relative phase difference. The objective is to estimate the relative phase $\theta = \smash{\Phi} - \phi$ by measuring the photon intensities at the output using the photodetectors on the right.}
\label{pic:mach_zehnder_interferometer}
\end{figure}

To illustrate the concepts of classical estimation theory, we use the example of the Mach-Zehnder interferometer (\textsc{mzi}), illustrated in Fig.~\ref{pic:mach_zehnder_interferometer}. It can be used to make high precision measurements of a relative phase difference between two beams of light derived from some optical input state. This phase referenced method has become a standard tool in estimation theory and has received considerable attention given its applications in enhanced phase estimations in optical interferometry~\cite{Dobrzanski2009_PRA, Pezze2014_AI}, frequency measurements~\cite{Bollinger1996_PRA, Huelga97}, and biosensors~\cite{Luff1998_JLT, Yang2001_OL}.

Consider an experiment to estimate the relative phase difference in the interferometer, where a single photon is sent into one input mode of the \textsc{mzi}, and the two output modes are monitored with photodetectors. Each run of the experiment provides two pieces of data $\smash{\x^{(1)}}$ and $\smash{\x^{(2)}}$ that measure the photon counts in each detector, $D_1$ and $D_2$ respectively. Assuming no losses in the interferometer and ideal detectors, the probability distribution function $\smash{p(\x^{(1)},\x^{(2)}\vert\theta)}$ has the following possible outcomes:
\begin{alignat}{3}
p(1,1\vert\theta) &= 0, \quad p(1,0\vert\theta) &&= \sin^2\left(\frac{\theta}{2}\right), \cr
p(0,0\vert\theta) &= 0, \quad p(0,1\vert\theta) &&= \cos^2\left(\frac{\theta}{2}\right),
\label{eqn:s_photon_prob_dist}
\end{alignat}
with $\theta=\Phi-\phi$ the phase difference that we are interested in. If we count $N_1$ photons in detector $D_1$ and $N_2$ in detector $D_2$, a suitable estimator for $\theta$ is
\begin{align}
\check{\theta} = \arccos\left(\frac{N_1 - N_2}{N}\right) ,
\label{eqn:estimator_single_photon}
\end{align}
where we denote the estimator of $\theta$ by $\check{\theta}$ (the notation $\smash{\hat{\cdot}}$ is reserved for quantum mechanical operators). The estimator is a function of the data $\check{\theta}(\x)$ and returns a value for $\theta$.
A generalisation of this to an $N$-photon Fock state in one arm of the \textsc{mzi} distributes the photons binomially over the output modes~\cite{Gerry2004_book}, with a more complicated corresponding estimator. If we inject a classical coherent state with intensity $I_0$ into one mode and measure the intensities $I_1$ and $I_2$ in detectors $D_1$ and $D_2$, the estimator becomes
\begin{align}
\check{\theta} = \arccos\left(\frac{I_1 - I_2}{I_0}\right) ,
\end{align}
which is the continuous version of equation~(\ref{eqn:estimator_single_photon}).

To calculate the precision of the measurement of $\theta$ in this experiment, we use the error propagation formula in Eq.~\eqref{eq:n304riwerofjdnk}, and the measurement operator is $M = \hat{n}_1 - \hat{n}_2$, with $\hat{n}_1$ and $\hat{n}_2$ the photon number operators for the output modes. Using the probability distribution in Eq.~\eqref{eqn:s_photon_prob_dist} we find that for $N$ photons sent into the interferometer
\begin{align}
 \braket{M} = -\cos\theta \quad\text{and}\quad (\Delta M)^2 = N \sin^2\theta\, .
\end{align}
This yields a precision of 
\begin{align}
 \delta\theta = \frac{\sqrt{N}\sin\theta}{\abs{N \partial_\theta \cos\theta}} = \frac{1}{\sqrt{N}}\, .
\end{align}
This is the so-called shot noise limit for interferometry.

\subsection{Fundamentals of estimation theory}\label{subsec:exp_values_covariance}
\noindent
The problem of estimating the value of a vector of parameters $\smash{\bm{\theta} = (\theta_1, \ldots, \theta_D)^\top}$ from a set of observed data $\smash{\x = (x_1, \ldots, x_{{D}})^\top}$ is formally addressed in parameter estimation theory. Here, $\top$ denotes the transpose. Owing to experimental uncertainties and errors, the inference of parameters is related to the measurement outcomes through some conditional probability distribution $\smash{p(\bm{\theta}\vert\x)}$ that constitutes a model of the physical system under consideration. The main question is how well we can estimate these parameters. In other words, {what is the best possible precision that we can achieve?} There are generally two possibilities for $\bm{\theta}$:

\begin{enumerate}
 \item $\bm\theta$ are not random, but unknown;
 \item $\bm\theta$ are random and unknown.
\end{enumerate}
In the first case we speak of \emph{Fisher estimation}, while in the second case we have \emph{Bayesian estimation}. An advantage of using Bayesian methods is that they do not rely on asymptotics to provide optimal performance, a property not enjoyed by Fisher estimation~\cite{McNeish2016_SEM}. The optimal performance for both methods coincide in the large sample scenario~\cite{asymptotic_theory_book_2005}. Research efforts in quantum metrology and estimation theory have been dominated by work in the Fisher regime. To reflect this, we focus our review mainly on the estimation of non-random parameters. The subtle difference between Fisher and Bayesian estimation will be covered in subsection~\ref{subsec:outlook_Bayesian}.

Given that the probability density function (\textsc{pdf}) of the data $\smash{p(\x\vert\bm{\theta})}$ is known (i.e., we can model the physical process), the set of parameters $\smash{\bm{\theta}}$ may be extracted from a set of observation data $\smash{\x}$ via an estimator $\smash{\check{\bm{\theta}}}$, which is a function of the observed data only. We will see that the estimators are also used to find the estimation errors. 

An important estimator for a non-random $\smash{\bm{\theta}}$ is the maximum likelihood estimator
\begin{align}
 \caron{\bm{\theta}}(\x) = \arg \max_{\bm{\theta}} \pr{\bm{\theta}|\x}\, ,
\end{align}
where the likelihood function $\pr{\bm{\theta}|\x}$ is the probability of $\bm\theta$ being the true value given the data set $\x$. This requires a model for the process that gives us the probability $\pr{\bm{\theta}|\x}$.

While the true value of $\smash{\bm{\theta}}$ is an array of numbers, its estimates are \emph{random variables}. This is due to the probabilistic nature of the data; two runs of an experiment with equal parameters $\smash{\bm{\theta}}$ will not generate equal data due to statistical fluctuations: $\smash{\x_1 \neq \x_2}$. Hence the estimates for both runs will differ from the actual values, and not be equal to each other. Given a very large measurement data set, an estimator that generates the true values of the parameters is referred to as a \emph{perfect estimator}.  In the next subsection, we introduce the covariance matrix as a natural measure of the performance of an estimator. 

\subsection{Expectation values and covariance}\label{subsec:classical_esti_geometry}
\noindent
A natural figure of merit to quantify the performance of an estimator is the variance of a parameter estimate with respect to its true value. Hence, we can characterise the estimation performance by searching for an estimator that has the smallest variance in parameter estimates. Although various other methods to characterise the performance exist, this is a natural choice that was first introduced by H. Cram{\' e}r and C. R. Rao~\cite{Helstrom73, Cramer1999_book}. In the remainder of this section we  assume that the measured data $\smash{\x}$ is continuous without loss of generality. 

For a multi-parameter estimation of $\smash{\bm{\theta}}$, we may define the natural optimality criterion as the difference between the estimator and the true value of $\bm\theta$, $\smash{\Delta {\bm{\theta}} = \check{\bm{\theta}} - \bm{\theta}}$. However, since this relies directly on the unknown true value $\bm\theta$, it is often more convenient to define the variation in the estimates of $\smash{\bm{\theta}}$ as
\begin{align}
\delta\check{\bm{\theta}} = \check{\bm{\theta}} -  \ex_{\x\vert\bm\theta}\left[\check{\bm{\theta}}(\x)\right],
\label{eqn:variation_estimate}
\end{align}
where 
\begin{align}\label{eq:expectation-fisher}
 \ex_{\x\vert\bm{\theta}} \left[ \caron{\bm{\theta}}(\x) \right] = \int  \dif\x \; \pr{\x\vert\bm{\theta}} \, \caron{\bm{\theta}}(\x)
\end{align}
is the \emph{expectation value} of the estimator $\check{\bm{\theta}}(\x)$ with respect to the probability distribution $\pr{\x\vert\bm{\theta}}$. When the estimator is unbiased (i.e., there is no systematic error in the estimator), the expectation value approaches the true value and $\delta\check{\bm{\theta}}$ reduces to $\Delta {\bm{\theta}}$ in the limit of large data sets.

We define the covariance matrix as
\begin{align}
 \cov(\check{\bm\theta}) & = \ex_{\x\vert\bm\theta} \left[ \delta\check{\bm{\theta}}(\x)\; \delta\check{\bm{\theta}}(\x)^\top \right] \cr &=  \ex_{\x\vert\bm\theta} \left[  \left( \caron{\bm\theta} - \ex_{\x\vert\bm\theta} \left[ \caron{\bm\theta} \right] \right) \left( \caron{\bm\theta} - \ex_{\x\vert\bm\theta} \left[ \caron{\bm\theta} \right] \right)^\top  \right]\, .
\end{align}
The covariance matrix depends on the parameters via the expectation value of the estimators.

The diagonal elements of the covariance matrix are the variances of the different parameters $\theta_j$ with $j\in \{1,\ldots,D\}$. Notice that the estimator's covariance is not the same as its mean square error matrix:
\begin{align}
\begin{split}
\ex_{\x\vert\theta}\left[(\Delta \check{\bm{\theta}})(\Delta \check{\bm{\theta}})^\top\right] &= \ex_{\x\vert\theta}\left[(\check{\bm{\theta}} - \bm{\theta})(\check{\bm{\theta}} - \bm{\theta})^\top\right]\\
&= \text{Cov}[\check{\bm{\theta}}] + b(\check{\bm{\theta}})^2,
\label{eqn:mean_square_Error}
\end{split}
\end{align}
where $\smash{b(\check{\bm{\theta}}) = \ex_{\x\vert\bm\theta}[\check{\bm{\theta}}(\x)] - \bm{\theta}}$ is the bias of the estimator. We see that the mean square error is equal to the variance if and only if we have an \emph{unbiased estimator}: $\smash{b(\check{\theta}_j) = 0}$. Unbiased estimators ensure that the average of the estimates converge to the true value of the parameter: $\smash{\braket{\check{\bm{\theta}}} = \bm{\theta}}$. While we include the effect that biased estimators have on the estimation precision in subsection\textcolor{red}{~\ref{subsec:biased estimators}}, we assume unbiased estimators for the rest of the review. In the next subsection we define the expectation value and covariance of estimators.

\subsection{Bounds on the covariance matrix}\label{subsec:classical_bounds}
\noindent
It is typically not possible to calculate the exact values of the covariance matrix. We can often only hope to place some limits on the \mse, the variance, and other quantities. In this section we will use the structure of the covariance matrix and arguments from information geometry to formulate generic bounds on the (elements of the) covariance matrix. 

For notational convenience we abbreviate $\ex_{\x\vert\bm\theta}$ by $\ex$ for the remainder of this section. 
For a positive semi-definite matrix product $XY$ we have that $\ex[XY] \geq 0$. We choose 
\begin{align}
 X = f - Ag \qquad\text{and}\qquad Y = X^\top \, ,
\end{align}
with $f$ a $D$-dimensional real vector of the same size as $\bm\theta$, $g$ an $R$-dimensional real vector, and $A$ an $R\times D$ matrix that depends only on $\bm\theta$. We then find 
\begin{align}
 \ex\left[ [f(\x,\bm\theta)-Ag(\x,\bm\theta)][f(\x,\bm\theta)-A g(\x,\bm\theta)]^\top  \right] \geq 0\, .
\end{align}
Since the expectation value is linear, we can expand this into
\begin{align}
 \ex\left[ff^\top\right]- \ex\left[fg^\top A^\top\right] -\ex\left[Agf^\top\right] +\ex\left[Agg^\top A^\top\right] \geq 0\, ,
\end{align}
and extract the matrix $A$, which is a constant with respect to the expectation: 
\begin{align}
 \ex\left[ff^\top\right] \geq \ex\left[fg^\top\right]A^\top + A \ex\left[gf^\top\right] - A \ex\left[gg^\top\right] A^\top \, .
\end{align}
When we redefine 
\begin{align}
 T = \ex\left[fg^\top\right] \qquad\text{and}\qquad G = \ex\left[gg^\top\right] \, ,
\end{align}
we arrive at 
\begin{align}\label{eq:bound}
 \ex\left[ff^\top\right] \geq T A^\top + A T^\top - A G A^\top \, .
\end{align}
This is a bound on the expectation value of $f f^\top$ (which at this point can be anything that is consistent with the general definition of $f$), and to make the bound as tight as possible we must maximise the right-hand side of Eq.~(\ref{eq:bound}).

Since $T$ and $G$ do not depend on $\x$ and $\bm\theta$ (they are averaged over), we can choose $A = TG^{-1}$ (where we require that the inverse of $G$ exists; this places a restriction on $g$). This leads to the compact form 
\begin{align}
 \ex\left[ff^\top\right] \geq T G^{-1} T^\top \, .
\end{align}
Next we choose $f = \caron{\bm\theta}(\x) - \ex[\check{\bm\theta}]$, so that $\ex[ff^\top] = \cov(\check{\bm\theta})$ and therefore 
\begin{align}\label{eq:bound-general}
 \cov(\check{\bm\theta}) \geq T G^{-1} T^\top \, .
\end{align}
This expression is valid for \emph{any} estimator $\caron{\bm\theta}(\x)$. The matrix $G$ is called the \emph{information matrix}. Different definitions of $g$ (and thus $G$) will produce different bounds that may have various advantages (computational, tightness, etc.). The matrix $G$ is the expectation of a projector $gg^\top$ that may not be full rank (and therefore has no inverse). This typically happens when the estimator does not have enough degrees of freedom and therefore cannot provide estimates of all parameters.

For the above choice of $f$ the $T$ matrix becomes
\begin{align}
 T & = \ex\left[ \left( \caron{\bm\theta}(\x) -\ex[\check{\bm\theta}] \right) g^\top(\x,\bm\theta) \right] \cr
 & = \ex\left[ \caron{\bm\theta}(\x) g^\top(\x,\bm\theta) \right] - \ex\left[  \bm\theta g^\top(\x,\bm\theta) \right]\, .
\end{align}
We restrict ourselves to choices of $g$ that satisfy the condition $\ex_{\bm\theta|\x}[ g^\top(\x,\bm\theta)]=0$.
Therefore, the first term in $T$ becomes
\begin{align}
\ex_{\x,\bm\theta}\left[ \caron{\bm\theta}(\x) g^\top(\x,\bm\theta) \right] = \ex_{\x}\left[ \caron{\bm\theta}(\x) \ex_{\bm\theta|\x}\left[ g^\top(\x,\bm\theta)\right] \right] = 0\, .
\end{align}
As a result, the bounds on the covariance matrix are determined by the estimator $\caron{\bm\theta}(\x)$ and our choice of the function $g(\x,\bm\theta)$. We can choose a variety of functions $g$ to obtain different bounds, with the most famous choice of $g$ leading to the Fisher information and the Cram\'er-Rao bound.

\subsection{The Cram\'er-Rao bound}
\noindent
The choice for $g(\x,\bm\theta)$ we consider here is
\begin{align}\label{eq:bnh4t0woidfjk}
 g(\x,\bm\theta) = \frac{\partial \ln \pr{\x\vert\bm\theta}}{\partial\bm\theta}\, ,
\end{align}
which requires that both the first and second derivative of $\pr{\x\vert\bm\theta}$ exists, and is absolutely integrable. The function $g(\x,\bm\theta)$ is a natural choice in that it is additive for independent samples due to the logarithm (since independent events multiply probabilities), and the derivative measures the rate of change of the probability distribution with respect to the parameter of interest. The intuition is that a fast changing probability distribution with $\theta$ will produce a clearer change in measurement outcomes $\x$ as we vary $\theta$.

As an example, we consider a single parameter $\theta$ such that $g$ is a scalar function. Then we can evaluate $T$ explicitly via partial integration:
\begin{align}
 T = \ex_{\x\vert\theta} \left[ \theta \frac{\partial \ln \pr{\x\vert\theta}}{\partial\theta} \right] = -1\, .
\end{align}
The information matrix $G$ becomes 
\begin{align}\label{eq:cfifirst}
 G = \ex_{\x\vert\theta} \left[ \left( \frac{\partial \ln \pr{\x\vert\theta}}{\partial\theta} \right)^2\right] \equiv I(\theta),
\end{align}
which is better known as the Fisher information $I(\theta)$. The classical Fisher information is generally dependent on $\theta$. If we have a model for the process under study, we can find $\pr{\x\vert\theta}$ and calculate the classical Fisher information directly. The variance in $\theta$ can then be bounded by 
\begin{align}
 (\delta\theta)^2 \geq I^{-1}\, .
\end{align}
This is the Cram\'er-Rao Bound (\crb). It is saturated when 
\begin{align}
 \caron{\theta}(\x) - \theta =  \frac{\partial \ln\pr{\x\vert\theta}}{\partial\theta}\, .
\end{align}

For a general $D$-parameter problem, the \crb\ is a matrix inequality
\begin{align}
 [\cov(\check{\bm\theta})]_{ij} \geq \left[ I(\bm\theta)^{-1}\right]_{ij}\, ,
 \label{eq:crb}
\end{align}
where the inequality means that $\cov(\check{\bm\theta}) - I(\bm\theta)^{-1}$ is a positive semi-definite matrix. This bound is typically attainable using a Maximum Likelihood estimator in the asymptotic regime of many independent samples. The Fisher information matrix is symmetric and positive, and it can be interpreted as the metric tensor in the parameter space. In particular, this means that when we re-parameterize the space and wish to estimate the parameters $\bm\vartheta = (\vartheta_1\ldots,\vartheta_D)$, with $\vartheta_j(\bm\theta)$ some function of the original parameters $\bm\theta$, the corresponding transformed Fisher information matrix becomes 
\begin{align}
 I(\bm\vartheta) = J^\top I(\bm\theta) J \quad\text{and}\quad J_{jk} = \frac{\partial\theta_j}{\partial\vartheta_k}\, ,
\end{align}
where $J$ is the Jacobian of the parameter transformation.

An important example for the Fisher information matrix is for a Gaussian (normal) distribution, which has a closed form~\cite{Kay1993}. Let the distribution be characterised by mean values $\bm\mu(\bm\theta)$ and covariance matrix $\bm\Sigma(\bm\theta)$:
\begin{align}
 \pr{\x\vert\bm\theta} = \frac{1}{\sqrt{(2\pi)^D \det \bm\Sigma}} \exp\left[ -\frac{1}{2} (\x -\bm\mu)^\top \bm\Sigma^{-1} (\x -\bm\mu) \right]\, .
\end{align}
The Fisher information matrix then takes the following closed form 
\begin{align}
 [I(\bm\theta)]_{jk} = \frac{\partial \bm\mu^\top}{\partial\theta_j} \bm\Sigma^{-1} \frac{\partial \bm\mu}{\partial\theta_k} + \frac12 \Tr{\bm\Sigma^{-1} \frac{\partial \bm\Sigma}{\partial\theta_j} \bm\Sigma^{-1}\frac{\partial \bm\Sigma}{\partial\theta_k}}\, .
\end{align}

Often, when estimating parameters there are several \emph{nuisance} parameters that must also be estimated. We are not intrinsically interested in these parameters, but the estimators of the parameters of interest depend on them. We can separate the tuple of parameters $\bm\theta$ into genuine and nuisance parameters, $\bm\theta = (\bm\theta_g , \bm\theta_n)$. The Fisher information matrix can then be written in block form:
\begin{align}
 I(\bm\theta) = 
 \begin{pmatrix}
  I(\bm\theta_g,\bm\theta_g) & I(\bm\theta_g,\bm\theta_n) \cr
  I(\bm\theta_n,\bm\theta_g) & I(\bm\theta_n,\bm\theta_n)
 \end{pmatrix}\, .
\end{align}
The \crb\ is still given by the inverse of $I$, but now we can use the inverse of a block matrix,
\begin{align} P_A
 \begin{pmatrix}
  A & D^\top \cr D & B
 \end{pmatrix}^{-1} P_A
 = \left( A - D^\top B D \right)^{-1} ,
\end{align}
where $P_A$ is the projection operator onto the subspace occupied by $A$, to establish how the nuisance parameters affect the \crb:
\begin{align}
 \cov(\check{\bm\theta}_g) \geq \left[ I(\bm\theta_g,\bm\theta_g) - I(\bm\theta_g,\bm\theta_n) I(\bm\theta_n,\bm\theta_n)^{-1} I(\bm\theta_n,\bm\theta_g)  \right]^{-1}\, .
\end{align}
In other words, the nuisance parameters \emph{lower} the Fisher information matrix compared to the Fisher information matrix for the genuine parameters alone, as expected.

There are other choices for the function $g(\x,\bm\theta)$ that lead to different bounds. For more details on classical and Bayesian bounds, see Van Trees and Bell (2007)\cite{Trees2017_book}.


\section{Geometry of estimation theory}\label{sec:classical_fi_geometry}
\noindent
The Fisher information in Eq.~(\ref{eq:cfifirst}) followed from our choice of $g(\x,\theta)$. In this section we will give an intuitive geometric derivation~\cite{Wootters81,Braunstein94} that will help us with the derivation of the quantum Fisher information in the next section. We relate parameter estimation to methods of distinguishing probability distributions, including the Fisher information and the relative entropy.

\subsection{The probability simplex}\label{subsec:qfi_qcrb_geometric}
\noindent
Any experiment used to infer a value of $\bm\theta$ will return different measurement outcomes $\x=(x_1,\ldots,x_D)$. Assuming that different values of $\bm\theta$ produce variations in the measurement outcomes (otherwise this particular measurement would not be useful in extracting a value of $\bm\theta$), we can posit a probability distribution $\pr{\x\vert\bm\theta}$, which may originate from some physical model. The problem of finding the value of $\bm\theta$ is then reduced to telling the difference between two probability distributions $\pr{\x\vert\bm\theta_A}$ and $\pr{\x\vert\bm\theta_B}$. In other words, how many times do we have to sample the system (i.e., what is $D$) in order to tell the difference between $\pr{\x\vert\bm\theta_A}$ and $\pr{\x\vert\bm\theta_B}$?

\begin{figure}[t!]
\begin{center}
\subfloat[{}{The simplex.}]{\includegraphics[width=0.45\columnwidth]{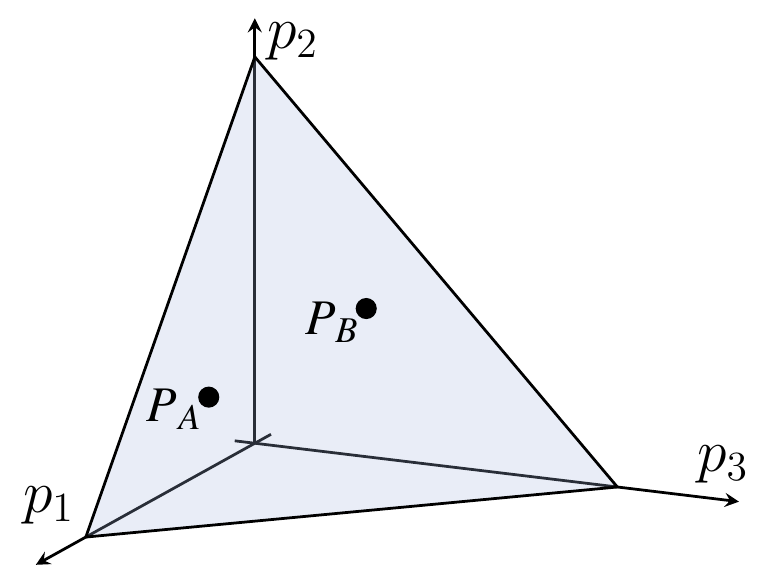}\label{fig:pd1a}} \hspace{10pt}
\subfloat[{}{Paths in the simplex.}]{\includegraphics[width=0.45\columnwidth]{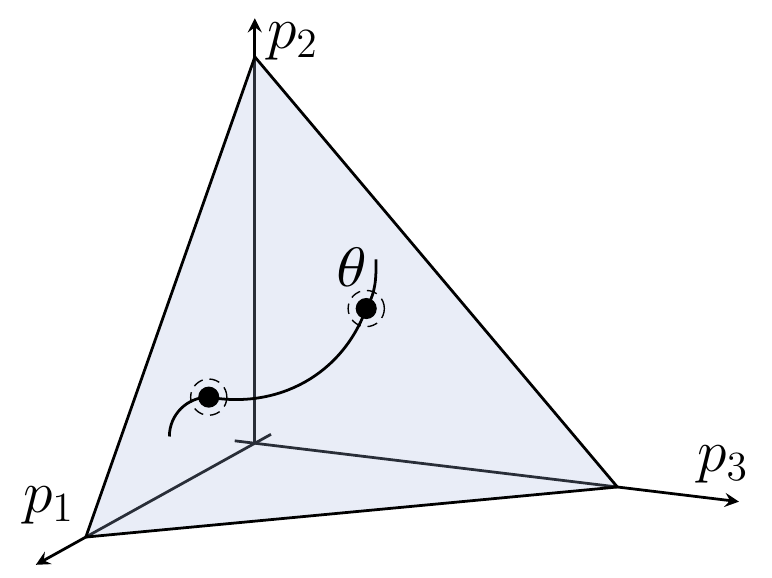} \label{fig:pd1b}}
\caption{Fig.~\ref{fig:pd1a}: the probability simplex where the measurement has three outcomes with probabilities $p_1$, $p_2$, and $p_3$. Every probability distribution is represented by a point in the simplex ($P_A$ and $P_B$). Fig.~\ref{fig:pd1b}: $\theta$ parametrises a path through the probability simplex.}
\label{fig:pd}
\end{center}
\end{figure}

The probability distributions $\pr{\x\vert\bm\theta}$ form a space called a \emph{probability simplex }(see Fig.~\ref{fig:pd} for a simple example). The probability distributions $\pr{\x\vert \theta}$ for a single parameter $\theta$ typically form a curve through the simplex that is parametrised by $\theta$. In order to tell how many measurements we need to make in order to distinguish two probability distributions on the curve we need some distance measure (a metric) on the simplex that fits naturally with statistics. This metric can then be used to tell how far away two distributions are from each other. In turn this will allow us to infer how many measurements we need to make to distinguish between the two distributions. In what follows we first specifically consider a single parameter $\theta$ for simplicity.

In its general discrete form we can write the infinitesimal distance $\dif s$ on the simplex in terms of incremental probability changes $\dif p$ and a metric $h$:
\begin{align}
 \dif s^2 = \sum_{jk} h_{jk}\, \dif p^j \dif p^k\, ,
\end{align}
where we used contravariant elements for the probability increments and the covariant form of the metric. The metric tensor obeys
\begin{align}\label{eq:048wuoeirdfj}
 h_{jk} h^{kl} = \delta_{jl}\, ,
\end{align}
where we use Einstein's summation convention and where $\delta_{jl}$ is the Kronecker delta. We have to derive a natural form for $h$. 

\subsection{The Fisher-Rao metric and statistical distance}\label{subsec:statistical_geometry}
\noindent
For this section, we follow the procedure in Bengtsson and $\dot{\mathrm Z}$yczkowski~\cite{Bengtsson2008} to formulate an appropriate metric on the space of probability distributions. A similar procedure is taken by Kok and Lovett~\cite{Kok2010}. A natural scalar product on the simplex and its dual space of classical random variables forms an expectation value:
\begin{align}
 \braket{A} = \sum_j A_j p^j\, .
\end{align}
The correlation between two classical random variables $A$ and $B$ is then 
\begin{align}
 \braket{AB} = \sum_j A_j B_k h^{jk} = \sum_j A_j B_j p^j\, .
\end{align}
Using the relation in Eq.~\eqref{eq:048wuoeirdfj} we find that 
\begin{align}
 h_{jk} = \frac{\delta_{jk}}{p^j}\, ,
\end{align}
which leads to the so-called \emph{Fisher-Rao metric} \textsc{(fr)}
\begin{align}
\dif s_{\rm FR}^2 = \sum_{jk}\dif p^j\dif p^k h_{jk} = \sum_j\frac{\dif p^j\dif p^j}{p_j} .
\label{eqn:fisher_rao_metric}
\end{align}
This defines the statistical distance between two probability distributions in the {probability simplex}. The generalisation of this metric for {continuous probability density functions} (\textsc{pdf}) can be written as~\cite{Bengtsson2008}
\begin{align}
h_{ab} = \frac{1}{4}\int_\Omega \dif \x\frac{\partial_a p(\x)\, \partial_bp(\x)}{p(\x)},
\label{eqn:continuous_fisher_rao_metric}
\end{align}
where $\Omega$ defines a {finite dimensional sub-manifold} of the probability simplex with coordinates $\theta^a$, and $\partial_a \equiv \partial/\partial \theta^a$. For simplicity we will mostly use the discrete form in the remainder of this section. 

\begin{figure}[t!] 
\begin{center}
\includegraphics[width =0.58\columnwidth]{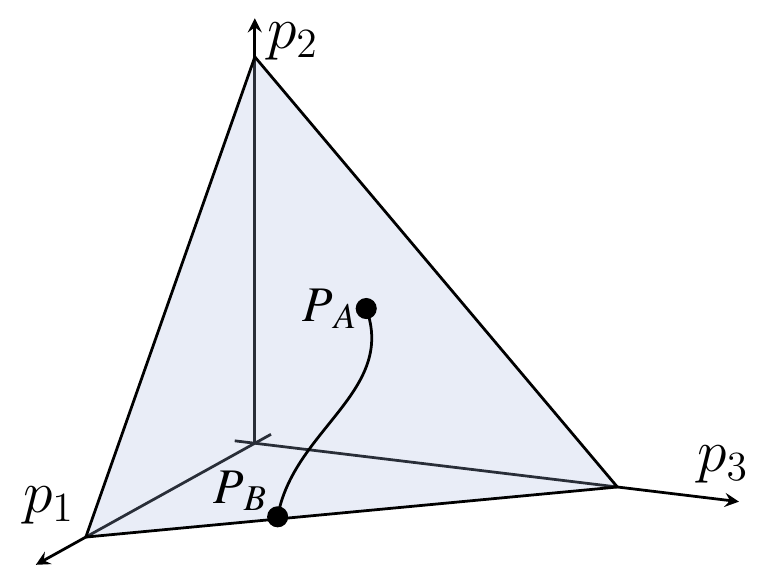}
\caption{The distance between probability distributions $P_A$ and $P_B$ diverges when one of them ($B$) lies on the hull of the simplex.}
\label{fig:pdinfty}
\end{center}
\end{figure}

The probability simplex with the \textsc{fr} metric exhibits strong curvature. Note that the statistical distance in Eq.~\eqref{eqn:fisher_rao_metric} diverges when one of the probabilities $p_j$ tends towards zero. This gives us a clue how to interpret the distance between two distributions: when the probability of one of the measurement outcomes is strictly zero, then obtaining that measurement outcome will allow us to infer \emph{with certainty} that the system is governed by the other probability distribution (see Fig.~\ref{fig:pdinfty}).

Next, we consider the displacement $\dif s$ in the probability simplex along a line element $\dif\theta$. We can write
\begin{align}
 \frac{\dif s^2}{\dif\theta^2} = \sum_j \frac{1}{p^j}\frac{(\dif p^j)^2}{\dif\theta^2} = \sum_j p_j \left( \frac{\partial\ln p^j}{\partial\theta} \right)^2\, .
\end{align}
Comparing this with Eq.~(\ref{eq:cfifirst}) we see that this is the Fisher information
\begin{align}\label{eq:t480u2wrufdjn}
 I(\theta) =  \left(\frac{\dif s}{\dif\theta}\right)^2 ,
\end{align}
and in the case of continuous data sets
\begin{align}
 I(\theta) = \int \dif\x\;  \pr{\x\vert\theta} \left( \frac{\partial\ln \pr{\x\vert\theta}}{\partial\theta} \right)^2 \, .
\end{align}
Therefore, the Fisher information measures how fast the probability distribution changes along paths parametrised by $\theta$. In order to tell the difference between two values $\theta$ and $\theta'$ a higher Fisher information will be beneficial. We can think of the Fisher information $I(\theta)$ as the average amount of information about $\theta$ in a single measurement.

For small finite distances induced by a shift $\delta\theta$, and starting at $\theta$, we can express the statistical distance as a Taylor expansion
\begin{align}
 s(\theta+\delta\theta) = s(\theta) + \delta\theta \left. \frac{\dif s}{\dif\theta} \right|_{\theta} + O(\delta\theta^2)\, .
\end{align}
Defining $\delta s = s(\theta+\delta\theta) - s(\theta)$, we obtain $\delta s = \delta\theta \sqrt{I(\theta)}$ up to first order in $\delta\theta$. We postulate that two probability distributions are distinguishable after $N$ measurements on independent identically prepared systems if the resulting distance crosses some threshold $\alpha$:
\begin{align}\label{eq:sd-req}
 N \delta s^2 \geq \alpha\, ,
\end{align} 
where usually we set $\alpha=1$. We can eliminate $\delta s$ from Eq.~(\ref{eq:sd-req}) to obtain
\begin{align}
 (\delta\theta)^2 \geq \frac{1}{N I(\theta)}\, ,
\end{align}
which is strongly reminiscent of the Cram\'er-Rao bound. The difference is that $\delta\theta$ here is the segment of the path in the probability simplex, rather than the variance in the estimator of $\theta$. Nevertheless, Eq.~(\ref{eq:sd-req}) is a powerful method for working out the number of measurements that are required to see a difference $\delta\theta$ in the data.

Eq.~\eqref{eqn:fisher_rao_metric} can also be re--expressed in a more convenient way if we introduce a new coordinate system $\smash{(x^j)^2 = p^j}$. This transforms the \textsc{fr} metric to the Euclidean metric:
\begin{align}
\dif s^2 = 4 \sum_j \dif x^j\dif x^j .
\label{eqn:fisher_rao_metric_2}
\end{align}
The factor 4 can be absorbed in a change of units, but we choose to keep it here since it will reappear in the quantum extension later on. While in classical estimation theory the occurrence of (real) probability amplitudes $x_i = \sqrt{p_i}$ is something of a curiosity, in the quantum extension to estimation theory this allows us to construct the Fubini-Study metric\cite{Facchi2010}.

\subsection{Relative entropy}\label{subsec:fi_entropy}
\noindent
In addition to distance measures, probability distributions are conveniently characterised by entropic functions. The Shannon entropy of a random variable $P$, 
\begin{align}
S({P})  = -\sum_{i=1}^N p_i\log_2 p_i,
\label{eqn:shannon_entropy}
\end{align}
measures the average amount of information in an event sampled from a system described by the probability distribution ${p_i}$. The units are bits, and in the remainder of this review, logarithms are base 2, unless stated otherwise. 

To compare two probability distributions, we can make use of the {Kullback-Leibler}, or \emph{relative} entropy
\begin{align}
D \infdiv{P}{Q}  = \sum_{i=1}^N p_i\log \frac{p_i}{q_i} .
\label{eqn:kl_entropy}
\end{align}
While this is not a metric since it is not symmetric under exchange of $p_i$ and $q_i$, it remains sufficient as a distinguishability measure between the two distributions. Eq.~\eqref{eqn:kl_entropy} describes the information gain when a prior distribution $Q$ is updated to the posterior distribution $P$. 

A Taylor expansion to first order of the logarithmic term of the relative entropy in Eq.~\eqref{eqn:kl_entropy} yields
\begin{align}
D\infdiv{P}{P + \dif P}  = \sum_{i=1}^Np^i\ln \frac{p^i}{p^i + \dif p^i} \approx \sum_i \frac{\dif p^i \dif p^i}{2p^i}.
\label{eqn:fisher_rao_metric_entropy}
\end{align}
The last term is identical to the statistical distance in Eq.~\eqref{eqn:fisher_rao_metric} up to a factor two. The Fisher information must therefore be closely related to the relative entropy $D$ in Eq.~(\ref{eqn:kl_entropy}). Assuming that the two distributions in the relative entropy are connected by a curve $\theta$, we may label the distributions by $\bm\theta$ and $\bm\theta'$. We find that the Fisher information matrix can be approximated by the second derivative of the relative entropy
\begin{align}\label{eq:hgiruwoesdjkn}
 [{I}(\bm\theta)]_{ij} &= \left(\frac{\partial^2}{\partial\theta'_i \partial\theta'_j}D \infdiv{\bm\theta}{\bm\theta'} \right)_{\theta'=\theta} \cr
 &= -\int \dif\x\; \pr{\x,\bm\theta} \frac{\partial^2 \log \pr{\x,\bm\theta}}{\partial\theta_i \partial\theta_j} .
\end{align}
Conversely, the relative entropy can be written in terms if the Fisher information matrix
\begin{align}
 D\infdiv{\bm\theta}{\bm\theta'} = \frac{1}{2}(\bm\theta'-\bm\theta)^\top\, {I}(\bm\theta)\, (\bm\theta'-\bm\theta)\, ,
\end{align}
up to higher order corrections in $(\bm\theta'-\bm\theta)$. The relative entropy has a number of advantages over the Fisher information in that it is not affected by changes in parameterisation, it can be used even if the distributions are not all members of a parametric family, and fewer smoothness conditions on the probability densities are needed.


\section{Single parameter quantum estimation}\label{sec:qe_theory}
\noindent
In this section we review single-parameter quantum estimation theory and derive the quantum Cram{\'e}r-Rao bound. We show how the quantum fisher information can be obtained as a limiting case of the classical Fisher information, and we provide an interpretation for the general parameter estimation scheme illustrated in Fig.~\ref{pic:parameter_estimation}. We give various closed forms for the quantum Fisher information. 

\noindent

\subsection{Quantum model of precision measurements}\label{subsec:quantum_esti_geometry}
\begin{figure}[t!] 
\begin{center}
\includegraphics[width =0.98\columnwidth]{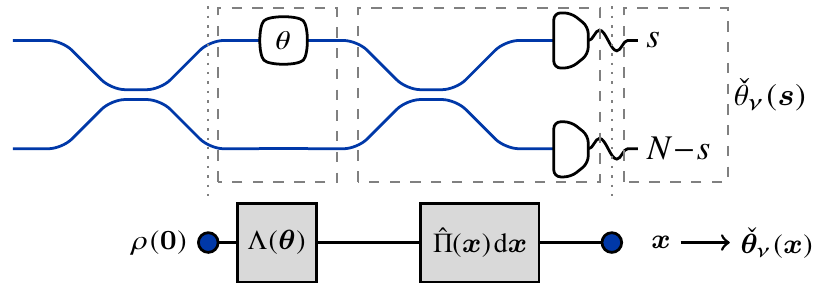}
\caption{General channel parameter estimation scheme with the possibility of adaptive control, shown in parallel with the phase estimation procedure using a Mach-Zehnder interferometer. The quantum channel $\smash{\Lambda({\theta})}$ imparts  parameters $\bm\theta$ on the input probe state. The probe is measured by an operator of the form $\smash{{\Pi}(\x)\dif \x}$ and distributes estimates according to Born's rule. By data processing the measurement outcomes, we obtain our estimate $\smash{\check{{\theta}}}$. The goal of quantum metrology is in finding both the optimal probe state and observable that minimises the covariance matrix of unbiased estimates.}
\label{pic:parameter_estimation}
\end{center}
\end{figure}

\noindent
Any estimation strategy is described through a probe preparation stage with state $\smash{\rho({0})}$, followed by an evolution that imprints the parameters of interest through a quantum channel $\smash{\rho({\theta}) = \Lambda[\rho({0})]({\theta})}$, and a measurement state by a self-adjoint observable $\smash{{X} = \int \dif\x \,  \x\; {\Pi}(\x)}$. This archetypal schema is illustrated in Fig.~\ref{pic:parameter_estimation} (in principle, a feedback mechanism can be included). For any given interaction, this protocol describes a two-step optimisation problem; an experimenter must make a suitable choice of probe state that is sensitive to changes in the parameters to assimilate maximal information, and they must make an appropriate measurement that maximises the information extracted from the probe. Analytically, this can be modelled by describing the evolved state of the system through $\smash{\rho({\theta})}$, and by associating a positive operator valued measure \textsc{(povm)} $\smash{{\Pi}(\x)\dif\x}$, which describes the measurement yielding the data $\x$. The probability distribution $\smash{\pr{\x\vert{\theta}}}$ is then given by Born's rule 
\begin{align}
 \pr{\x\vert{\theta}} \dif \x =\Tr{{\Pi}(\x)\rho({\theta})} \dif \x,
\label{eqn:born_rule}
\end{align}
where  $\smash{\int \dif \x \, {\Pi}(\x) = {\mathbbm{I}}}$. Born's rule gives the probability distribution function (\textsc{pdf}) that distributes the measurement outcomes $\smash{\x}$, given the parameterisation $\smash{{\theta}}$. The state captures uncertainties associated with the state-preparation procedure, while the \textsc{povm} captures those associated with the measurement stage. Together with Born's rule, they model the probabilistic nature of the measurement data. 

The bounds in the previous section were derived for probability distributions $\pr{\x\vert\theta}$. This immediately generalises to the case where the probability distribution results from some quantum mechanical process described above through Born's rule. A natural question is {what is the best possible precision in $\theta$ that can be obtained from (many copies of) $\rho(\theta)$?} By optimising over the probe states and all measurement strategies, the result to this question is the quantum Cram\'{e}r-Rao bound (\textsc{qcrb}), which lower bounds the variance of any unbiased estimator that maps measured data from quantum measurements to parameter estimations~\cite{Helstrom1976}. It is of fundamental interest since it can be regarded as an intrinsic property of the system, and is determined entirely by the quantum Fisher information  (\textsc{qfi}), which depends only on the state $\rho(\theta)$. In the following section, we define the quantum mechanical version of the Fisher information. We can derive this via geometrical arguments on the probability space similar to section~\ref{sec:classical_fi_geometry}.

\subsection{The quantum Fisher information}\label{subsec:quantum_fi_geometry}
\noindent
We will find an expression for the \qfi\ inspired by the above derivation of the classical Fisher information \cite{Braunstein94,Braunstein95}. Notable contributions to this extension were made by Wootters\cite{Wootters81}, Hilgevoord and Uffink\cite{Hilgevoord1991_FoP}, and Braunstein and Caves\cite{Braunstein94}.  We restrict ourselves again to the case of a single parameter $\theta$. In the quantum mechanical case the expectation value of an operator is given by the Born rule
\begin{align}
 \braket{A} = \tr{\rho A}\, ,
\end{align}
where $\rho$ is the quantum state of the system and $A$ is a self-adjoint operator. In order to find the single parameter \qfi\ we again define a metric via the correlation between two observables $A$ and $B$. There is, however, a complication. Since observables in quantum mechanics generally do not commute, the product $AB$ is often not self-adjoint and cannot be considered an observable: $(AB)^\dagger = B^\dagger A^\dagger = BA \neq AB$. A natural remedy for this problem is to use the anti-commutator $\{A,B\} = AB+BA$ as the observable for the correlation:
\begin{align}\label{eq:bghuo04woesjk}
 \Braket{\mbox{$\frac12$}\{A,B\}} & = \frac12 \Tr{\rho \{A,B\}} =  \frac12 \Tr{A \{\rho,B\}} \cr & \equiv \Tr{A\, \mathscr{R}_\rho(B)}\, ,
\end{align}
where we included a factor $\frac12$ for normalisation, and we defined the super-operator 
\begin{align}
 \mathscr{R}_\rho(B) \equiv \frac12 \{ \rho, B\}\, ,
\end{align}
which plays the role of the metric with raised indices. We can find the statistical distance by constructing the lowering operator $\mathscr{L}_\rho = \mathscr{R}_\rho^{-1}$ such that $\mathscr{L}_\rho (\mathscr{R}_\rho(B)) = B$. Explicit verification shows that $\mathscr{L}_\rho$ satisfies
\begin{align}\label{eq:sld-metric}
 \mathscr{L}_\rho(B) = 2\sum_{jk} \frac{B_{jk}}{p_j+p_k} \ket{j}\bra{k}\, ,
\end{align}
where $\{\ket{j}\}$ is the eigenbasis of $\rho$, with eigenvalues $p_j$, and $B_{jk}$ is the matrix element of $B$ corresponding to $j$ and $k$. It is similarly easy to show that $\mathscr{R}_\rho (\mathscr{L}_\rho(B)) = B$, which proves that $\mathscr{R}_\rho$ and $\mathscr{L}_\rho$ are each others' inverse operations. Note that in this form the set of pure states is excluded from the space of density matrices. However, we will show in section \ref{subsubsec:sld_info} that we can also include pure states.

The lowering operator allows us to construct a scalar product between density operators $\tr{\sigma_1 \mathscr{L}_\rho(\sigma_2)}$. In particular, for small changes in the density operator $\dif\rho$ we can construct the infinitesimal quantum statistical distance $\dif s_Q$
\begin{align}\label{eq:gu480woesj}
 \dif s_Q^2 = \Tr{\dif\rho \mathscr{L}_\rho(\dif\rho) }\, .
\end{align}
The \qfi\ is then given by the change of the quantum statistical distance along the curve $\theta$: 
\begin{align}\label{eq:qfi_lowering}
 I_Q(\theta) \equiv \left( \frac{\dif s_Q}{\dif \theta} \right)^2 = \Tr{\partial_\theta \rho\, \mathscr{L}_\rho(\partial_\theta\rho) }\, , 
\end{align}
where again $\partial_\theta \equiv \partial/\partial\theta$. 

The \qfi\ in Eq.~(\ref{eq:qfi_lowering}) is a Riemannian metric on the quantum state space. Nagaoka~\cite{Nagaoka82} and Braunstein and Caves~\cite{Braunstein94} show that this quantum Fisher information can be attained by a judicially chosen measurement. In other words, consider a measurement $M$ expressed in terms of its \povm\ elements:
\begin{align} 
 M = \int \dif\x\; m(\x)\; {\Pi}(\x)\, ,
\end{align}
where $m(\x)$ is the real eigenspectrum of $M$. For the optimal choice of $M$ the \qfi\ in Eq.~(\ref{eq:qfi_lowering}) coincides with the Fisher information in Eq~(\ref{eq:t480u2wrufdjn}), where $\pr{\x\vert\theta} = \tr{{\Pi}(\x) \rho(\theta)}$ is the probability distribution over the measurement outcomes of $M$. The two necessary and sufficient conditions for $M$ to be optimal are that for all $\x$ 
\begin{align}
\begin{split}
(1) & \qquad \Im \Tr{\rho {\Pi}(\x) \mathscr{L}_\rho(\partial_\theta\rho)} = 0 \, , \cr
(2) & \qquad \frac{\sqrt{{\Pi}(\x)}\sqrt{\rho}}{\tr{{\Pi}(\x) \rho}} = \frac{\sqrt{{\Pi}(\x)}\, \mathscr{L}_\rho(\partial_\theta\rho)\, \sqrt{\rho}}{\tr{\rho {\Pi}(\x) \mathscr{L}_\rho(\partial_\theta\rho)}} \, ,
\end{split}
\end{align}
where Im denotes the imaginary part.

The \qfi\ has a number of interesting properties~\cite{Toth14,Yu13}. First, it is convex in the quantum states This means that for any two states $\rho_1$ and $\rho_2$ we have 
\begin{align}
 I_Q(p_1 \rho_1 + p_2 \rho_2) \leq p_1 I_Q(\rho_1) + \leq p_2 I_Q(\rho_2) \, ,
\end{align}
with probabilities $p_1$ and $p_2$ such that $p_1+p_2 = 1$. Moreover, the \qfi\ is additive for independent measurements
 \begin{align}
  I_Q(\rho_1 \otimes \rho_2,\theta) = I_Q(\rho_1,\theta) + I_Q(\rho_2,\theta) \, , 
 \end{align}
as well as for direct sums:
 \begin{align}
  I_Q(p_1\rho_1 \oplus p_2 \rho_2,\theta) = p_1 I_Q(\rho_1,\theta) + p_2 I_Q(\rho_2,\theta) \, .  
 \end{align}
Second, for unitary evolutions $U = \exp(-i\theta G)$ generated by a Hermitian operator $G$, the \qfi\ does not depend on the position along the orbit of $U$:
 \begin{align}
  I_Q(U\rho U^\dagger, G) = I_Q(\rho,U^\dagger G U) = I_Q(\rho,G) \, .
 \end{align}
It does not increase under \textsc{cptp} maps $\mathscr{E}[\rho]$ that do not depend on the parameter of interest:
  \begin{align}
  I_Q(\rho,\theta) \geq I_Q\left(\mathscr{E}[\rho],\theta\right) ,
 \end{align} 
and tracing out a subsystem cannot increase the \qfi:
 \begin{align}
  I_Q(\rho,G_1\otimes \identity_2) \geq I_Q(\tr[2]{\rho},G_1) \, .
 \end{align}
 Adding white noise to the state of $N$ particles, each described in a $d$-dimensional Hilbert space, is equivalent to mixing in the identity matrix, such that 
 \begin{align}
  \rho = p \ket{\psi}\bra{\psi} + (1-p) \frac{\identity}{d^N}\, . 
 \end{align}
The \qfi\ then becomes
 \begin{align}
  I_Q(\rho,\theta) = \frac{p^2}{p+\frac{1-p}{2} d^{-N}} I_Q(\psi,\theta) \, .
 \end{align} 
Third, we can write the \qfi\ for a unitarily evolved state in terms of the generator $G$ as
\begin{align}
  I_Q(\theta) & = 4 \sum_{j,k} p_j \abs{\braket{j|G|k}}^2 - 8 \sum_{j,k} \frac{p_j p_k}{p_j+p_k} \abs{\braket{j|G|k}}^2 \\
  & = \braket{G^2} - 8 \sum_{j,k} \frac{p_j p_k}{p_j+p_k} \abs{\braket{j|G|k}}^2\, ,
\end{align} 
where we used the diagonal form of the density operator $\rho = \sum_j p_j \ket{j}\bra{j}$. The \qfi\ does not depend on the diagonal elements of $G$, so we can write
 \begin{align}
  I_Q(\rho,G) = I_Q(\rho,G+D)\, ,
 \end{align}
 where $D$ is any diagonal matrix. Another way to express the \qfi\ for mixed states is via purifications~\cite{Escher2011}:
\begin{align}\label{eq:h9842woirufhsk}
 I_Q (\rho,\theta) = \min_{\Psi} \left( \Braket{\dot\Psi|\dot\Psi} - \Abs{\Braket{\dot\Psi|\Psi}}^2\right)\, ,
\end{align}
where $\ket{\Psi}$ is a purification of $\rho$. Another form is due to Fujiwara and Imai~\cite{Fujiwara08}:
\begin{align}\label{eq:fvnooreisjdkxc}
 I_Q (\rho,\theta) = \min_{\Psi} \Braket{\dot\Psi|\dot\Psi} \, .
\end{align}
Eqs.~\eqref{eq:h9842woirufhsk} and \eqref{eq:fvnooreisjdkxc} are equivalent for the purification that achieves the minimum. In other words, $\abs{\braket{\dot\Psi|\Psi}}^2=0$ for the minimal purification state $\ket{\Psi_{\rm min}}$.

\subsection{Distance measures in quantum estimation}\label{subsec:quantum_fi_distance}
\noindent
We momentarily return to the role of distance measures in the definition of the \qfi, and its relation to other distance functions in classical and quantum parameter estimation. 

The classical statistical distance induces a curvature in the probability simplex that can be removed by introducing probability amplitudes, as shown in Eq.~(\ref{eqn:fisher_rao_metric_2}). Introducing a complex phase in the amplitudes, $x_j \to x_j {\rm e}^{i\phi_j}$, allows us to relate these amplitudes to normalised vectors $\ket{\psi}$ in a complex Hilbert space $\mathscr{H}$ whose distance to other vectors in $\mathscr{H}$ is then given by the \emph{angle} between them. In infinitesimal form, this gives the Wootters distance~\cite{Wootters81,Hubner92}
\begin{align}
 \dif s^2_{\rm W} = \left( \arccos \abs{\braket{\psi|\psi+\delta \psi}} \right)^2 ,
\end{align}
and the metric is called the Fubini-Study metric on the space of projectors $R(\mathscr{H}) = \mathscr{H}\otimes \mathscr{H}$. The pullback metric from $R(\mathscr{H})$ to $\mathscr{H}$ is given by~\cite{Facchi2010}
\begin{align}
 h_{\rm FS} =  \frac{\braket{\partial_\theta\psi|\partial_\theta\psi}}{\braket{\psi|\psi}} - \frac{\braket{\partial_\theta\psi|\psi}\braket{\psi|\partial_\theta\psi}}{\braket{\psi|\psi}^2} ,
\end{align}
where $\ket{\partial_\theta\psi} \equiv \partial_\theta\ket{\psi}$. Assuming $\braket{\psi|\psi}=1$ then up to a factor 4 this equals the \qfi\ for pure states\cite{Fujiwara94}
\begin{align}\label{eq:hge0w4reosjndk}
 I_Q(\theta) = 4 h_{\rm FS} = 4\left[ \braket{\partial_\theta\psi|\partial_\theta\psi} - \abs{\braket{\partial_\theta\psi|\psi}}^2 \right]\, .
\end{align}
When we extend the Fubini-Study metric to density operators in $R(\mathscr{H})$ we obtain the Bures metric~\cite{Amari93, Amari2016_book,Facchi2010} given in Eq.~\eqref{eq:gu480woesj}
\begin{align}
 4 \dif s_{\rm Bures}^2 =   \dif s_Q^2 =  \Tr{\dif\rho \mathscr{L}_\rho(\dif\rho) }\, .
\end{align}
The origin of the factor 4 is the same as in Eq.~\eqref{eqn:fisher_rao_metric_2}.

The Wootters distance between two quantum states $\ket{\psi}$ and $\ket{\phi}$ is closely related to the fidelity $F = \abs{\braket{\psi|\phi}}^2$ between the states, $\dif s^2_{\rm W} = ( \arccos\sqrt{F} )^2$. The corresponding fidelity for mixed states is the Uhlmann fidelity\cite{Uhlmann76}
\begin{align}
 F \left(\rho, \sigma\right) = \left(\text{Tr} \sqrt{\sqrt{\rho}\sigma\sqrt{\rho}}\right)^2 ,
\label{eqn:quantum_fieldity}
\end{align}
and we can express the quantum statistical distance $\dif s_Q^2$ as 
\begin{align}
 \dif s_Q^2(\rho,\rho+\dif\rho) = 8 \left[ 1-\sqrt{F(\rho,\rho+\dif\rho)} \right] .
\end{align}
The \qfi\ can then be written in terms of the quantum fidelity as~\cite{Braunstein94} 
\begin{align}\label{eq:gh94weuohisd}
 I_Q(\theta) = \frac{8}{\delta\theta^2} \left[ 1-\sqrt{F(\rho,\rho+\delta\rho)} \right] ,
\end{align}
which will allow for analytic expressions in a variety of cases.

\subsection{The Symmetric Logarithmic Derivative}\label{subsubsec:sld_info}
\noindent
The classical Fisher information is the expectation value of the squared derivative of the logarithm of the probability distribution, as shown in Eq.~\eqref{eq:cfifirst}. In quantum estimation theory, we can define a similar quantity, now an operator $L_\theta$, that is called the \emph{symmetric logarithmic derivative}~\cite{Helstrom67,Helstrom68} (\sld), which is equal to the lowering operator in Eq.~\eqref{eq:sld-metric}  of the derivative of $\rho$
\begin{align}\label{eq:sld_identity}
 L_\theta = \mathscr{L}_\rho (\partial_\theta \rho)\, .
\end{align}
Moreover, the \sld\ is implicitly defined by the relation
\begin{align}
\partial_\theta\rho = \frac{1}{2} ( \rho L_\theta + L_\theta \rho) \, .
\label{eqn:904ureihdbfdfsjeo}
\end{align}
The symmetric form of this definition is directly related to the symmetrized definition of the correlation between quantum observables $\frac12\braket{\{A,B\}}$ in Eq.~\eqref{eq:bghuo04woesjk} via the identification with $\mathscr{L}_\rho (\partial_\theta \rho)$, which is a metric operator derived directly from the inner product $\frac12\braket{\{A,B\}}$. Some intuition for the definition of $L_\theta$ can be gained from the classical logarithmic derivative $L_{\rm cl} = \partial_\theta \log \pr{\x|\theta}$, which gives the relation $\partial_\theta \pr{\x|\theta} = \pr{\x|\theta} L_{\rm cl}$ (Note that $L_{\rm cl}$ is equal to the function $g(\x,\theta)$ in Eq.~\eqref{eq:bnh4t0woidfjk}). Replacing the classical probability distribution $\pr{\x|\theta}$ with a density operator introduces an ambiguity in the operator order of $\rho$ and $L_\theta$, which is resolved by taking the anti-commutator in Eq.~\eqref{eqn:904ureihdbfdfsjeo}. 

To prove relation Eq.~\eqref{eq:sld_identity}, we write $L_\theta$ in the eigenbasis of $\rho$:
\begin{align}\label{eq:bgwoejkdf}
 L_\theta = \sum_{jk} L_{jk} \ket{j}\bra{k} \quad\text{and}\quad \rho = \sum_l p_l \ket{l}\bra{l}\, ,
\end{align}
and construct the operator form of $L_\theta$
\begin{align}
 \frac12 \left( L_\theta \rho + \rho L_\theta \right) 
 & =  \frac12 \sum_{jk} \left( p_j + p_k  \right) L_{jk} \ket{j}\bra{k} \cr
 & = \sum_{jk} (\partial_\theta \rho)_{jk} \ket{j}\bra{k} \, .
\end{align}
Each matrix element of $\partial_\theta \rho$ must  match that of $\frac12(L_\theta \rho + \rho L_\theta)$, and therefore we have 
\begin{align}\label{eq:sldeigenbasis}
 L_{jk} = \frac{2 (\partial_\theta \rho)_{jk}}{p_j+p_k}\, .
\end{align}
Substituting this back into $L_\theta$ we see that the \sld\ in Eq.~(\ref{eq:sldeigenbasis}) takes the same form as $\mathscr{L}_\rho (\partial_\theta \rho)$ in Eq.~(\ref{eq:sld-metric}), and the identity in Eq.~(\ref{eq:sld_identity}) is proved.
This identification unifies the geometric interpretation of the \textsc{qfi} with its interpretation as a limiting case of the classical Fisher information in the next subsection.

The \qfi\ can now be written in terms of the \sld\ $L_\theta$ by noting that
\begin{align}\label{eq:qfi}
 I_Q(\theta) & = \Tr{(\partial_\theta \rho) \mathscr{L}_\rho(\partial_\theta \rho)} = \Tr{(\partial_\theta \rho) L_\theta} \cr & = \frac12 \Tr{\left( L_\theta \rho + \rho L_\theta \right) L_\theta} \cr & = \Tr{\rho L_\theta^2}\, . 
\end{align}
In the eigenbasis of the density operator $\rho$ and using Eq.~(\ref{eq:qfi_lowering}), the \qfi\ can then be written as 
\begin{align}\label{eq:qfieigenbasis}
 {I_Q(\theta) = 2 \sum_{jk} \frac{\abs{\braket{j|(\partial_\theta \rho)|k}}^2}{p_j + p_k}}\, .
\end{align}
The sum extends over all $j$ with nonzero $p_j$. For vanishing probabilities $p_j$, the \textsc{qfi} becomes ill-defined and we have to find an alternative way to define it~\cite{Safranek2017_PRA,Seveso2019_arxiv}, such as through Eq.~\eqref{eq:hge0w4reosjndk}. Alternatively, we review a regularisation procedure in section~\ref{sec:research_applications} which can be used to determine the \textsc{qfi} for pure states from expressions valid for mixed states. From Eq.~\eqref{eq:qfieigenbasis}, we observe that the \qfi\ is dependent on the quantum state and its derivative only, and not on the measurement that is performed. In this sense, the \qfi\ is a property of the state. 

We also note that the \qfi\ is the expectation value of the square of the \sld\ (i.e., its second moment), which prompts us to ask what is the first moment of $L_\theta$. Using the fact that $\tr{\partial_\theta \rho}=0$, it is straightforward to show that $\tr{\rho L_\theta}=0$. This leads us to the important relations
\begin{align}\label{eq:sld-props}
 \braket{L_\theta} = 0 \qquad\text{and}\qquad I_Q(\theta) = \braket{L_\theta^2} = (\Delta L_\theta)^2\, ,
\end{align}
where $(\Delta A)^2 \equiv \braket{A^2}-\braket{A}^2$ is the variance of an operator $A$.

Next, we will show that the \sld\ form of the \qfi\ in Eq.~\eqref{eq:qfi} originates from the maximisation over all possible measurements in an estimation procedure\cite{Nagaoka82,Braunstein94}. This connects the geometric interpretation of the \qfi\ as the Bures metric in the space of density operators to the statistical interpretation of the \qfi\ as the maximum amount of information about $\theta$ that can be extracted on average in an optimal measurement. Using the Born rule and the fact that the \sld\ is traceless, the classical Fisher information can be written as~\cite{Paris2009_IJQI}
\begin{align}
I(\theta) = \int \dif \x \, \frac{\left(\text{Re}\left\{\Tr{\rho(\theta) {\Pi}(\x) L_\theta}\right\}\right)^2}{\Tr{{\Pi}(\x)\rho(\theta)}},
\label{eqn:multiparameter_classical_fisher_born}
\end{align}
Next, we  maximise this quantity over all possible \textsc{povm}s $\smash{{\Pi}(\x)}$. Given the complex vectors $\alpha, \beta \in \mathbbm{C}^2$, and $\text{Re}[\alpha]\text{Re}[\beta] \leq \norm{\alpha\beta}$, we develop Eq.~\eqref{eqn:multiparameter_classical_fisher_born} into
\begin{align}
I(\theta) &\leq \int \dif \x \, \Norm{\frac{\Tr{\rho(\theta) {\Pi}(\x) L_\theta}}{\sqrt{\Tr{{\Pi}(\x)\rho(\theta)}}}}^2\cr
&= \int \dif \x \, \Norm{\Tr{\frac{\sqrt{\rho(\theta)}\sqrt{{\Pi}(\x)}}{\sqrt{\Tr{{\Pi}(\x)\rho(\theta)}}} \sqrt{{\Pi}(\x)}L_\theta\sqrt{\rho(\theta)}}}^2 \!\! , ~ \;
\label{eqn:multiparameter_quantum_fisher}
\end{align}
where equality holds if and only if $\smash{\text{Im}[\tr{\smash{\rho(\theta) {\Pi}(\x) L_\theta}}] = 0}$, i.e. if the vectors lie in the real space $\mathbbm{R}^2$. This requires the \textsc{sld} to be Hermitian. Introducing $\smash{(\Delta L_\theta)^2}$ as the variance of the \textsc{sld}, we use the Schwartz inequality to obtain:
\begin{align}
\begin{split}
I(\theta) &\leq \int \dif \x \; \Tr{\sqrt{{\Pi}(\x)}L_\theta\sqrt{\rho(\theta)}\sqrt{{\Pi}(\x)}L_\theta\sqrt{\rho(\theta)}},\\
&= \int \dif \x \; \Tr{\rho(\theta)L_\theta{\Pi}(\x)L_\theta},\\
&= \Tr{\rho(\theta)L_\theta^2},
\label{eqn:multiparameter_quantum_fisher_2}
\end{split}
\end{align}
where the final equality used the vanishing trace property of the \textsc{sld}. This completes the maximisation of the \textsc{cfim} over all possible measurements. It shows that the classical Fisher information for any measurement is upper bounded by the quantum Fisher information (\textsc{qfi})
\begin{align}
I(\theta) \leq I_Q(\theta) = \Tr{\rho(\theta)L_\theta^2} = \Tr{\partial_\theta\rho(\theta)L_\theta}.
\label{eqn:classical_quantum_fisher_infos}
\end{align}
Note that the \textsc{qfi} is independent of the \textsc{povm} and is a function of the state only. 

As we observe from Eq.~\eqref{eq:bgwoejkdf}, the \textsc{sld} for mixed states often requires diagonalising the density matrix. For arbitrarily large $s$-dimensional states, this becomes increasingly difficult. To address this difficulty, alternative methods at determining $\smash{L_\theta}$ have been developed. For example, it has been shown that evaluating $\smash{L_\theta}$ is isomorphic to solving a set of linear algebraic equations~\cite{Ercolessi2013}.

The implicit definition of the \textsc{sld} in Eq.~\eqref{eqn:904ureihdbfdfsjeo} is a basis-independent Lyapunov matrix equation that has the general solution~\cite{Paris2009_IJQI}
\begin{align}
L_\theta = 2\int_0^\infty \dif s \, \exp\left[-\rho(\theta)s\right]\partial_\theta\rho(\theta) \exp\left[-\rho(\theta)s\right].
\label{eqn:lyapuniv_representation_sld}
\end{align}
When $\rho$ is not full rank we can still define $L_\theta$ in this way, but some care needs to be taken in order to show that the \qfi\ is still well-defined~\cite{Liu2016_JPA}. The \sld\  is not uniquely defined when the state is not full-rank, since the part of the operator acting on the null-space of $\rho$ is not specified. The \qfi\ based on this expression for the \sld\ can then be written as 
\begin{align}
 I_Q(\theta) = 2\int_0^\infty \dif s \; \Tr{(\partial_\theta \rho) e^{-\rho s} (\partial_\theta \rho) e^{-\rho s}},
 \label{eqn:lyapunov_based_qfi}
\end{align}
The Lyapunov representation proves to be very useful for scenarios in which a periodic nature is observed for the anti-commutator of the density matrix and its partial derivative~\cite{Liu2016_JPA}. 

When the evolution imparting the parameter $\theta$ onto the quantum state $\rho(\theta)$ is a unitary transformation of the form $U = \exp(-i\theta G)$ with $G$ a Hermitian observable and the eigenvalues of $\rho$ independent of $\theta$, we can use the Lyapunov form to find a particularly elegant expression for the \qfi. 
\begin{align}
 I_Q(\theta) & = 2 \int_0^\infty  \dif s\; \Tr{(\partial_\theta \rho) e^{-\rho s} \partial_\theta \rho e^{-\rho s}} \cr
 & = -2 \int_0^\infty  \dif s\;\Tr{\left([G,\rho] e^{-\rho s}\right)^2} \cr 
 & \leq -2 \int_0^\infty  \dif s\;\Tr{\left([G,\rho] e^{-\rho s}\right)}^2\;
\end{align}
Next, we recall that the Schwarz inequality for the trace is given by 
\begin{align}\label{eq:Schwarztrace}
 \Abs{\Tr{AB^\dagger}}^2 \leq \Tr{AA^\dagger} \Tr{BB^\dagger} \, ,
\end{align}
and we can bound the \qfi\ by
\begin{align}\label{eq:qfivariance}
 I_Q(\theta) & \leq -2 \int_0^\infty \Tr{[G,\rho]^2} \Tr{e^{-\rho s}}^2\; ds \cr
 & = -2 \int_0^\infty \Tr{(G\rho - \rho G)^2} \Tr{e^{-\rho s}}^2\; ds \cr
 & = 4 \left( \Tr{G^2 \rho^2} - \Tr{(G\rho)^2}\right)  \int_0^\infty \left( \sum_k e^{-sp_k} \right)^2\; ds \cr
 & \leq 4 (\Delta G)^2\, ,
\end{align}
where the last inequality can be obtained by evaluating the traces in the eigenbasis of $\rho$. When the probe state is pure ($\rho^2 = \rho$), the integral in the last line evaluates to 1, and we obtain
\begin{align}\label{eq:bgh49weofsj}
 I_Q(\theta,\psi) = 4 (\Delta G)^2\, ,
\end{align}
which can often be found analytically.

Finally, we address the issue that the \textsc{sld}, as defined in Eq.~\eqref{eq:sldeigenbasis} becomes singular for pure states. Nevertheless, there is a simple expression for the \qfi\ for pure states, as shown in Eq.~\eqref{eq:bgh49weofsj}. We can also find a simple expression for the \sld\ when $\rho$ is pure. Using $\smash{\rho^2 = \rho}$, differentiating with respect to the parameters $\smash{\theta}$ and comparing with the definition Eq.~\eqref{eqn:904ureihdbfdfsjeo} we arrive at
\begin{align}
L_\theta = 2\partial_{\theta}\rho = 2i [\rho,G]\, ,
\label{eqn:sld_final_form_pure}
\end{align}
where we relate the \sld\ to the von Neumann equation of motion describing the dynamics of the system. 

The calculation of the \textsc{qfi} for any physical system is at the heart of quantum metrology and is typically a difficult task. Determining the \textsc{qfi} using the \textsc{sld} operator is particularly suited to unitary quantum metrology. It is less suited for noisy processes, where the calculation involves complex optimisation procedures~\cite{Sarovar2006_JPA, Escher2011}. To address this, an extended Hilbert space approach may be taken where information about the parameter is obtained by observing both the system and its environment~\cite{Escher2012_PRL}. This method prescribes the \textsc{qfi} in terms of the state evolving Hamiltonian, and is well suited to many physical implementations of parameter estimations, including open quantum systems~\cite{Chin12, Kolodynski2013_NJP, Alipour14, Dobrzanski2014_PRL}.

\subsection{The quantum Cram\'er-Rao bound}
\noindent
In this section we will derive \qcrb\ (or the \emph{Helstrom} bound) for a single parameter $\theta$. We follow Helstrom's original derivation~\cite{Helstrom67} that employs the \sld. First, we consider a measurement $M$ on a system in state $\rho$ that serves as an estimator for $\theta$. In other words, the expectation value $\tr{\rho M}$ provides the estimate $\caron{\theta}(\x)$ with a possible bias $b(\theta) = \tr{\rho(M - \theta)}$. We can take the derivative of the bias with respect to $\theta$ to obtain
\begin{align}\label{eq:qcrb67oeritfjkg}
 \partial_\theta b(\theta) = \Tr{(M-\theta) \partial_\theta \rho} - 1\, ,
\end{align}
where we assume that $M$ does not itself depend on $\theta$. This is an important caveat that we will return to when we wish to determine the optimal estimator in section~\ref{subsubsec:the_qfim}. 
We next square Eq.~(\ref{eq:qcrb67oeritfjkg}), and together with the \sld\ and the Schwarz inequality for the trace we get
\begin{align}
 (1+\partial_\theta b)^2 & = \Tr{(M-\theta) \partial_\theta \rho}^2 \cr
 & = \frac12 \Tr{(M-\theta) \left( L_\theta \rho + \rho L_\theta \right)}^2 \cr 
 & = \left(\Re\Tr{(M-\theta) \rho L_\theta}\right)^2 \cr
 & \leq \Abs{\Tr{(M-\theta) L_\theta \rho}}^2 = \Abs{\Tr{ \sqrt{\rho}(M-\theta) L_\theta \sqrt{\rho}}}^2 \cr
 & \leq \Tr{\rho L_\theta^2} \Tr{\rho (M-\theta)^2}\, .
\end{align}
Identifying $ \tr{\rho (M-\theta)^2}$ with the \mse\ in $\theta$, we arrive at the quantum Cram\'er-Rao bound (\qcrb)
\begin{align}
 \var\theta & = \ex_\rho \left[(M-\theta)^2\right] \cr & \geq \frac{ (1+\partial_\theta b)^2}{\tr{\rho L_\theta^2}} =  (1+\partial_\theta b)^2 I_Q^{-1}(\theta) \, ,
\end{align}
where we have identified $\tr{\rho L_\theta^2}$ with the \qfi\ $I_Q(\theta)$. 
Note that for types of bias with negative derivatives the \mse\ appears to be better than in the case of an unbiased estimator ($b=0$). This can cause some confusion when comparing the \mse\ with a pre-calculated value of the \qfi. When $N$ independent measurements are made using an unbiased estimator, additivity of the \qfi\ implies that the resulting bound is given by 
\begin{align}\label{eq:qcrbsingle}
 \var\theta \geq \frac{1}{N I_Q}  \, .
\end{align}
This is the form of the \crb\ for a single parameter that is mostly used.

The \qcrb, together with the bound in Eq.~(\ref{eq:qfivariance}) on the \qfi, leads immediately to a familiar result. For a single shot measurement ($N=1$) Eq.~(\ref{eq:qcrbsingle}) becomes
\begin{align}
 \var\theta \geq \frac{1}{4(\Delta G)^2}\, ,
\end{align}
where we set $\hbar=1$.
When we define $\var\theta = (\delta\theta)^2$, this leads to~\cite{Braunstein95}
\begin{align}\label{eq:ursingle}
 \delta\theta\, \Delta G \geq \frac{1}{2}\, ,
\end{align}
and can be interpreted as an uncertainty relation for a quantity $\theta$ and its generator of translations $G$. While Heisenberg's uncertainty relations are typically derived for conjugate observables like position and momentum, Eq.~(\ref{eq:ursingle}) allows us to define uncertainty relations between energy and time, or angular momentum and rotation angles where Robertson inequalities cannot be constructed due to a lack of self-adjoint operators for time and rotation angles.

The next question to address is how to find the optimal estimator that saturates the \qcrb. First, does there exist a measurement for which the \qfi\ equals the classical Fisher information? And if so, what is this measurement? To answer these questions, we recall the error propagation formula from Eq.~\eqref{eq:n304riwerofjdnk}:
\begin{align}
 \var\theta = \frac{(\Delta M)^2}{\abs{\dif\braket{M}/\dif \theta}^2}\, .
\end{align}
This formula relates the variance in the parameter $\theta$ to the variance of the operator $M$ that is used to estimate $\theta$. Helstrom \cite{Helstrom68} states that the \crb\ is saturated {if and only if} $L_\theta = k(\theta) (M-\theta)$, with $k(\theta)$ some function that does not depend on $M$. Choosing
\begin{align}\label{eq:rahgfskbj}
 M = \theta\, \identity + \frac{L_\theta}{I_Q(\theta)}\, ,
\end{align}
we can prove that $\var\theta$ saturates the \crb. For this choice the bias vanishes: $b(\theta) = \tr{\rho (M-\theta)} = 0$. From Eq.~(\ref{eq:sld-props}) we calculate that 
\begin{align}
 \braket{M} = \theta \qquad\text{and}\qquad \frac{\dif\braket{M}}{\dif\theta} = 1\, ,
\end{align}
and 
\begin{align}
 \braket{M^2} = \Braket{\theta^2 +2\theta \frac{L_\theta}{I_Q} + \frac{L_\theta^2}{I_Q^2}} = \theta^2 + \frac{1}{I_Q}\, .
\end{align}
From this, we find that $\var\theta = I_Q^{-1}$, which saturates Eq.~(\ref{eq:qcrbsingle}). So there indeed does exist an estimator that saturates the \qcrb, but it generally depends on the unknown parameter $\theta$. 

It was shown by Braunstein and Caves~\cite{Braunstein94} that the optimal measurement for $\theta$ is a von Neumann measurement that consists of projections onto the eigenstates of the \sld\ $L_\theta$. However, $L_\theta$ generally depends on  the unknown value of $\theta$, and it may not be possible to choose the optimal estimator at the outset~\cite{BarndorffNielsen2000}. Since the \qcrb, like the classical \crb, is an asymptotic bound on the variance of $\theta$, many measurements must be made before the bound is saturated, and this allows for adaptive measurements that converge to the optimal measurement~\cite{Nagaoka82,asymptotic_theory_book_2005,Fujiwara2006}.

\subsection{Biased Estimators}\label{subsec:biased estimators}
\noindent
Helstrom \cite{Helstrom67} constructed a quantum version of the Cram\'er-Rao bound for a single variable $\theta$ from the \sld. First, we consider a measurement $M$ on a system in state $\rho$ that serves as an estimator for $\theta$. In other words, the expectation value $\tr{\rho M}$ provides the estimate $\caron{\theta}(x)$ with a possible bias $b(\theta) = \tr{\rho(M - \theta)}$. We can take the derivative of the bias with respect to $\theta$ to obtain

\begin{align}\label{eq:qcrb67}
 \partial_\theta b(\theta) = \Tr{(M-\theta) \partial_\theta \rho} - 1\, ,
\end{align}
where we assume that $M$ does not itself depend on $\theta$. This is an important caveat that we will return to when we wish to determine the optimal estimator. 
We next square Eq.~(\ref{eq:qcrb67}), and together with the \sld\ and the Schwarz inequality for the trace we get
\begin{align}
 (1+\partial_\theta b)^2 & = \Tr{(M-\theta) \partial_\theta \rho}^2 \cr
 & = \left(\Re\Tr{(M-\theta) \rho L_\theta}\right)^2 \cr
 & \leq \Abs{\Tr{(M-\theta) L_\theta \rho}}^2 \cr
 & \leq \Tr{\rho L_\theta^2} \Tr{\rho (M-\theta)^2}\, .
\end{align}
Identifying $ \tr{\rho (M-\theta)^2}$ with the \mse\ in $\theta$, we arrive at the quantum Cram\'er-Rao bound (\qcrb)
\begin{align}
 \var\left[\theta\right] = \ex_\rho \left[(M-\theta)^2\right] \geq \frac{(1+\partial_\theta b)^2}{I_Q} \, .
\end{align}
Note that for types of bias with negative derivatives the \mse\ appears to be better than in the case of an unbiased estimator ($b=0$). This can cause some confusion when comparing the \mse\ with a pre-calculated value of the \qfi. 

\subsection{The role of entanglement}\label{subsec:entanglement_for_quantum_metrology}
\noindent
Consider an experiment that estimates a parameter $\theta$. The experiment is repeated $N$ times under identical conditions, and each time the average information that is extracted about $\theta$ is given by the \qfi. Since the Fisher information is additive for independent measurements, the total information in the $N$ experiments is $N I_Q$, leading to the \crb\ in Eq.~\eqref{eq:qcrbsingle}. The Root Mean Square Error (\textsc{rmse}) then behaves as
\begin{align}\label{eq:hh420ru3eiojds}
 \delta\theta \geq \frac{1}{\sqrt{N I_Q}}\, .
\end{align}
The square-root scaling $1/\sqrt{N}$ of $\delta\theta$ is called the Standard Quantum Limit (\textsc{sql}), or shot-noise limit. This is the best possible performance for a classical estimation procedure, i.e., estimation procedures that do not employ entangled probe states.

To see how we can improve over Eq.~\eqref{eq:hh420ru3eiojds}, we consider the case where the parameter is imparted on the quantum state via the unitary evolution $\exp(-i\theta G)$, such that the \qfi\ takes the form
\begin{align}
 I_Q^{\rm (max)} (\theta) = 4(\Delta G)^2\, .
\end{align}
To maximise the \qfi\ is therefore to maximise $(\Delta G_N)^2$ over the state of the $N$ physical systems. Clearly, when each  experiment is independent, the variances $(\Delta G_N)^2$ add, such that $(\Delta G_N)^2 = N(\Delta G)^2$ and we recover the \textsc{sql} in Eq.~\eqref{eq:hh420ru3eiojds}. However, we can also prepare the $N$ systems in a suitable entangled state. This will allow us to increase $(\Delta G_N)^2$ substantially, scaling instead with $N^2$. The \qcrb\ then becomes
\begin{align}
 \var\theta \geq \frac{\alpha}{N^2}\, ,
\end{align} 
where $\alpha$ is some constant, typically of order unity. This leads to an \textsc{rmse} that scales with $N^{-1}$:
\begin{align}
 \delta\theta \sim \frac{1}{N}\, .
\end{align} 
This is called the \emph{Heisenberg limit}, and it is the ultimate limit for quantum parameter estimation~\cite{Giovannetti2006_PRL}. 

To see how such a precision can be achieved, we consider the optical \textsc{noon} state~\cite{Bollinger1996_PRA,Huelga97,Boto00,Lee2002_JMO,Kok2002_PRA}
\begin{align}\label{eq:noon}
 \ket{\psi} = \frac{\ket{N,0} + \ket{0,N}}{\sqrt{2}}\, ,
\end{align}
where $\ket{N}$ is the $N$-photon Fock state. This is a two-mode entangled state that is extremely challenging to make in the lab~\cite{Walther04,Mitchell04}, but it serves as a clear proof of principle. Assuming a simple phase shift $\theta$ in the second mode, the \textsc{noon} state evolves to
\begin{align}
 \ket{\psi(\theta)} = \frac{\ket{N,0} + {\mathrm e}^{iN\theta}\ket{0,N}}{\sqrt{2}}\, ,
\end{align}
where each photon in the second mode picks up a phase ${\mathrm e}^{iN\theta}$. We calculate the \qfi\ using the expression in Eq.~\eqref{eq:hge0w4reosjndk}
\begin{align}
I_Q(\theta) & = 4 \braket{\partial_\theta\psi(\theta)\vert\partial_\theta\psi(\theta)} - 4\Abs{\Braket{\partial_\theta\psi(\theta)\vert\psi(\theta)}}^2\, . 
\end{align}
The derivative of the state is given by 
\begin{align} 
 \ket{\partial_\theta \psi(\theta)} = -\frac{iN {\mathrm e}^{iN\theta}}{\sqrt{2}} \ket{0,N}\, ,
\end{align}
and the \qfi\ becomes $I_Q = N^2$, as required. To complete the estimation procedure, practical measurement observables were proposed by Pryde \emph{et al.}~\cite{Pryde2003_PRA} and Cable \emph{et al.}~\cite{Cable2007_PRL}.  A similar argument can be constructed using Greenberger-Horne-Zeilinger (\textsc{ghz}) states~\cite{Greenberger89}.

\begin{figure}[t!]
\begin{center}
\includegraphics[width =0.72\columnwidth]{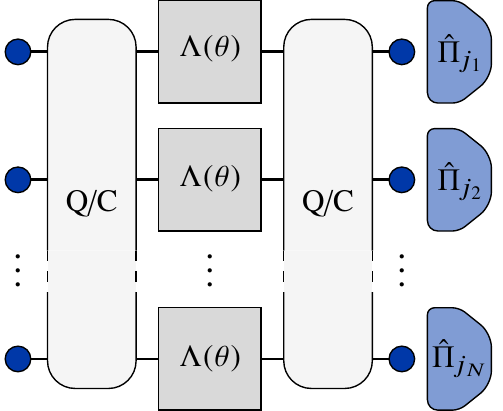}
\caption{General parallel parameter estimation strategies. The initial probe states, $\smash{\rho(0)}$ are represented by the circles on the left. The probe states may be entangled before passing through an evolving channel, illustrated by the grey central squares. The state passes through another potentially entangling evolution and is finally measured using local \povm s ${\Pi}_j$, represented by the blue polygons on the right. This setup allows for four classes of estimation, where the input states many be entangled or not (Q/C on the left), and where the input state may be subjected to entangled or separable measurement observables (Q/C on the right).}
\label{pic:parameter_estimation_strategies_parralllel}
\end{center}
\end{figure}

To understand more generally how quantum entanglement can improve the estimation precision, Giovannetti, Lloyd and Maccone classified various metrology approaches~\cite{Giovannetti2006_PRL}. They unified parallel and adaptive sequential strategies into a general framework shown in Fig.~\ref{pic:parameter_estimation_strategies_parralllel}. The unitary evolutions $U$ imparting the parameter are again of the form $U=\exp(-i\theta G)$. This classification allows for classical or quantum state preparation and measurement procedures, resulting in four different classes of experiments: classical states and classical measurements (\textsc{cc}), classical states and quantum measurements (\textsc{cq}), quantum states and classical measurements (\textsc{qc}), and both quantum states and measurements (\textsc{qq}). Here, we understand by ``classical'' that the input state is separable, and the measurement projects onto separable \textsc{povm} elements. Since the \qcrb\ is determined by the quantum state via the \qfi, no quantum entangling strategies at the measurement stage can introduce further enhancements to the estimation procedure than what is already present in the quantum state. Therefore the \textsc{cc} and \textsc{cq} strategies will always yield at best the \textsc{sql}~\cite{Giovannetti2011_NP}. Any resolution enhancements must then be sourced from the probe preparation. Bound entanglement can also surpass the shot noise limit \cite{Hyllus12,Czekaj15}.

Toth~\cite{Toth12} showed that for qubits, genuine multi-partite entanglement is required to achieve the Heisenberg limit. Considering three possible parameters $\theta_x$, $\theta_y$ and $\theta_z$ generated by the Pauli operators $\sigma_x$, $\sigma)y$, and $\sigma_z$, the following general results hold:
\begin{enumerate}\itemsep=0pt
 \item for $N$-qubit separable states the \qfi\ is bounded by $\sum_j I_Q(\theta_j) \leq 2N$ and for a single parameter $I_Q(\theta_j) \leq N$;
 \item for general $N$-qubit quantum states $\sum_j I_Q(\theta_j) \leq N (N+2)$,
 \item for $k$-producible states, where a pure state is $k$-pro\-du\-ci\-ble if it is a tensor product state of at most $k$ qubits, $I_Q(\theta_j) \leq nk^2 + (N-nk)^2$, where $n$ is the integer part of $N/k$.
 \item the sum of the \qfi's for each parameter is bounded by 
 \begin{align}
  \sum_j I_Q(\theta_j) \leq nk(k+2) + (N-nk)(N-nk+2) \!\!\!\!\!\!\!
 \end{align}
 if $N-nk \neq 1$, and 
 \begin{align}
  \sum_j I_Q(\theta_j) \leq nk(k+2)+2 
 \end{align}
 if $N-nk = 1$;
 \item for multi-partite quantum states with $M$ unentangled particles $\sum_j I_Q(\theta_j) \leq M +(N - M)(N -  M +2)$. 
\end{enumerate}
The broader question of how useful quantum states are for quantum metrology was answered by Oszmaniec \emph{et al.}~\cite{Oszmaniec16}, who showed that pure states chosen randomly from the symmetric subspace typically achieve the optimal Heisenberg scaling without the need for local unitary optimisation. Further, an explicit non-random choice of symmetric probe states with error correction capabilities has recently been demonstrated useful for robust metrology~\cite{Ouyang19_arxiv}. In this work, it was shown that if the probe state lies within the code space of certain permutation-invariant quantum codes~\cite{Ouyang2014_PRA}, a precision enhancement is possible even in the presence of noise. Other symmetric states have been considered for robust quantum metrology~\cite{Koczor19_arxiv}. These studies reflect the current area of intense research on error correction inspired robust metrology, where quantum error correction is not applied. This removes the requirement for feed-forward and error-correction, which reduces the difficulty and practicality of implementing practical quantum metrology. These methods are motivated by the near term emergence of noisy quantum devices; the so-called noisy intermediate-scale quantum (\textsc{nisq}) era. We review fault-tolerant quantum metrology methods later in section~\ref{subsec:noisy_metrology}.

The entanglement requirements above can be turned on their head, such that a \qfi\ in excess of a certain value indicates that at least genuine $k$-partite entanglement must be present in the quantum state. This is a so-called \emph{entanglement witness}~\cite{Shimizu05,Guhne09,Pezze09,Apellaniz17_PRA}. In other words, the difference in precision scaling can be used to deduce whether entanglement is present in the probe state. Relaxing the Heisenberg limit to $\delta\theta \sim 1/N^{1-\epsilon}$ for any $\epsilon > 0$, the amount of entanglement required can be made arbitrarily small~\cite{Augusiak16}. 

The bounds established by Toth on the \qfi\ holds for qubits undergoing unitary evolutions $\exp(-i \theta_j \sigma_j/2)$. The question remains whether other types of evolution can lead to a different scaling. Indeed, when the generator of translations in $\theta$ is a multi-particle Hamiltonian, the estimation precision in an experiment using $N$ particles can scale with $N^{-m}$ for some $m>1$, or even $2^{-N}$, as shown by Boixo \emph{et al}.~\cite{Boixo2008_PRL} and Roy and Braunstein~\cite{Roy2008_PRL}, respectively. These results cannot be compared directly with the Heisenberg limit, since the evolution in those generalised scaling laws is generated by fundamentally different physical processes than the typical single-particle evolution $\exp(-i \theta G)$. 

One way to understand these limits is via the query complexity of the estimation procedure~\cite{Zwierz2010_PRL}. In the standard parameter estimation procedure each particle evolves according to the effective Hamiltonian $G$. However, for systems with a bi-partite Hamiltonian $H_{jk} = G_j \otimes G_k$ each evolution requires a \emph{pair} of particles. Each pair is now a \emph{query} of the parameter $\theta$, and for $N$ particles there are $Q=\frac12 N(N-1)$ queries. This structure generalises to multi-partite Hamiltonians. The precision limit always scales at most linearly with the number of queries, not the number of particles~\cite{Zwierz2010_PRL,Zwierz:2011hra,Zwierz:2012eb,Hall:2012cr}.

\subsection{Non-entangling strategies}\label{subsec:nonentangling_strategies}
\noindent
The entanglement strategies in the previous section refer to systems consisting of distinguishable particles. A different situation arises in quantum optics, where at least in principle, non-entangled states can achieve sub-shot noise precision. In particular, the squeezed vacuum can be used to suppress the fluctuations due to shot noise~\cite{Caves1981_PRD,Pezze08}. Other states that have been used to attain the Heisenberg limit are the class of entangled coherent states (\textsc{ecs})~\cite{Jing2014_CTP,Joo2012_PRA,Ono2010_PRA}. For phase estimations in a Mach-Zehnder interferometer, \textsc{ecs} have demonstrated better precision scalings than \textsc{noon} states~\cite{Joo2011_PRL}. Even in a lossy interferometer, \textsc{ecs} can still beat the shot-noise limit for modest loss rates~\cite{Zhang2013_PRA}.

Nevertheless, generating highly entangled states is practically difficult for two main reasons. First, the photonic overhead increases exponentially with the number of entangled modes and second, the fidelity decreases due to decoherence processes~\cite{Wang2011_PRL}. Furthermore, despite the results in the previous section the use of quantum entanglement as a resource is still poorly understood. Specifically, the performance of \textsc{noon} states for optical phase imaging performs worst in comparison with the class of other states including entangled coherent states, entangled squeezed coherent states, and entangled squeezed vacuum~\cite{Zhang17}. This suggests that mode entanglement alone is not sufficient to provide certain precision enhancements. Indeed, too much entanglement is detrimental to attaining the Heisenberg scaling in the estimation of unitarily generated parameters~\cite{Baumgratz2016_PRL}. Similarly, the estimation precision of $n$ optical phase differences can be enhanced by using an $n$-mode entangled state as input in a multi-mode interferometer~\cite{Humphreys13, Liu2016_JPA2,Yue2014_SR}. However, this precision enhancement has also been matched using separable states with equal number of modes~\cite{Knott2016_PRA}. This has been demonstrated theoretically for spatial distinguishability of different light emitters~\cite{Sidhu2017_PRA}. Alternative approaches to achieve quantum enhanced measurements have been investigated~\cite{Braun2018_RMP}. These methods rely on the use of quantum correlations, and nontrivial Hamiltonian extensions. 

\subsection{Optimal estimation strategies}\label{subsec:optimal_implementation}
\noindent
Once the fundamental limits to the precision of parameter estimations have been determined, a natural question that arises is given that all classical noise has been eliminated, how can we identify the measurement(s) that practically saturate these bounds?  Optimal measurements can be constructed from the eigenstates of the \textsc{sld}~\cite{Paris2009_IJQI}. In almost all cases, determining the measurement that corresponds to this theoretical description is difficult. Generally it depends on the parameters that we would like to estimate. 

Adaptive strategies have been suggested to circumvent the parameter dependence of the optimal measurements~\cite{Pirandola2018_NP}. Wiseman showed that feedback control of the phase of a local oscillator can approximate the measurement of the phase quadrature in an optical mode~\cite{Wiseman95}. This was demonstrated experimentally by Armen \emph{et al.}~\cite{Armen02} By adaptively changing the phase in one arm of a Mach-Zehnder interferometer, nearly optimal measurement of the relative phase given $N$ input photons can be achieved~\cite{Berry00, Berry2002_PRA}. This technique was extended to narrowband squeezed beams by Berry and Wiseman~\cite{Berry06,Berry13}. Fujiwara proved that a sequence of maximum likelihood estimators is asymptotically efficient for adaptive quantum parameter estimation~\cite{Fujiwara06}. Okamoto \emph{et al.}\ use this technique to estimate the phase between left- and right-handed circular polarisation of single photons~\cite{Okamoto12}. For a review of quantum feedback control techniques, see Serafini~\cite{Serafini12}. Palittapongarnpim and Sanders proposed tests to see whether adaptive strategies in quantum metrology are robust against phase noise~\cite{Palittapongarnpim19}.


Achieving the Heisenberg limit is state dependent. However, the probe state chosen should be tailored to achieve the best practical precision for a specific parameter. For example, squeezed light is routinely used for phase estimations~\cite{Ligo2013_NP}. A natural question to ask is what is the optimal probe state that maximises the \qfi\ for a parameter estimation protocol? This was answered by Braunstein, Caves and Milburn~\cite{Braunstein95} and Giovannetti, Lloyd and Maccone~\cite{Giovannetti2006_PRL} in the context of unitary evolutions $\exp(-i\theta G)$. The optimal probe state is an equal superposition of eigenstates corresponding the minimum and maximum eigenvalues of the generator $G$. These states are generally difficult to prepare.

Only very few experiments have reported a Heisenberg limit scaling for parameter estimates~\cite{Higgins2007_N, Higgins2009_NJP}. This is generally due to two factors. First, achieving the \textsc{sql} is already practically difficult since it requires eliminating all non-intrinsic system noises. Second, state entanglement of multipartite systems is challenging to realise due to their increasing susceptibility to environmental losses with increasing particle number. For example, the path-entangled \textsc{noon} states in Eq.~\eqref{eq:noon} can be shown to achieve the Heisenberg limit resolution scaling for phase measurements in optical interferometers~\cite{Kok2010}, but for larger photon number the loss of a single photon becomes increasingly likely, and this completely destroys the capability of measuring the parameter $\theta$.

Even modest Markovian noise reduces the Heisenberg limit scaling achievable by highly entangled states to scalings proportional to the \textsc{sql}~\cite{Huelga97, Kolodynski2010_PRA, Escher2011}. Common decoherences include depolarisation, dephasing and amplitude damping. Owing to the difficulty and stabilisation of highly entangled states, alternative approaches to achieve quantum enhanced measurements have been investigated~\cite{Braun2018_RMP}. These methods rely on the use of quantum correlations and identical particles such as photons.

\subsection{Numerical approaches}\label{subsec:numerical_estimation}
\noindent
As has been shown so far, if the \textsc{qfi} is known, the fundamental precision bound is known and the optimal measurement strategy can be determined. Often however, it is not possible to find the \textsc{qfi} analytically, for example when probe states with a large rank are used. In those cases a numerical approach may be better suited. 

Saturating the \textsc{qcrb} requires a suitable choice of estimator, which can be found numerically. Unfortunately, the numerical method required to determine a well-behaved and efficient estimator depends on the estimation problem, since the procedure will generally depend on how the parameters are encoded in the state. However, a common procedure used is maximum likelihood estimation, which given its simplicity, has found widespread use in estimation theory. The maximum likelihood procedure attempts to find the values of the parameters $\smash{\bm{\theta}}$ that maximise the log likelihood function $\smash{\ln[p(\bm{x}\vert\bm{\theta})]}$. In some circumstances this may be as simple as taking the derivative of the likelihood function and equating it to zero to find the maximum. However, this is not always a straightforward operation and alternative methods for obtaining the maximum likelihood estimator must be used.

Numerically, the maximum likelihood estimator may be implemented via an iterative scoring algorithm~\cite{Kay1993}. Defining the $k$th iteration of the estimator by $\smash{\check{\bm{\theta}}^{(k)}}$, the scoring algorithm proceeds according to the iterative equation
\begin{align}
\check{\bm{\theta}}^{(k+1)}(\x) =  \check{\bm{\theta}}^{(k)}(\x) + I_Q^{-1}\bigl[\check{\bm{\theta}}^{(k)}\bigr]\frac{\partial \ln \left[p(\x\vert\bm{\theta})\right]}{\partial\bm{\theta}}\Bigr|_{\bm{\theta} = \check{\bm{\theta}}^{(k)}}\, .
\label{eqn:scoring_algo}
\end{align}
Based on any information on the system, by taking an initial guess of the parameters $\smash{\check{\bm{\theta}}^{(0)}}$, successive iterations of the scoring algorithm generate estimates that more closely approximates the true value. For open quantum systems the Markov chain Monte Carlo integration and Metropolis Hastings algorithm are better suited than the scoring algorithm~\cite{Gong2017_arxiv}. 


\section{Multi-parameter quantum estimation}\label{sec:multi_parameter_est}
\noindent
In this section, we review enhanced quantum parameter estimation of multiple parameters simultaneously. Many practical high-precision estimation protocols require a multi-parameter estimation approach. This includes the estimation of multiple phases~\cite{Vidrighin2014_NC,Humphreys13,Berry15,Crowley2014_PRA}, characterisation of multidimensional fields~\cite{Baumgratz2016_PRL,Pang2014_PRA,Yao14_PRA}, and Hamiltonian tomography~\cite{Zhang14_PRL,Kura18_PRA}. 

Multi-parameter quantum metrology raises two important questions. First, what is the attainability of the multi-parameter \textsc{qcrb}. If the \textsc{sld}s for each parameter are mutually compatible---that is they  commute with each other---then a simultaneous, optimal estimate for all of the parameters can be made in their common eigenbasis. If the optimal measurements corresponding to the different \textsc{sld}s for each parameter do not commute, a compromise between the estimation precision for each parameter must be addressed.  Second, what is the tradeoff between estimation precision enhancement and the physical resources used to attain it~\cite{Imai07_JPA}? Specifically, is it better to estimate a tuple of parameters simultaneously or sequentially? Addressing these questions will help in the design of novel estimation schemes that propel precisions closer to the Heisenberg limit.

In this section we will first derive the \qfi\ matrix for multiple parameters, and construct the corresponding Cram\'er-Rao bound. We consider alternatives to the \qfi, based on the right logarithmic derivative, as well as the Kubo-Mori information and the Wigner-Yanase skew information. We conclude this section with a discussion of the Holevo bound and the general attainability of the \qcrb.

\subsection{The quantum Fisher information matrix}\label{subsec:qfim_intro}
\noindent
The \qfi\ in section \ref{sec:qe_theory} produces a real positive number associated with a single parameter $\theta$ that can be written as an inner product
\begin{align}
 I_Q (\theta) = \Tr{\partial_\theta \rho\, \mathscr{L}_\rho(\partial_\theta \rho)}\, .
\end{align}
For multiple parameters $\bm\theta = (\theta_1,\ldots,\theta_D)$ the \qfi\ becomes a matrix, since the generalisation to the $D$-dimensional parameter space creates a natural two-form~\cite{Amari93}
\begin{align}
 [I_Q (\bm\theta)]_{jk} = \Tr{\partial_j \rho\, \mathscr{L}_\rho(\partial_k \rho)}\, ,
\end{align}
where $j,k = \{1,2,\ldots,D\}$. If we furthermore associate a new \sld\ $L_j = \mathscr{L}_\rho (\partial_j \rho)$ with each parameter $\theta_j$:
\begin{align}
 \partial_j \rho = \frac12 \left( L_j \rho + \rho L_j \right)\, ,
\end{align}
where $\partial_j \equiv \partial/\partial\theta_j$, then in terms of the \sld s the \qfi\ matrix becomes
\begin{align}\label{eq:mpqfi}
 {[I_Q(\bm\theta)]_{jk} = \frac12 \Tr{\rho \{L_j, L_k\}}}\, .
\end{align}
The anti-commutator appears due to the possibility of a nonzero commutator between $L_j$ and $L_k$. Since the optimal estimator for $\theta_j$ is given by the projectors along the eigenvectors of $L_j$, it is clear that in general non-commuting $L_j$ and $L_k$ will cause trouble for the simultaneous estimation of $\theta_j$ and $\theta_k$. Nevertheless, the \qfi\ matrix is well-defined, and we can write for the matrix elements
\begin{align}
 [I_Q (\bm\theta)]_{lm} = \sum_{jk} \frac{2}{p_j + p_k} \braket{j|(\partial_l \rho)|k}\braket{k|(\partial_m \rho)|j} \, ,
\end{align}
which again is hard to calculate in general, and does not include pure states. 

When $\rho = \ket{\psi}\bra{\psi}$ is a pure state and the evolution of the parameters is given by $\exp(-i\theta_j G_j)$ with $G_j$ the self-adjoint generators of $\theta_j$, the state $\ket{\psi}$ obeys the Schr\'odinger-like equations
\begin{align}
 i\, \partial_j \ket{\psi} = G_j \ket{\psi}\, .
\end{align}
The \sld\ $L_j$ can then be written as 
\begin{align}
 L_j & = 2\partial_j \rho = 2\left( \ket{\smash{\partial_j\psi}}\bra{\psi} + \ket{\psi}\bra{\smash{\partial_j\psi}}\right)\, ,
\end{align}
with $\ket{\smash{\partial_j \psi}} = \partial_j\ket{\psi}$. Moreover, since $\ket{\psi}\bra{\smash{\partial_j\psi}} = - \ket{\smash{\partial_j\psi}}\bra{\psi}$, we obtain 
\begin{align}\label{eq:pureqfimatrix}
 [I_Q (\bm\theta)]_{jk} & = 4 \Re\left( \braket{\partial_j\psi |\partial_k\psi} - \braket{\partial_j\psi|\psi}\braket{\psi|\partial_k\psi}\right)\, .
\end{align}
This is the multi-parameter quantum Fisher information for pure states and simple unitary evolution. We obtain a particularly useful form for $I_Q (\psi,\bm\theta)$ when we relate the derivatives of $\ket{\psi}$ to the generators $G_j$ associated with $\theta_j$. Eq.~(\ref{eq:mpqfi}) becomes
\begin{align}\label{eq:qfigenS}
 [I_Q (\bm\theta)]_{jk} & = 4 \left(\braket{\psi |\frac12\{G_j,G_k\}|\psi} - \braket{\psi|G_j|\psi}\braket{\psi|G_k|\psi}\right) \cr
 & \equiv 4 \cov_S(\mathbf{G})_{jk}\, ,
\end{align}
where we defined $\mathbf{G} = (G_1,\ldots,G_D)$, and 
\begin{align}\label{eq:covG}
  \cov_S(\mathbf{G})_{jk} = \braket{\psi |\frac12\{G_j,G_k\}|\psi} - \braket{\psi|G_j|\psi}\braket{\psi|G_k|\psi}
\end{align}
is the symmetrized covariance matrix for $\mathbf{G}$. This can also be written in terms of a non-symmetrized covariance matrix $\cov(\mathbf{G})$ according to
\begin{align}
 \cov_S(\mathbf{G})_{jk} = \frac{1}{2}\left[ \cov(\mathbf{G})_{jk} +  \cov(\mathbf{G})_{kj} \right]\, ,
\end{align}
where $\cov(\mathbf{G})_{jk} = \braket{\psi|G_j G_k|\psi} - \braket{\psi|G_j|\psi} \braket{\psi|G_k|\psi} $.  An alternative form for $I_Q (\bm\theta)$ is then
\begin{align}\label{eq:qfigenNS}
 [I_Q (\bm\theta)]_{jk} = 4 \Re \cov(\mathbf{G})_{jk}\, .
\end{align}
When all $G_j$ commute with each other, the covariance matrix $\cov(\mathbf{G})$ is real. The expressions in Eqs.~(\ref{eq:qfigenS}) and (\ref{eq:qfigenNS}) are the multi-parameter generalisations of Eq.~(\ref{eq:bgh49weofsj}). The multi-parameter \qfi\ is a manifestly symmetric positive semi-definite matrix, and like the classical Fisher information matrix it transforms as a tensor under re-parameterization. Defining again a new set of variables $\smash{\bm{\vartheta}}$ through some invertible transformation Jacobian matrix, $\smash{{J}}$, such that  $\smash{\bm{\vartheta} = {J \bm\theta}}$, then the \textsc{qfim} for the new parameters may be written
\begin{align}
I_Q(\bm{\vartheta}) = {J}^\top I_Q(\bm{\theta}) {J}, \quad \text{with} \quad {J}_{kl} = \frac{\partial\theta_k}{\partial\vartheta_l}\, .
\label{eqn:jacobian_trans_qfim}
\end{align}

We can bound the \qfi\  for general mixed states by the covariance matrix, just as we did for a single parameter in Eq.~(\ref{eq:qfivariance}). We start with the general definition in Eq.~(\ref{eq:mpqfi}) and note that we can modify the derivative of $\rho$ according to
\begin{align}
 \partial_j \rho = -i [G_j,\rho] = -i [G_j - \braket{G_j},\rho] \equiv -i [\Delta G_j,\rho] \, ,
\end{align}
where we defined $\braket{G_j} = \tr{\rho\, G_j}$. Using $\Delta G \equiv G - \braket{G}$, the \sld\ for $\theta_k$ is then 
\begin{align}\label{eq:bgh498weosiuf}
 \mathscr{L}_\rho (\partial_k \rho) & = -2i \sum_{lm} \frac{[\Delta G_k,\rho]_{lm}}{p_l+p_m} \ket{l}\bra{m} \cr
 & = 2i \sum_{lm} \frac{p_l - p_m}{p_l + p_m} \braket{l|\Delta G_k|m} \ket{l}\bra{m} \, .
\end{align}
Similarly, we find
\begin{align}
 \partial_j \rho & = -i \sum_{n} p_n \left( \Delta G_j \ket{n}\bra{n} - \ket{n}\bra{n} \Delta G_j \right)  \, .
\end{align}
Putting this together in Eq.~(\ref{eq:qfivariance}), we obtain
\begin{align}\label{eq:ghrpws}
 [I_Q(\bm\theta)]_{jk} = 2 \sum_{lm} (p_l+p_m) \left( \frac{p_l - p_m}{p_l + p_m} \right)^2 \braket{l|\Delta G_k|m}  \braket{m|\Delta G_j|l}\, .
\end{align}
Writing this in symmetric form and noting that 
\begin{align}
  \left( \frac{p_l - p_m}{p_l + p_m} \right)^2 \leq 1 \quad\text{and}\quad \sum_{lm} (p_l+p_m) A_{lm} = 2 \sum_{lm} p_l A_{lm} \, ,
\end{align}
for any symmetric matrix $A$ and probabilities $0\leq p_m \leq 1$, we infer that 
\begin{align}
 [I_Q(\bm\theta)]_{jk} &\leq 4 \Tr{\rho \frac12\{ \Delta G_j,\Delta G_k\}}\, .
\end{align}
When $\rho$ is a pure state, the trace reduces to the matrix element $\cov_S(\mathbf{G})_{jk}$ in Eq.~(\ref{eq:covG}). We can therefore define a more general symmetrized covariance matrix for the generators of translation
$\cov_S(\mathbf{G})_{jk} =  \tr{\rho \frac12\{ \Delta G_j,\Delta G_k\}}$.
Similarly it is easy to show that we can cast the \qfi\ matrix in non-symmetric form using the real part of a suitably generalised non-symmetric covariance matrix for the generators.

Returning for the moment again to the \qfi\ matrix for pure states, another useful form for $I_Q (\bm\theta)$ in Eq.~(\ref{eq:pureqfimatrix}) is \cite{Gammelmark14}
\begin{align}\label{eq:gammelmark}
 [I_Q (\bm\theta)]_{jk} = 4 \partial_j^{(1)} \partial_k^{(2)} \left. \log\abs{\braket{\psi(\bm\theta^{(1)})|\psi(\bm\theta^{(2)})}} \right|_{\bm\theta^{(1)} = \bm\theta^{(2)} = \bm\theta}.
\end{align}
This can be easily shown by explicitly evaluating the derivatives. The expression then reduces to Eq.~(\ref{eq:pureqfimatrix}).

In Lyapunov form, the \qfi\ for multiple parameters becomes 
\begin{align}\label{eq:nvgur4809weiojs}
 [I_Q(\bm\theta)]_{jk} = 2\int_0^\infty \dif s \; \Tr{(\partial_j \rho) e^{-\rho s} (\partial_k \rho) e^{-\rho s}}\, .
\end{align}
We can use this form to construct an analytic expression for the \qfi\ without having to evaluate the integral \cite{Safranek18}. It makes use of the concept of \emph{vectorisation} of matrices, denoted by $\vect A$ of a matrix $A$, where the columns of a matrix are put below each other in a single column:
\begin{align}\label{eq:30i49eordjfxc}
 [I_Q(\bm\theta)]_{jk} = 2 \vect (\partial_j \rho)^\dagger \left( \rho^* \otimes \identity + \identity \otimes \rho \right)^{-1} \vect (\partial_k \rho)\, .
\end{align}
Similarly, we can construct the \sld:
\begin{align}\label{eq:iu4ierjkd}
 \vect L_j = 2 \left( \rho^* \otimes \identity + \identity \otimes \rho \right)^{-1} \vect (\partial_j \rho)\, .
\end{align}
These expressions are valid for finite-dimensional systems, and can be calculated directly based on matrix forms of the density matrix $\rho$ and its derivatives $\partial_j \rho$. When $\rho$ is pure, the inverse $( \rho^* \otimes \identity + \identity \otimes \rho )^{-1}$ does not exist. However, we can circumvent this problem by using instead the density matrix 
\begin{align}
 \rho_\epsilon = (1-\epsilon) \rho + \epsilon\, \frac{\identity}{d} \, ,
\end{align}
where $d$ is the dimension of the Hilbert space of $\rho$. Calculating the \qfi\ and \sld\ based on $\rho_\epsilon$, and taking the limit of $\epsilon\to0$ then retrieves the forms of the \qfi\ and \sld\ for pure states \cite{Safranek18}. 

\subsection{The quantum Cram\'er-Rao bound}\label{subsec:qcrb_another}
\noindent
Now that we have the \qfi\ matrix in terms of the \sld s $L_j$ associated with $\theta_j$, as given in Eq.~(\ref{eq:bgh498weosiuf}), we can derive the multi-parameter quantum Cram\'er-Rao bound. This inequality was originally first derived by Helstrom \cite{Helstrom73,Helstrom68} and is occasionally referred to as the Helstrom bound. We start with the following identity for \emph{unbiased} estimators
\begin{align}
 \Tr{(\partial_j \rho) (M_k - \theta_k)} = \delta_{jk}\, ,
\end{align}
where $M_k$ is the estimator for $\theta_k$ and $\delta_{jk}$ is the Kronecker delta. Using the \sld, this can be written as
\begin{align}
  \delta_{jk} = \frac12 \Tr{(\rho L_j + L_j \rho) (M_k - \theta_k)} \, .
\end{align}
Next, we introduce two real-valued vectors $\y$ and $\z$ such that
\begin{align}
  \sum_{k} z_j y_k & = \frac12 \sum_{jk} \Tr{z_j (\rho L_j + L_j \rho) (M_k - \theta_k) y_k} \cr
  & = \Re \Tr{\rho \left( \sum_j z_j L_j\right) \left( \sum_k y_k (M_k - \theta_k) \right)}\, .
\end{align}
We can square both sides of the equation, and note that $(\Re u)^2 \leq \abs{u}^2$, with equality if and only if $\Im u = 0$:
\begin{align}\label{eq:youghaldbfjsdv}
  (\z^\top\y)^2  \leq \Abs{\Tr{\rho \left( \sum_j z_j L_j\right) \left( \sum_k y_k (M_k - \theta_k) \right)}}^2\, ,
\end{align}
where $\z^\top\y$ is the standard dot product between two real-valued vectors. Using the Schwarz inequality for traces in Eq.~(\ref{eq:Schwarztrace}), we can write Eq.~(\ref{eq:youghaldbfjsdv}) as
\begin{align}
  (\z^\top\y)^2 & \leq \Tr{\rho \left( \sum_j z_j L_j\right)^2} \Tr{\rho \left( \sum_k y_k (M_k - \theta_k) \right)} \cr
  & \equiv (\z^\top I_Q \z) (\y^\top \cov(\bm\theta)\, \y) \, , 
\end{align}
where we identified the \qfi\ matrix and defined the covariance matrix
\begin{align}
  \cov(\bm\theta)_{jk} = \Tr{\rho (M_j - \theta_j) (M_k - \theta_k)} \, .
\end{align}
For the choice of $\z = I_Q^{-1}\y$, we obtain the inequality
\begin{align}
 \y^\top \cov(\bm\theta)\, \y \geq \y^\top I_Q^{-1} \y\, .
\end{align}
This bound is valid for any real vector $\y$, and therefore simplifies to 
\begin{align}\label{eq:mpqcrb}
 \cov(\bm\theta) \geq I_Q^{-1},
\end{align}
where the matrix inequality $A\geq B$ means that $A-B$ is a positive semi-definite matrix. This is the famous multi-parameter quantum Cram\'er-Rao bound. 
Given that the \textsc{qfi} elements transform as a metric tensor, the \textsc{qfi} for the vector of parameters $\smash{\bm{\vartheta}}$ can be written in terms of the \textsc{qfi} for $\smash{\bm{\theta}}$, and the \qcrb\ becomes
\begin{align}
\cov(\check{\bm\vartheta}) \geq I_Q(\bm{\vartheta})^{-1} = J^{-1} I_Q(\check{\bm\theta})^{-1} [J^\top]^{-1}.
\label{eqn:qcrb_for_functions}
\end{align}
where $J$ is the transformation Jacobian with matrix elements $\smash{J_{kl} = \partial\theta_k/\partial\vartheta_l}$.

We can immediately infer the \mse\ of a parameter $\theta_j$ as
\begin{align}
 \var \theta_j \geq [I_Q^{-1}]_{jj}\, ,
\end{align}
since the variances are the diagonal elements of the covariance matrix. Note that $[I_Q^{-1}]_{jj} \geq [I_Q]_{jj}^{-1}$, since the \qfi\ is a positive definite matrix. Given a covariance matrix and a positive-definite risk matrix $\mathbf{R}$, we can balance the precision of the various parameters. This leads to the inequality
\begin{align}
 C \equiv \tr{\mathbf{R} \cov(\bm\theta)} \geq \Tr{\mathbf{R} I_Q^{-1}} \equiv C_Q\, .
\end{align}
The \qcrb\ for a single parameter can in principle be achieved asymptotically by a suitable measurement. The question is whether the multi-parameter \qcrb\ can be attained. We explore this in the next subsection.

\subsection{Saturating the quantum Cram\'er-Rao bound}\label{subsubsec:the_qfim}
\noindent
The the optimal unbiased quantum estimators that saturate the \textsc{qcrb} takes the form
\begin{align}
{M}_j(\bm{\theta}) = {\theta_j}\mathbbm{1} + [\bm{I}_Q(\bm{\theta})^{-1}\cdot\mathbf{L}]_j,
\label{eq:optimal_measurements}
\end{align}
which form a set of self-adjoint operators~\cite{Paris2009_IJQI}. They are linear combinations of the \sld s $\mathbf{L} = (L_1,\ldots,L_D)$. Determining the measurement ${M_j}(\bm{\theta})$ is typically a difficult task, since it depends on $\bm{\theta}$. To overcome this difficulty, adaptive measurements have been suggested~\cite{Berry00, Berry2002_PRA}. An important caveat to the multi-parameter \qcrb\ is that the \textsc{qcrb} for multiple parameters is generally not saturable, since the optimal observables in Eq.~\eqref{eq:optimal_measurements} may not be compatible. It is easy to see that this may occur when the \textsc{sld}s associated with the parameters do not commute. However, there is a bit more to it than that. 
 
To explore the attainability of the multi-parameter \qcrb, we consider a more general bound derived by Holevo~\cite{Holevo82,Nagaoka82}.
\begin{align}\label{eq:hcrb}
 C \geq \min_{\{M_j\}} \left( \Tr{\mathbf{R}\; \Re J} + \Norm[1]{\sqrt{\mathbf{R}}\; (\Im J) \sqrt{\mathbf{R}}} \right) \equiv C_H\, ,
\end{align}
where $\bm{R}$ denotes a positive definite weight matrix, and $J_{jk} = \tr{{M}_j {M}_k \rho(\bm\theta)}$ and $\norm[1]{\cdot}$ is the trace norm \cite{Ragy16}. The bound in Eq.~\eqref{eq:hcrb} is the Holevo Cram{\'e}r Rao bound (\textsc{hcrb}) and defines a scalar lower bound on the weighted mean square error and represents the best precision attainable with global measurements on an asymptotically large number of identical copies of a quantum state~\cite{Guta2006_PRA,Hayashi2008_JMP,Yamagata2013_AS,Yang2019_CMP}. 

Despite its importance for practical metrology, the \textsc{hcrb} has seen limited use in multi-parameter quantum metrology. This is due to the non-trivial optimisation over a set of observables and the implementation of global measurements is a difficult task. However, few results that use the \textsc{hcrb} do exist for qubit state estimation~\cite{Suzuki16}, two-parameter estimation with pure states~\cite{asymptotic_theory_book_2005} and two-parameter displacement estimation with two-mode Gaussian states~\cite{Bradshaw2017_PLA,Bradshaw2018_PRA}. A recent study by Albarelli \emph{et al.} have investigated the numerical tractability of calculating the \textsc{hcrb} for multi-parameter estimation problems~\cite{Albarelli19}.


The inequality in Eq.~\eqref{eq:hcrb} follows~\cite{Nagaoka82} from the lemma that given the identity $\int_\Omega f(\omega) M(\dif \omega) = F$ we have 
\begin{align}\label{eq:084urouhsjasnd}
 \int_\Omega \abs{f(\omega)}^2 M(\dif\omega) \geq F F^\dagger\, ,
\end{align}
where $M$ is an observable with outcomes in $\Omega$, $f(\omega)$ a complex function on $\Omega$, and $F$ an operator. We choose $f(\check{\bm\theta}) = \sum_j \xi_j (\check{\theta}_j - \theta_j)$ and $F = \sum_j \xi_j M_j$, substitute them into Eq.~\eqref{eq:084urouhsjasnd} and take the expectation value with respect to $\rho$. Taking into account a risk matrix $\mathbf{R}$ and optimising over all $M_j$ we obtain Eq.~\eqref{eq:hcrb}.

The Holevo form of the \qcrb\ can be attained when the statistical model involves the broad class of Gaussian state shifts where the parameters are encoded in shifts of the first moment\cite{Holevo82,Hayashi2008_JMP,Kahn09,Yamagata2013_AS}. When we choose the  measurements $M_j$ derived from the \sld s in Eq.~(\ref{eq:optimal_measurements}) and substitute them into $C_H$, we find that the first term becomes the standard \qcrb: $\min_{\{M_j\}} \tr{\mathbf{R}\, \Re J} \to \tr{\mathbf{R}\, I_Q^{-1}}$, and the second term becomes $\norm[1]{\mathbf{R}\, \Im I_Q^{-1}}$. Since $I_Q^{-1}$ is a real positive definite matrix, this term vanishes, and $C_H$ reduces to the \qcrb\ based on the \sld s. The significance of Eq.~\eqref{eq:hcrb} is that for non-commuting \sld s there may be a different set of observables that outperform the $M_j$ in Eq.~\eqref{eq:optimal_measurements}. 

Next, we note that $\Im J$ can be written in terms of the commutator 
\begin{align}
  \Im J_{jk} = \frac12 \tr{\rho [M_j,M_k]}\, . 
 \end{align}
 Assuming $\mathbf{R}$ and $I_Q$ strictly positive matrices, and noting that $\tr{\rho [M_j,M_k]}=0$ implies that $\tr{\rho [L_j,L_k]}=0$, a necessary and sufficient condition for the saturability of the multi-parameter \qcrb\ is then~\cite{Matsumoto2002_JPA,Monras2011_PRA,Gill2012_arxiv,Ragy16,Vaneph2013_QMQM,Vidrighin2014_NC,Crowley2014_PRA,Suzuki16}
\begin{align}\label{eqn:asymp_attainability_criteria}
 \Tr{\rho [L_j,L_k]} = 0\, .
\end{align}
This is of course a weaker condition than requiring that the \sld s commute directly. The condition in Eq.~\eqref{eqn:asymp_attainability_criteria} is necessary and sufficient for unitary evolutions on pure states~\cite{Matsumoto2002_JPA}, which is equivalent to requiring the existence of commuting generators that generate the evolution of the probe. For mixed states, the demands to realise optimal simultaneous estimation are more involved. Specifically, we require the existence of a single probe state that maximises the \textsc{qfi} for all values of $\smash{\bm{\theta}}$, a compatible measurement that ensures saturability of the \textsc{qcrb}, and a diagonal \textsc{qfim}, which would allow independent estimations of each parameter~\cite{Ragy16}. Alternative methods to provide better precision bounds may involve collective measurements over many independent copies of the system, which is experimentally challenging.

From the above discussion we see that the \textsc{sld} operator plays a pivotal role in quantum estimation theory. For a multi-parameter estimation problem, finding the \textsc{sld} for each parameter in $\smash{\bm{\theta}}$ is sufficient to inform whether a simultaneous, efficient estimation can be performed. It also prescribes the optimal estimator that saturates the \textsc{qcrb}; the fundamental limit to estimation precisions allowed by quantum mechanics. We therefore turn our attention to find a functional form for the \textsc{sld}. 

\subsection{Simultaneous versus sequential estimation}
\label{subsec:sim_parrallel_comparison}

\noindent
Multi-parameter quantum estimation is important for modeling a wider class of physical systems. For example, it may be necessary to infer the value of a parameter by estimating a set of related but different parameters. Also, there are examples where knowledge of multiple parameters are required, such as for microscopy, optical, electromagnetic, and gravitational field imaging. One approach for the estimation of multiple parameters is to prepare individual optimal probe and measurements schemes for each parameter. However, this is generally challenging to implement experimentally. It would also be unsuitable for sensing dynamically evolving probes. Instead, a natural approach would be to simultaneously estimate each parameter at the same time.  The \textsc{qcrb} matrix bounds the precision of simultaneous multi-parameter estimates, and it could in principle be faster to implement the measurements simlutaneously with fewer resources. For example, in the case of estimating $D$ phases, Humphreys \emph{et al.}~\cite{Humphreys2013_PRL} have demonstrated that simultaneous estimation provides an intrinsic $\mathcal{O}(D)$ precision improvement over the best quantum scheme for individual measurements devised by Lee, Kok, and Dowling~\cite{Lee2002_JMO} using \textsc{noon} states. A similar advantage has been demonstrated by Baumgratz and Datta for multi-field estimation~\cite{Baumgratz2016_PRL}. Within the literature, simultaneous strategies are also referred to as parallel estimation strategies. Even for large photon losses in phase imaging applications, simultaneous estimation schemes can provide a constant factor advantage over individual schemes. This has seen a surge of recent work focused on yielding quantum enhanced sensing from simultaneous estimation of multiple parameters~\cite{Fujiwara1994_METR, Monras2011_PRA, Genoni2013_PRA, Vidrighin2014_NC, Yao14_PRA, Kok2017_PRA}.

The principal difficulty of simultaneous multi-parameter estimation arises from the non-attainability of the precision bounds when the infinitesimal generators associated with the parameters do not mutually commute~\cite{Helstrom74,Belavkin1976_TMP,Fujiwara2001_PRA2}. This can lead to the requirement of non-Hermitian measurements to saturate precision bounds~\cite{Yuen73}, which would be difficult to realise. We also saw in the previous section that the necessary condition to saturate the \textsc{qcrb} asymptotically is a zero expectation value of the commutator of the \textsc{sld}s, as shown in Eq.~\eqref{eqn:asymp_attainability_criteria}. If this condition holds, there exists a common optimal measurement basis for simultaneous estimates. This case is well studied with general results derived for the optimal estimation of multiple phases. For example, Macchiavello calculates the cost associated with the estimation and finds it increases with increasing number of phase measurements~\cite{Macchiavello2003_PRA}. Ballester demonstrated that entanglement in the probe states and measurements provides no advantage for estimating multiple phases~\cite{Ballester2004_PRA2}. If however, Eq.~\eqref{eqn:asymp_attainability_criteria} does not hold, there is no simultaneous eigenbasis for the optimal simultaneous estimation of all parameters. One option to counter this would be to search for alternative informative bounds. However, this depends strongly on the specific estimation problem. For example, Genoni \emph{et al.}\ find that for the estimation of complex parameters, the so-called right logarithmic derivative is more informative~\cite{Genoni2013_PRA} (see section~\ref{subsec:rld_multipara}). We have already seen alternative bounds based on the \textsc{sld} and the Holevo bound.

Several studies have illuminated progress towards simultaneous estimation of parameters with non-commuting \textsc{sld}s. Even when the condition in Eq.~\eqref{eqn:asymp_attainability_criteria} is not satisfied, a precision advantage may still be granted when estimating all of the parameters simultaneously~\cite{Ragy16}. For non-commuting unitary operations, it has been shown that entanglement in both the probe states and measurements can attain the Heisenberg limit~\cite{Ballester2004_PRA,Imai07_JPA}. Constructing the optimal simultaneous projective measurements requires knowledge of the \textsc{povm}s that saturate the \textsc{qcrb}. Knowledge of this provides an instructive guide for the design of multi-parameter estimation experiments and has been heavily investigated. For single parameter estimation, finding an optimal measurement for which the equality $I(\theta) = I_Q(\theta)$ holds is always possible~\cite{Braunstein94, Braunstein95}. Unfortunately, this does not trivially extend to multi-parameter, where few analytic results are known for specific physical systems. The first is the estimation of single qubit mixed states~\cite{Bagan2006_PRA,Hayashi2008_JMP}. Simultaneous optimal estimation has also been considered for the class of pure states. Determining the attainability of the \textsc{qcrb} for pure states was considered by Matsumoto. Specifically, if
\begin{align}
\text{Im} \left[\Braket{\partial_l\psi_\theta|\partial_m\psi_\theta}\right] = 0 \; \forall \; l, m
\label{eqn:weak_commutativity_theorem}
\end{align}
and the \textsc{qfim} is invertible, then a measurement scheme exists for which the classical Fisher information saturates the \textsc{qfim}~\cite{Matsumoto2002_JPA}. This is weaker constraint than the requirement of commuting Hamiltonians for each parameter $\smash{[{H}_j, {H}_k] = 0}$ for all $j$, $k$ used for the estimation of multi-dimensional fields~\cite{Baumgratz2016_PRL}. Further conditions on projective measurements for optimal phase-like estimation schemes using pure states has been discussed by Pezze \emph{et al.}~\cite{Pezze2017_PRL}. First, assuming we have a pure probe state $\ket{\psi_\theta}$ evolving under unitary dynamics and satisfying Eq.~\eqref{eqn:weak_commutativity_theorem}, then the projectors $\{\ket{\Gamma_k}\bra{\Gamma_k}\}$ constructed from the probe state and from vectors on the orthogonal subspace always saturates the \textsc{qcrb}, which can be experimentally realised. Pezze \emph{et al.}\ also demonstrate that there always exists a set of projectors nonorthogonal to the probe such that $\braket{\Gamma_k|\psi_\phi} = 0$ for all $k$ that saturates $I(\theta) = I_Q(\theta)$ if and only if~\cite{Pezze2017_PRL}
\begin{align}
\text{Im} \left[\Braket{\partial_l\psi_\theta|\Gamma_k}\Braket{\Gamma_k|\psi_\theta}\right] = \Abs{\Braket{\psi_\theta|\Gamma_k}}^2 \text{Im}\left[\Braket{\partial_l\psi_\theta|\psi_\theta}\right].
\label{eqn:theroem2}
\end{align}
for all $l = 1,2, \ldots, D$ and all $k \neq 1$. Finally, any general projective measurement for pure probe states under unitary dynamics is optimal if
\begin{align}
\lim_{\phi\to\theta} \frac{\text{Im} \left[\Braket{\partial_l\psi_\phi|\Gamma_k}\Braket{\Gamma_k|\psi_\phi}\right]}{\Abs{\Braket{\Gamma_k|\psi_\phi}}} = 0,
\label{eqn:theroem1}
\end{align}
for all indices $l$, $m$ and all $k$ for which $\braket{\Gamma_k|\psi_\phi}=0$ and if Eq.~\eqref{eqn:theroem2} is fulfilled for all indices $l$ and all $k$ for which $\braket{\Gamma_k|\psi_\phi}\neq0$. These conditions are necessary and sufficient for saturating the multi-parameter \textsc{qcrb} provided the \textsc{qfim} is invertible. Further conditions for when the information content from simultaneous estimation of multiple parameters matches or exceeds that from separable measurements for general probe states was analysed by Ragy \emph{et al.}~\cite{Ragy16}.

Generally, the conditions for optimal simultaneous measurements may be hard to realise. This makes it difficult or impossible to saturate the \textsc{qcrb} matrix, and several tradeoffs are considered. The first  deals with the probe states, since the optimal probe is generally different for each of the $D$ parameters~\cite{Yuan2016_PRL,Szczykulska2017_QST}. The second tradeoff deals with the choice of measurements for each parameter, as the optimal measurements for different parameters are usually incompatible~\cite{Gill2000_PRA}. To impose meaningful quantifiers of these tradeoffs, a cost function is introduced to assign a figure of merit. A general cost function, based on the Fisher information for a particular choice of probe state and measurement strategy is
\begin{align}
C_{{\Pi}}(I_Q) = \Tr{I_Q(\bm{\theta})I(\bm{\theta})^{-1}},
\label{eqn:performance_quantifier}
\end{align}
where $\smash{I(\bm{\theta})^{-1}}$ is the inverse of the classical Fisher information for a given \textsc{povm}, ${\Pi}$. This quantifier enables optimisation over probe states and measurement strategies to address general tradeoffs in multi-parameter estimation schemes with non-commuting \textsc{sld}s. Although any positive definite matrix can be used as the weight matrix in the cost function in Eq.~\eqref{eqn:performance_quantifier}, the \textsc{qfim} is a natural choice for state estimation since it maximises the average fidelity between the estimated state and the actual state~\cite{Bagan2006_PRA}. Notice that the cost function is lower bounded by $\smash{C_{{\Pi}}(I_Q) \geq D}$, with equality applicable for measurements that are simultaneously optimal for all parameters. This cost function has been used to determine general bounds for thermometry of distant black bodies~\cite{Pearce17}.

Using this strategy, a tradeoff between how well different parameters may be estimated is often considered. This approach has been explored by Crowley \emph{et al.}\ for the estimation of phase $\phi$ and loss $\eta$ in optical interferometry, where saturating the \textsc{qcrb} for both parameters simultaneously is impossible~\cite{Crowley2014_PRA}. However, in many instances there is a single measurement observable for both $\phi$ and $\eta$ that saturates the Holevo \crb~\cite{Albarelli19}.

For a general two-mode pure probe state with definite photon number $N$, Demkowicz-Dobrzanski \emph{et al.}\ show that in lossy interferometers the density matrix for the evolved state can be written as the direct sum~\cite{Dobrzanski2009_PRA} $\rho(\bm{\theta}) = \oplus_{j=0}^N p_j \ket{\psi_j}\bra{\psi_j}$. An intuition behind the origin for the tradeoff between precisions in phase and loss estimation can be understood by noting that the derivatives $\smash{\partial_\phi\ket{\smash{\psi_j}}}$ and $\smash{\partial_\eta\ket{\smash{\psi_j}}}$ are the same up to an imaginary constant~\cite{Crowley2014_PRA}. Hence, estimating $\bm{\theta}=(\phi,\eta)^\top$ corresponds to measuring along the axes given by $\smash{\ket{\smash{\psi_j}} \pm \partial_\phi\ket{\smash{\psi_j}}}$ and $\smash{\ket{\smash{\psi_j}} \pm i \partial_\phi\ket{\smash{\psi_j}}}$~\cite{Fujiwara2001_PRA2}. Using a similar quantifier as in Eq.~\eqref{eqn:performance_quantifier}, the \textsc{qfi} for phase measurements can be saturated at the cost of the following diminished information on the loss parameter~\cite{Crowley2014_PRA}
\begin{align}
{I}(\eta) = \left[I_Q(\bm{\theta})\right]_{\eta\eta} - \frac{1}{4\eta^2}\left[I_Q(\bm{\theta})\right]_{\phi\phi}.
\label{eqn:trade_off_Crowley}
\end{align}
For the same probe state, a similar estimation trade-off has been explored for the estimation of phase and phase diffusion for different levels of noise~\cite{Szczykulska2017_QST}. 

Besides optimising over probe states and measurement strategies, a further optimisation over multi-parameter encoding channels has been considered in the literature. The two channel schemes that examine this optimisation is parallel and sequential estimation schemes. An important question then is which scheme is better suited for different quantum information tasks. In quantum estimation, based on general precision bounds for single parameters in the presence of noise, parallel schemes perform better than sequential feedback schemes~\cite{Escher2011,Dobrzanski2014_PRL}. For Hamiltonian parameter estimation on $D$-dimensional systems, sequential feedback has been demonstrated to provide an order of $\mathcal{O}(D+1)$ improvement over parallel schemes~\cite{Yuan2016_PRL}. In quantum channel discrimination, the comparison of these two strategies has drawn a lot of attention~\cite{Acin2001_PRL,DAriano2001_PRL,Duan2007_PRL}. In this scenario, sequential feedback schemes outperform parallel schemes, where the use of ancillary systems was necessary~\cite{Chiribella2008_PRL,Harrow2010_PRA}. However, the use of ancillary systems or entanglement was relaxed in a separate protocol by Duan \emph{et al.}, who developed the optimal sequential scheme to discriminate between different quantum channels~\cite{Duan2009_PRL}.

\subsection{The Right Logarithmic Derivative}\label{subsec:rld_multipara}
\noindent
The classical Fisher information is the unique Riemannian metric on the space of classical probability distribtions that has the property of contraction under coarse graining. However, in the quantum-mechanical case the possibility of non-commuting operators breaks this uniqueness, and instead we obtain a family of metrics~\cite{Petz96b}. It is therefore interesting to explore some alternative quantum extensions of the Fisher information.

The \qfi\ and the \sld\ are derived from the symmetrised correlation $\frac12\braket{\{A,B\}}$ between two observables $A$ and $B$ (see Eq.~\eqref{eq:bghuo04woesjk}). This correlation can be interpreted as an instance of an inner product in the space of linear operators
\begin{align}
 (A,B)_\rho \equiv \frac12 \Tr{\rho\left( B A^\dagger + A^\dagger B \right)}\, ,
\end{align}
which also applies to operators $A$ and $B$ that are not self-adjoint. As shown in section~\ref{sec:qe_theory}, this inner product defines the metric operator $\mathscr{L}_\rho$ that is used to construct the \sld\ and the \qfi. Our choice of the inner product was motivated by the possible non-commutativity of $A$ and $B$. The \qfi\ matrix elements are defined in terms of this inner product as
\begin{align}
 [I_Q]_{jk} = (L_j,L_k)_\rho\, ,
\end{align}
with the usual definition for the \sld s.

However, there are also other definitions for the inner product we could use. For example, we could define the inner product as~\cite{Yuen73,Fujiwara94}
\begin{align}
 \braket{\!\braket{A,B}\!}_\rho \equiv \Tr{\rho B A^\dagger}\, .
\end{align}
This leads to a different metric, and correspondingly to a different logarithmic derivative. Following the same procedure as in section~\ref{sec:qe_theory}, the inner product leads to a raising operator $\mathscr{R}_\rho^{(R)}(B) = \rho B$. Since the lowering operator is defined as the inverse of the raising operator, we find
\begin{align}\label{eq:vh3498wouhrsd}
 \mathscr{L}_\rho^{(R)} (\rho B) = B\, .
\end{align}
When we let the lowering operator act on the derivative $\partial_j \rho$, we obtain the so-called \emph{Right} Logarithmic Derivative~\cite{Belavkin72,Yuen73,Helstrom74} (\rld)
\begin{align}
  R_j \equiv \mathscr{L}_\rho^{(R)} (\partial_j \rho) \, ,
  \end{align} 
which from Eq.~\eqref{eq:vh3498wouhrsd} can be written as
\begin{align}
  \partial_j \rho = \rho R_j = R_j^\dagger \rho\, .
\end{align} 
The information matrix corresponding to these \rld s is given by 
\begin{align}
 [I_R]_{jk} = \Braket{\!\Braket{R_j,R_k}\!}_\rho = \Tr{R_j^\dagger \rho R_k }\, .
\end{align}
This inner product is a different way to define a metric in the space of linear operators, leading to an alternative way of measuring the distance between density matrices along curves parametrised by $\bm\theta$. The question is when and how this is useful for metrology.

We can derive a bound on the covariance matrix for $\bm\theta$ analogous to the \qcrb, but instead based on the \rld. Unbiased estimators $M_j$ for $\theta_j$ obey $\tr{\rho M_j} = \theta_j$, and we find
\begin{align}\label{eq:urhkfj}
 \Tr{(\partial_k \rho) (M_j - \theta_j)} = \Tr{\rho (M_j - \theta_j) R_k^\dagger} = \delta_{jk}\, .
\end{align}
The fact that the \rld\ involves a simple product of operators $R_j$ and $\rho$ means that we can left- and right-multiply Eq.~(\ref{eq:urhkfj}) by two {complex} vectors $\mathbf{y} = (y_1,\ldots,y_D)$ and $\mathbf{z} = (z_1,\ldots,z_D)$, respectively:
\begin{align}
 \sum_{jk} \Tr{y_j^* \rho (M_j-\theta_j) R_k^\dagger z_k} = \sum_j y_j^* z_j\, .
\end{align}
We separate the trace into a product of operators 
\begin{align}
 \sqrt{\rho} \sum_j y^*_j (M_j - \theta_j) \qquad\text{and}\qquad \sqrt{\rho} \sum_k z^*_j R_j\, ,
\end{align}
and use the Schwarz inequality for the trace to obtain the following inequality:
\begin{align}
 \Abs{\mathbf{y}^\dagger \mathbf{z}}^2 \leq \left( \sum_{jk} z_j^* \Tr{\rho R_j R_k^\dagger} z_k \right) \left( \sum_{jk} y_j^* \cov(\bm\theta)_{jk} y_k \right)\, .
\end{align}
We can choose $\mathbf{y}$ and $\mathbf{z}$ any way we want. Using $[I_R]_{jk} = \tr{\rho R_j R_k^\dagger}$ and setting $\mathbf{z} = I_R^{-1}\mathbf{y}$, the resulting inequality is
\begin{align}
 \mathbf{y}^\dagger \cov(\bm\theta)\, \mathbf{y} \geq \mathbf{y}^\dagger I_R^{-1} \mathbf{y}\, .
\end{align}
Again, since this must be true for any complex-valued vector $\y$, we obtain the inequality
\begin{align}\label{eq:rldbound}
 \cov(\bm\theta) \geq  I_R^{-1} \, .
\end{align}
Clearly, the matrix $I_R$ plays a role similar to the \qfi\ $I_Q$, but now using the \rld. Expressed using a cost matrix $\mathbf{R}$, the bound based on the \rld\ becomes 
\begin{align}\label{eq:bgrw08uosfi}
 \tr{\mathbf{R} \cov(\bm\theta)} \geq \Tr{\mathbf{R} I_R^{-1}} \equiv C_R \,.
\end{align}
There are instances in which the this bound is tighter (i.e., more informative) that the \qcrb, or $C_R > C_Q$.

For a single parameter the \qcrb\ based on the \sld\ is always higher than the bound based on the \rld~\cite{Helstrom1976}. To show this, we write the \rld\ in the diagonal basis of the density operator. First, we use the definitions
\begin{align}
 \partial_\theta \rho = \sum_{mn} (\partial_\theta \rho)_{mn} \ket{m}\bra{n}\, .
\end{align}
and
\begin{align}
 R_\theta = \sum_{jk} R_{jk} \ket{j}\bra{k}\, .
\end{align}
The \rld\ can then be written as 
\begin{align}
 R_\theta = \sum_{jk} \frac{(\partial_\theta \rho)_{jk}}{p_j} \ket{j}\bra{k}\, .
\end{align}
From this it follows that 
\begin{align}\label{eq:ght2038wriuhf}
 I_R & = \Tr{(\partial_\theta \rho) R_\theta} \cr & = \sum_{jk} \frac{\Abs{\braket{j|(\partial_\theta \rho)|k}}^2 }{p_j}  = \sum_{jk} \frac{\Abs{\braket{j|(\partial_\theta \rho)|k}}^2 }{p_k} \cr
 & = \frac12 \sum_{jk} \left( \frac{1}{p_j} + \frac{1}{p_k} \right) \Abs{\Braket{j|(\partial_\theta \rho)|k}}^2\, .
\end{align}
We can now compare this with the \qfi\ for a single parameter as in Eq.~(\ref{eq:qfieigenbasis}), and note that 
\begin{align}
  \frac{2}{p_j + p_k} \leq \frac12 \left(\frac{1}{p_j} + \frac{1}{p_k} \right)\, .
\end{align}
This means that for single parameter estimation problems $I_Q \leq I_R$ (when $R_\theta$ exists), and the \qcrb\ based on the \sld\ is never less than the \rld-based bound. Therefore, the \rld\ bound is interesting only when we consider multiple parameters. 

Another constraint is that the \rld\ does not exist for pure states. Nevertheless, a meaningful bound may still be constructed, relying on the fact that the non-existence of $R_j$ for pure states does \emph{not} imply the non-existence of $I_R^{-1}$~\cite{Fujiwara94}. Fujiwara gives two examples. First, $I_R^{-1}$ for the quadrature measurements $(q,p)$ of a pure coherent state $\ket{\alpha}$ with complex amplitude $\alpha$ and frequency $\omega$ is given by
\begin{align}\label{eq:fvgnjhruw4oeis}
 I_R^{-1} = \frac{1}{2}
 \begin{pmatrix}
  \omega & i \\ -i & \omega^{-1}
 \end{pmatrix}\, .
\end{align}
The information in Eq.~\eqref{eq:fvgnjhruw4oeis} is interesting, since the associated bound in Eq.~\eqref{eq:bgrw08uosfi} takes the form of the additive uncertainty relation~\cite{Maccone14} for the rescaled position and momentum
\begin{align}
 \Delta\tilde{q}^2 +  \Delta\tilde{p}^2 \geq 1\, ,
\end{align}
where $\tilde{q} = q/\sqrt{\omega}$ and $\tilde{p} = \sqrt{\omega}\,p$. This is in contrast to the \qcrb\ using the \qfi\ for the evolution $U = \exp(-i\tilde{q} \tilde{p})$ with $\tilde{p}$ the momentum operator, which recovers the standard form of the uncertainty relation $\Delta\tilde{q} \, \Delta\tilde{p} \geq \frac12$.

Second, for a measurement of a rotation $R(\theta,\phi)$ of a spin-$j$ system in a state $\ket{j}$ oriented along the $z$-axis the information $I_R^{-1}$ becomes
\begin{align}
 I_R^{-1} = \frac{1}{\sin^2\theta}
 \begin{pmatrix}
  \sin^2\theta & -i\sin\theta \\ i\sin\theta & 1
 \end{pmatrix}\, ,
\end{align}
where $\theta$ and $\phi$ are the usual spherical coordinates. Again, this information translates via Eq.~\eqref{eq:bgrw08uosfi} into an additive uncertainty relation for the two rotation angles
\begin{align}
 \Delta\theta^2 + \sin^2\theta\; \Delta \phi^2 \geq 1\, .
\end{align}
Both these examples outperform the standard \qcrb, and provide the most informative bounds~\cite{Yuen73,Holevo82,Nagaoka82,Maccone14}.  Suzuki provides an explicit formula for the bound on two-parameter estimation using qubits~\cite{Suzuki16}.

Finally, the \textsc{rld} is generally not Hermitian. Consequently, the optimal estimator derived from these two quantities may not correspond to a physical \textsc{povm}. Despite this, the use of non-self adjoint operators have been demonstrated to saturate the multi-parameter \textsc{qcrb}~\cite{Yuen73,asymptotic_theory_book_2005}.

\bigskip

\noindent
It is interesting to consider how the \rld\ bound relates to the Holevo form of the \qcrb\ in Eq.~\eqref{eq:hcrb}. Nagaoka~\cite{Nagaoka82} defines the \rld\ bound in Eq.~\eqref{eq:bgrw08uosfi} as 
\begin{align}
 C_R = \Tr{\mathbf{R}\, \Re I_R^{-1}} + \Norm[1]{\sqrt{\mathbf{R}}\, (\Im I_R^{-1}) \sqrt{\mathbf{R}}}\, ,
\end{align}
which takes the form of Eq.~\eqref{eq:hcrb} when the observables $M_j$ are chosen such that $J = I_R^{-1}$. These are generally not the same obervables found through minimisation, and we therefore find 
\begin{align}
 C_H \geq \max\{ C_Q, C_R \}\, ,
\end{align}
and the Holevo form is the most informative bound on the covariance matrix of $\bm\theta$.

\subsection{Kubo-Mori information}\label{subsec:KMI}
\noindent
When the quantum system under consideration is in a thermal state, a more natural \crb\ can be found based on so-called Kubo-Mori information. The density operator of a thermal state can be written in exponential form as 
\begin{align}
 \rho = {\rm e}^{-\beta H -\theta A}\, ,
\end{align}
where $\beta = 1/k_B T$, with $k_B$ Boltzmann's constant, $T$ the temperature, $H$ the Hamiltonian, and we absorbed the partition function in $H$. The term $\theta A$ corresponds to a force on the system generated by $A$. 

The quantum relative entropy for two states $\rho$ and $\sigma$ is given by~\cite{Petz94} 
\begin{align}\label{eq:gh93woesjdn}
 D\infdiv{\rho}{\sigma} = \tr{\rho\log\rho} - \tr{\rho\log\sigma}\, .
\end{align}        
The quantum relative entropy for two states ${\rm e}^{-\beta H-\theta A}$ and ${\rm e}^{-\beta H-\theta' B}$ then becomes
\begin{align}
 D \infdiv{\rho(\theta,A)}{\rho(\theta',B)} & = 
 \tr{{\rm e}^{-\beta H -\theta A}(\beta H +\theta' B)} \cr & \phantom{=}~
 - \tr{{\rm e}^{-\beta H -\theta A}(\beta H +\theta A)}\, . 
\end{align}
Taking the derivative of the quantum relative entropy with respect to $\theta$ and $\theta'$, we obtain the relation
\begin{align}
 \frac{1}{\beta^2} \frac{\partial^2 D \infdiv{A}{B}}{\partial\theta \partial{\theta'}} & = \braket{A}\braket{B} \\
 & \phantom{=}~ - \frac{1}{\beta} \int_0^\beta \Tr{{\rm e}^{-\beta H} {\rm e}^{x H} A {\rm e}^{-x H}B} \dif x\, ,\nonumber
\end{align}
when $\theta'\to\theta$. This motivates the Bogoliubov inner product for thermal states:
\begin{align}\label{eq:vwoijsknxcv}
 (A,B)_\rho^{\rm (B)} = \int_0^1 \Tr{\rho^x A^\dagger \rho^{1-x} B}  \dif x\, ,
\end{align}
which can be used to construct a new logarithmic derivative~\cite{Petz93} called the Bogoliubov Logarithmic Derivative ({\sc bld})
\begin{align}
 \partial_\theta \rho =  \int_0^1 \rho^x B_\theta\; \rho^{1-x}\; \dif x\, .
\end{align}
and 
\begin{align}
 B_\theta = \int_0^\infty (x+\rho)^{-1} (\partial_\theta \rho)\, (x+\rho)^{-1}  \dif x\, ,
\end{align}
leading to the Kubo-Mori information 
\begin{align}\label{eq:gh94weosjd}
 I_{\rm KM} (\theta) = (B_\theta,B_\theta)_\rho^{\rm (B)} = \int_0^1 \Tr{\rho^x B_\theta^\dagger \rho^{1-x} B_\theta } \dif x\, .
\end{align}
The form of the Bogoliubov inner product in Eq.~\eqref{eq:vwoijsknxcv} originates in the theory of linear response of thermal systems using the canonical correlation between $A$ and $B$
\begin{align}\nonumber
 \int_0^1 \Tr{ \rho^x (A - \tr{\rho A}) \rho^{1-x} (B - \tr{\rho B}) } \dif x\, ,
\end{align}
and therefore the use of $I_{\rm KM}(\theta)$ in Eq.~\eqref{eq:gh94weosjd} can be considered more natural for statistical physics applications than the \qfi~\cite{Hayashi02}. For further details, see chapter 4 of Kubo, Toda, and Hashitsume~\cite{Kubo85}. 

A Cram\'er-Rao-type inequality can be derived when we consider the covariance of $M_\theta = \check{\theta}-\theta$, such that 
\begin{align}
 \cov_B (\theta) = \int_0^\infty \Tr{\rho^x M_\theta \rho^{1-x} M_\theta} \dif x\geq  I_{\rm KM}^{-1}(\theta)\, .
\end{align}
For a single parameter $\theta$, the Kubo-Mori information is equal to 
\begin{align}\label{eq:bveriuskfvjksxjnc}
 I_{\rm KM}(\theta) = \sum_{jk} \frac{\log p_j - \log p_k}{p_j + p_k} \Abs{\braket{j|\partial_\theta \rho |k}}^2 \,.
\end{align}
This is a more informative bound than the one based on the \rld\ in Eq.~\eqref{eq:ght2038wriuhf}, but less informative than the \qcrb\ based on the \sld\ in Eq.~\eqref{eq:qfieigenbasis}~\cite{Petz93}. The Kubo-Mori information gives the bound for consistent superefficient estimators~\cite{Hayashi02}, i.e., estimators that outperform maximum likelihood estimation. The Kubo-Mori information plays a central role in the construction of uncertainty relations between energy and temperature in quantum thermodynamics~\cite{Miller18}.

\subsection{Wigner-Yanase skew information}\label{subsec:WYI}
\noindent
We have seen in Eqs.~\eqref{eq:qfieigenbasis}, \eqref{eq:ght2038wriuhf}, and \eqref{eq:bveriuskfvjksxjnc} that we can write $I_Q$, $I_R$, and $I_{\rm KM}$ in terms of the matrix elements of the derivative of the density operator $\partial_\theta \rho$ in the basis of $\rho$:
\begin{align}
 I_Q (\theta) & = \sum_{jk} \frac{2}{p_j + p_k} \Abs{\braket{j|(\partial_\theta \rho)|k}}^2 \, ,\\
 I_R (\theta) & =  \sum_{jk} \left( \frac{1}{2p_j} + \frac{1}{2p_k} \right) \Abs{\Braket{j|(\partial_\theta \rho)|k}}^2 \, , \\
 I_{\rm KM} (\theta) & =  \sum_{jk} \frac{\log p_j - \log p_k}{p_j + p_k} \Abs{\braket{j|(\partial_\theta \rho) |k}}^2 \, .
\end{align}
The different types of quantum information are determined by the functions $c(p_j,p_k)$ of the eigenvalues of $\rho$ in the above equations. Morozova and Chentsov (and extended by Petz) proposed these functions as a classification of monotone Riemannian metrics on matrix spaces~\cite{Morozova,Petz96b}, and they include another interesting example:
\begin{align}
 I_{\rm WY} (\theta) & =  \sum_{jk}\frac{4}{\bigl(\sqrt{p_j}+\sqrt{p_k}\bigr)^2} \Abs{\braket{j|(\partial_\theta \rho) |k}}^2 \, .
\end{align}
This is the Wigner-Yanase skew information~\cite{Wigner63}. When the parameter $\theta$ is generated unitarily by the self-adjoint operator $G$, the form becomes
\begin{align}
 I_{WY} (G) & = -\frac12 \Tr{[\sqrt{\rho},G]^2} \cr 
 & =  \Tr{G^2 \rho} - \Tr{G{\rho}^{\frac12} G {\rho}^{\frac12}}\, ,
\end{align}
which no longer relies on finding the spectrum of $\rho$. In the limit of pure states, $I_{\rm WY}$ reduces to the variance of $G$. For unitary evolutions $\exp(-i\theta G)$, is was shown~\cite{Luo03} that $I_Q(\theta) = 8 I_{\rm WY}(G)$.

Wigner and Yanase sought to capture the sense in which observables that commute with some conserved quantity $G$ are easier to measure than observables that do not~\cite{Wigner63}. The skew information they propose measures the information in $\rho$ about not-so-easy-to-measure observables, e.g., it provides a measure of non-commutativity between $\rho$ and $G$. Their definition ensures that $I_{\rm WY}$ is convex, additive, and is non-increasing over time for open quantum systems. 

As with most other metrics in this review, the Wigner-Yanase skew information can be used to derive uncertainty relations between observables $A$ and $B$. Furuichi and Yanagi define the correlation between observables as~\cite{Furuichi12}
\begin{equation}
 {\rm Cor}(A,B) \equiv \Tr{\rho AB} - \Tr{\sqrt{\rho}A\sqrt{\rho}B}\, ,
\end{equation}
and prove that this is bounded by 
\begin{equation}
 {\rm Cor}\,(A,B)^2 \geq I_{\rm WY} (A) \,  I_{\rm WY} (B)\, .
\end{equation}
There are many other uses for the Wigner-Yanase skew information, including as an entanglement witness~\cite{Chen05}. 

Another way to understand the Wigner-Yanase skew information is as a measure of the \emph{quantum} fluctuations in a state $\rho$ for an observable $A$, as opposed to the variance $(\Delta A)^2$, which measures both the classical and the quantum fluctuations:
\begin{align}
 (\Delta A)^2 = -\frac12 \Tr{[\rho^a,A] [\rho^{1-a},A]} + \Tr{ \rho^a \delta A \rho^{1-a}\delta A}\, , \qquad
\end{align}
where $a\in[0,1]$ and $\delta A = A - \braket{A}$. The first term is the Wigner-Yanase skew information generalised by Dyson, which reduces to $I_{\rm WY}(A)$ when $a = \frac12$. The second term is the classical uncertainty in $A$ given the state $\rho$. A common approach~\cite{Li11,Frerot16,Miller18} is to average the quantum fluctuations over the possible values of $a$:
\begin{align}
 -\frac12 \int_0^1\Tr{[\rho^a,A] [\rho^{1-a},A]}  \dif a\ = \Braket{A^2} - (A,A)_\rho^{({\rm B})}\, .
\end{align}
This relates the average to the Bogoliubov inner product in Eq.~\eqref{eq:vwoijsknxcv}, from which the Kubo-Mori information was derived. Hence the Bogoliubov inner product is intimately related to the classical fluctuations of $A$ given the state $\rho$. It is easily verified that for pure states $\smash{(A,B)_\psi^{\rm (B)}} = \braket{A}\braket{B}$ and the classical uncertainty vanishes.

\subsection{Bayesian quantum estimation theory}\label{subsec:outlook_Bayesian}
\noindent
So far we have considered Fisher estimation theory, where $\bm\theta$ is an unknown parameter that is not random. However, often the parameter $\bm\theta$ can itself be a random variable. This is handled by the Bayesian approach to probability theory. In this case we have some probability distribution $\pr{\bm\theta}$ over $\bm\theta$, called the \emph{prior}. Bayesian quantum metrology using prior information was considered by Demkowicz-Dobrza{\'{n}}ski~\cite{Dobrzanski2011_PRA}, and Macieszczak, Fraas and Demkowicz-Dobrza{\'{n}}ski~\cite{Macieszczak14}.

Let $\pr{\x,\bm\theta}$ be the \emph{joint} probability distribution of $\x$ and $\bm\theta$. Bayes' theorem then states that $\pr{\x,\bm\theta} = \pr{\x|\bm\theta}\, \pr{\bm\theta}$. We can change the natural expectation value to
\begin{align}\label{eq:expectation-bayes}
 \ex_{\x,\bm\theta} \left[ f \right] = \int  \dif\x\, \dif\bm\theta\; \pr{\x,\bm\theta} \, f(\x,\bm\theta)\,,
\end{align}
for some integrable function $f(\x,\bm\theta)$. The Bayesian covariance matrix becomes
\begin{align}
 \cov(\check{\bm\theta}) = \ex_{\x,\bm\theta} \left[  \left( \caron{\bm\theta} - \ex_{\x,\bm\theta} \left[ \caron{\bm\theta} \right] \right) \left( \caron{\bm\theta} - \ex_{\x,\bm\theta} \left[ \caron{\bm\theta} \right] \right)^\top  \right]\, .
\end{align}
This no longer depends directly on $\bm\theta$ via the probability distribution $\pr{\x\vert\bm\theta}$, and it incorporates the prior information about $\bm\theta$. Classically, we can switch between Fisher and Bayesian estimation by simply replacing the expectation values throughout, even though the interpretations of the resulting quantities will be subtly different in important ways. For example, the classical \mse\ will depend on the true value of $\bm\theta$, while the Bayesian \mse\ will have no dependence on $\bm\theta$ at all.

The optimal estimator $\check{\theta}_B$ that minimises the Bayesian \mse\ of a parameter $\theta$ is the mean of the posterior probability distribution~\cite{Kay1993} $\pr{\theta|\x}$
\begin{align}
 \check{\theta}_B(\x) = \int \dif\theta\; \theta\;  \pr{\theta|\x}\,  .
\end{align} 
This estimator changes as additional data $\x$ is obtained. In addition, the choice of a prior probability distribution $\pr{\theta}$ is important for a successful Bayesian parameter estimation. For more details on Bayesian estimation in the classical domain, see Kay (1993)~\cite{Kay1993}.

To derive the Bayesian form of the quantum Fisher information we use again the density matrix of a system $\rho(\theta)$ evolved according to a parameter $\theta$ of interest. There is now a prior probability distribution $\pr{\theta}$ over $\theta$, so that we can define a prior-weighted density operator as:
\begin{align}
 \overline{\rho} = \int \dif\theta\, \pr{\theta} \rho(\theta)\, . 
\end{align}
Similarly, we can define the \emph{posterior mean operator}
\begin{align}
 \overline{\theta\rho} = \int \dif\theta\, \pr{\theta} \theta \rho(\theta)\, . 
\end{align}
Just like the \qfi\ was given by the scalar product of $\partial_\theta \rho$ with itself, $I_Q(\theta) = (\partial_\theta \rho,\partial_\theta \rho)_\rho = \Tr{\partial_\theta \rho\, \mathscr{L}_\rho (\partial_\theta \rho)}$, we define a new Bayesian information $I_B$ based on $\overline{\theta\rho}$ using the same inner product that originated from the symmetrised operator correlation in Eq.~\eqref{eq:bghuo04woesjk}:
\begin{align}
 I_B \equiv \bigl(\overline{\theta\rho},\overline{\theta\rho}\bigr)_{\bar{\rho}} = \Tr{\overline{\theta\rho}\, \mathscr{L}_{\bar{\rho}} (\overline{\theta\rho})}\, 
\end{align}
A new symmetric logarithmic posterior mean operator $S_\theta \equiv \mathscr{L}_{\bar{\rho}} (\overline{\theta\rho})$ can be constructed from
\begin{align}\label{eq:ujy5etdr}
 \overline{\theta\rho} = \frac12 \left\{ \overline{\rho} , S_\theta\right\}\, , 
\end{align}
which plays a similar role to the \sld\ in the ordinary quantum Cram\'er-Rao bound. Using that $A = \frac12 \{ \rho , B\}$ has a general solution~\cite{Personick71}
\begin{align}
 B = 2\int_0^\infty \dif s\; e^{-\rho s} A e^{-\rho s}\, ,
\end{align}
we find
\begin{align}\label{eq:bg0weoshudjl}
 S_\theta = 2\int_0^\infty \dif s\; e^{-\bar{\rho} s} \bigl( \overline{\theta\rho} \bigr) e^{-\bar{\rho} s} \, .
\end{align}
The quantum Bayesian information $I_B$ then takes the form
\begin{align}
 I_B = 2 \int_0^\infty  \dif s \; \Tr{  \bigl( \overline{\theta\rho} \bigr) e^{-\bar{\rho} s} \bigl( \overline{\theta\rho} \bigr) e^{-\bar{\rho} s} } = \Tr{\overline{\rho}\, S_\theta^2} \, .
\end{align}
Contrary to the \qfi, which has units of $\theta^{-2}$, this information has units of $\theta^2$. 

Personick \cite{Personick71} showed how these definitions leads to a minimum \mse\ for the situation where we have a prior distribution $\pr{\theta}$ over the parameter $\theta$. The \mse\ for $\theta$ is obtained via a measurement of some observable $M$:
\begin{align}\label{eq:204hrweknsd}
 \cov_{\rm B}(M) & \equiv \int \dif\theta\; \pr{\theta} \Tr{\rho(\theta) \left(M - \theta \identity\right)^2} \cr
 & = \Tr{\overline{\rho} M^2} - 2 \Tr{\left(\overline{\theta\rho}\right) M} + \int  \dif\theta\; \pr{\theta}\, \theta^2 \cr
 & = \ex_\theta\left(\theta^2\right) - \Tr{\left(\overline{\theta\rho}\right) M}\, ,
\end{align}
where $\overline{\rho}$ and $\overline{\theta\rho}$ are defined as above, and we used the subscript ``B'' to emphasize this is a Bayesian covariance. Our task is to find the optimal operator $M$ that achieves the minimal \mse. It is given by Eq.~(\ref{eq:bg0weoshudjl}), and $M$ is Hermitian and unique \cite{Personick71}, at least in the single-parameter case. The proof by Personick proceeds by considering 
\begin{align}
 \cov_{\rm B}(M) \leq \cov_{\rm B}(M + \epsilon H)\, ,
\end{align}
with $\epsilon\geq 0$ and $H$ and arbitrary Hermitian operator. Using the definition in Eq.~(\ref{eq:204hrweknsd}) and expanding the square then leads to the condition that 
\begin{align}
 \Tr{H \left( \overline{\rho} M + M \overline{\rho} - 2 \overline{\theta\rho} \right)} = 0\, ,
\end{align}
and from Eq.~\eqref{eq:ujy5etdr} the identification $M = S_\theta$ follows. The Bayesian quantum Cram\'er-Rao bound then becomes
\begin{align}
 (\Delta\theta)^2_B \geq (\Delta\theta)^2_p - I_B\, ,
\end{align}
where $(\Delta\theta)^2_p$ is the variance of the prior distribution
\begin{align}
 (\Delta\theta)^2_p = \int \dif\theta \; \pr{\theta} \theta^2,
\end{align}
and the quantity $(\Delta\theta)^2_p - I_B$ is the quantum Allan variance~\cite{Chabuda19}. 

This construction was used by Macieszczak \emph{et al}.\ for Bayesian quantum frequency estimation \cite{Macieszczak14}, leading to the introduction of the quantum Allan variance by Chabuda, Leroux, and Demkowicz-Dobrza\'nski~\cite{Chabuda19} and their application of the quantum Allan variance to the precision of atomic clocks.  Rubio and Dunningham describe quantum metrology in the presence of limited data using this formalism~\cite{Rubio19_NJP}. Bern\'ad, Sanavio and Xuereb use the quantum Bayesian estimation technique to estimate the nonlinear opto-mechanical coupling strength~\cite{Bernad18} and the matter-field coupling strength in the dipole approximation~\cite{Bernad19}. Rz\k{a}dkowski and Demkowicz-Dobrza\'nski apply Bayesian techniques to discrete phases \cite{Rzadkowski17}.

The multi-parameter form of $\cov_\text{B}$ for $\bm\theta = (\theta_1,\dots,\theta_D)$ with estimator observables $\bm{S} = (S_1,\ldots,S_D)$ is
\begin{align}
 \cov_{\rm B}(\bm{S})_{jk} = &\ex_{\bm\theta}\left(\theta_j\theta_k\right) - \Tr{\overline{\theta_j\rho}\, S_k} - \Tr{\overline{\theta_k\rho}\, S_j} \\
 &+ \frac12\Tr{\overline{\rho} \{S_j,S_k\}}\, ,
\end{align}
where $j,k\in\{1,\ldots,D\}$. Furthermore, we can prove that
\begin{align}
 \Tr{\overline{\theta_j\rho}\, S_k} + \Tr{\overline{\theta_k\rho}\, S_j} = \Tr{\overline{\rho} \{S_j,S_k\}}\, ,
\end{align}
and we can therefore introduce the \emph{quantum Bayes information} matrix 
\begin{align}
 [I_B(\bm{S})]_{jk} = \Tr{\overline{\theta_j\rho}\; \mathscr{L}_{\bar{\rho}} \left(\overline{\theta_k\rho}\right)} = \frac12 \Tr{\overline{\rho}\, \{ S_j,S_k\}}\, .
\end{align}
The proof of this expression for the matrix follows from the fact that the Bayesian estimation case has the same geometric structure as the Fisher estimation case. The multi-parameter Bayesian minimum \mse\ bound can then be written as 
\begin{align}
 \cov_\text{B}(\bm{S})_{jk} \geq \ex_{\bm\theta}\left(\theta_j\theta_k\right) - [I_B(\bm{S})]_{jk}\, .
\end{align}
A similar result for the multi-parameter Bayesian \textsc{mse} was independently obtained by Rubio and Dunningham~\cite{Rubio2019_arxiv}.


\section{Special cases of quantum estimation}\label{sec:research_applications}
\noindent
There are a number of special applications in quantum parameter estimation. First, in quantum optics we very often deal with Gaussian states, and these admit closed forms for the \qfi\ and the \sld s. Second, we consider the case when the evolution of the probe state does not take the form of a simple unitary $U= \exp(-i\theta G)$.

\subsection{Estimation using Gaussian states}\label{subsubsec:gaussian_state}
\noindent
The practical implementation of quantum parameter estimation often involves Gaussian quantum states, due to their ubiquity, relative easy of preparation, and their admittance of closed form analytical expressions for the \qfi\ and the \sld. Of particular importance is the use of Gaussian states in the context of quantum optics, where coherent, squeezed, and thermal states all form part of the Gaussian family of states. 

Consider a system comprised of $n$ modes of a bosonic field. The Hilbert space $\smash{\mathscr{H} = \otimes_{k=1}^n\mathscr{H}_k}$ is the tensor product of the infinite dimensional Fock space of each mode $k$, that is spanned by the number basis $\smash{\{\ket{m}_k\}}$, with $m$ a natural number. The creation and annihilation operators for each mode are defined by 
\begin{align}
 \hat{a}_k \ket{m}_k & = \sqrt{m}\ket{m-1}_k \cr \hat{a}_k^\dagger \ket{m}_k & = \sqrt{m+1}\ket{m+1}_k\, ,
\end{align}
obeying the commutation relation $[a_k,a_j^\dagger] = \delta_{jk}$, with all other commutators zero. In the phase space formalism, a conventional approach is to arrange these operators into a vectorial operator $\smash{\hat{\bm{a}} := (\hat{a}_1, \hat{a}_1^\dagger, \ldots, \hat{a}_n, \hat{a}_n^\dagger)^\top}$. The vector then satisfies the compact commutation relation
\begin{align}
\left[\hat{\bm{a}}_i, \hat{\bm{a}}_j\right] = \bm{\Omega}_{ij}
\label{eqn:compact_commutations}
\end{align}
where the matrix $\smash{\bm{\Omega} = \oplus_{j=1}^n i\sigma_y}$ satisfies $\smash{\bm{\Omega}^\top = -\bm{\Omega}=\bm{\Omega}^{-1}}$, and $\smash{\sigma_y}$ is the Pauli $y$ matrix. An equivalent description of bosonic systems can be written in terms of the dimensionless canonical quadrature field operators:
\begin{align}
\hat{q}_j = \frac{1}{\sqrt{2}}\left(\hat{a}_j+\hat{a}^\dagger_j\right), \quad \text{and} \quad \hat{p}_j = -\frac{i}{\sqrt{2}}\left(\hat{a}_j - \hat{a}_j^\dagger\right).
\label{eqn:quadrature_field_operators}
\end{align}
These observables act similar to the position and momentum operators of the quantum harmonic oscillator, and satisfy the canonical commutation relations $\smash{[\hat{q}_j,\hat{p}_k] = i\delta_{jk}}$. Introducing the vector of operators $\smash{\hat{\bm{R}} = (\hat{q}_1,\hat{p}_1,\ldots,\hat{q}_n,\hat{p}_n)^\top}$, we similarly re-write the canonical commutation relations in the compact form
\begin{align}
\left[\hat{\bm{R}}_j,\hat{\bm{R}}_k\right] = i \bm{\Omega}_{jk}.
\label{eqn:vector_canonical_operators}
\end{align}

A Gaussian $n$-mode quantum state $\rho$ is a state that is completely characterized by $n$ first moments $\bm{\lambda}$ and an $n\times n$ second moment matrix $\bm\Sigma$ such that 
\begin{align}
\bm{\lambda} = \Tr{\rho\, \hat{\bm{a}}} = \braket{\hat{\bm{a}}}, \quad \bm{\Sigma}_{jk} = \Tr{\rho \left\{\hat{\tilde{a}}_j, \hat{\tilde{a}}_k\right\}},
\end{align}
with the zero mean operators $\smash{\hat{\tilde{a}}_j = \hat{a}_j - \braket{\hat{a}_j}}$, and $\{{A},{B}\} = {A}{B} + {B}{A}$ the anti-commutator of operators ${A}$ and ${B}$. The characteristic function of $\rho$ is defined as 
\begin{align}
 {\chi(\bm{\zeta}) = \Tr{{\rho\exp[-\hat{\bm{a}}^\top\bm{\Omega}\,\bm{\zeta}]}}}\, ,
\end{align}
where $\smash{\bm{\zeta} = (\zeta_1, \zeta_1^*, \ldots, \zeta_n, \zeta_n^*)^\top} \in \mathbbm{R}^{2n}$. For a Gaussian density matrix with first moment $\smash{\bm{\lambda}}$ and second moment $\smash{\bm{\Sigma}}$, the corresponding characteristic function is also Gaussian:
\begin{align}
\chi(\bm{\zeta})= \exp\left[\frac{1}{2}\bm{\zeta}^\top (\bm{\Omega}\bm{\Sigma}\bm{\Omega}^\top)\bm{\zeta} - i(\bm{\Omega}\bm{\lambda}^\top\bm{\zeta})\right],
\label{eqn:characteristic_function}
\end{align}
A Gaussian unitary transformation is any trace preserving quantum channel that preserves the Gaussian nature of the characteristic function. 

A common pure Gaussian state is the vacuum $\smash{\rho_\text{in} = \ket{0}\bra{0}}$, while for mixed states it is the thermal state:
\begin{align}
\rho_\text{th} = \sum_{n=0}^\infty \frac{\overline{n}^n}{(1 + \overline{n})^{1 + n}}\ket{n}\bra{n},
\label{eqn:thermal_state}
\end{align} 
where $\smash{\overline{n} = \braket{\hat{n}}}$ is the expectation of the photon number $\smash{\hat{n}}$. Thermal radiation exhibits a Gaussian characteristic function with zero mean $\smash{\bm{\lambda}_\text{th} = \bm{0}}$ and covariance $\smash{\bm{\Sigma}_\text{th} = (2\overline{n} +1)\mathbbm{1}}$. Any single-mode Gaussian state can be written in terms of a displaced squeezed thermal state~\cite{WeedbrookRMP_2012}:
\begin{align}
\rho_\text{G} = {S}_\epsilon{D}_\alpha \rho_\text{th} {D}^\dagger_\alpha{S}^\dagger_\epsilon = {U}\rho_\text{th}{U}^\dagger,
\label{eqn:general_gaussian_state}
\end{align} 
where the Gaussian unitaries
\begin{align}
{D}_\alpha &= \exp\left[\alpha\hat{a}^\dagger - \alpha^*\hat{a}\right],\\
{S}_\epsilon &= \exp\left[\frac{1}{2}(\epsilon^*\hat{a}^2 - \epsilon\hat{a}^{\dagger 2})\right],
\label{eqn:gaussian_unitary_set}
\end{align} 
are defined as the displacement operator and single mode squeezing operator, respectively. The ordering of a displacement and squeezing operations can be reversed according to
\begin{align}
{D}_\alpha{S}_\epsilon={S}_\epsilon{D}_\gamma,
\label{eqn:holstein_primakoff_trans}
\end{align}
with $\smash{\gamma = \alpha\cosh r_\epsilon - \alpha^*\exp[i\vartheta_\epsilon]\sinh r_\epsilon}$. States described by Eq.~\eqref{eqn:general_gaussian_state} are readily prepared in laboratories and their unitary evolutions realised by use of lasers, linear optical elements, and squeezing mediums~\cite{Ferraro2005_arxiv}. The multi-mode generalisation of Eq.~\eqref{eqn:general_gaussian_state} can be found using the matrix generalisation~\cite{Braunstein2005_PRA} of the operators in Eq.~\eqref{eqn:gaussian_unitary_set}.

Both the mean and second moment of a Gaussian quantum state may depend on parameters $\bm\theta$ that we wish to estimate. To bound the precision of the estimation procedure, we aim to calculate the \qfi\ of Gaussian quantum states.
This is not always an easy task. All of the work described in the previous sections relied on knowledge of the density matrix, and for Gaussian states this is most conveniently addressed in the phase-space formalism by mapping transformations of the state to transformations of the moments~\cite{Adesso2014_OSID}. 

The first work in this direction was completed by Pinel \emph{et al.}~\cite{Pinel2012_PRA} who derived the expression for the ultimate limit to parameter estimations using pure Gaussian states of arbitrarily many modes. A more general approach to finding the \textsc{qfi} and the \textsc{sld} for Gaussian states is through the mean displacement and covariance matrix. This approach has been taken by a number of authors. The first result in this direction was by Monras in 2013, who derived the form of the \textsc{sld} for general Gaussian states evolving under Gaussian unitaries~\cite{Monras2013_arxiv}. By taking an \textsc{sld} ansatz that is quadratic in the quadrature operators, the \textsc{sld} and \textsc{qfi} are written as an infinite series solution to the Stein-equation. A similar approach by Jiang confirmed Monras' result and gave the \textsc{sld} for states in exponential form in terms of the generator and its moments~\cite{Jiang2014_PRA}. Gao and Lee followed an alternative method to derive the \textsc{sld} and the \textsc{qfi} for multi-mode Gaussian states. The necessity of inverting relatively large matrices is a drawback of this method~\cite{Gao2014_EPJD}. Expressions for the \textsc{qfim} for multi-mode Gaussian states were reported by \v{S}afr\'anek, Lee, and Fuentes~\cite{Safranek2015_NJP}. A unification of these results and a resolution of the problematic divergence behaviour of the \textsc{qfim} for pure states has been addressed~\cite{Safranek2018_JPA}. Specifically, by applying the regularisation procedure $\mathsf{R}$ to expressions of the \textsc{qfim} valid for mixed Gaussian states only, the \textsc{qfim} for Gaussian states with any number of pure modes can be immediately determined. Defining $I_Q[\rho_\text{G}; \{\bm{\lambda}, \bm{\Sigma}\}]$ as the \textsc{qfim} for a mixed Gaussian state with mean $\bm{\lambda}$ and covariance $\bm{\Sigma}$, the regularisation procedure $\mathsf{R}$ for some regularisation parameter $\xi > 1$ proceeds according to
\begin{align}
\mathsf{R}: \lim_{\xi \to 1} I_Q\left[\rho'_\text{G}; \{\bm{\lambda}, \xi \bm{\Sigma}\}\right].
\label{eqn:pure_gauss_regularisation}
\end{align}
The resulting value is the correct \textsc{qfim} for any Gaussian state and does not suffer from divergences. A similar regularisation has been also been used to regularise the \textsc{qfi} for non-Gaussian states~\cite{Safranek2017_PRA}. Marian and Marian gave a full analysis of the \qfi\ for two-mode Gaussian states~\cite{Marian16,Marian12}.

Due to the structure of Gaussian states, the \textsc{sld} is at most quadratic in the bosonic (or equivalently the canonical) mode operators. Hence, for the \sld\ of parameter $\theta_k$ we make the following Ansatz:
\begin{align}
 L_k = \frac{1}{2}A^{(k)}_{\alpha\beta}\left\{ \hat{\tilde{\bm{a}}}^\alpha, \hat{\tilde{\bm{a}}}^\beta\right\} + B^{(k)}_\alpha \hat{\tilde{\bm{a}}}^\alpha - \frac{1}{2}\Tr{A^{(k)}\bm{\Sigma}},
\label{eqn:monras_sld_form}
\end{align}
for the \textsc{sld} for Gaussian states, where Greek indices imply Einstein summation convention. Notice that the first term in Eq.~\eqref{eqn:monras_sld_form} is composed of bilinear contributions (i.e., a product of a creation operator with an annihilation operator) that corresponds to linear unitary operations such as beam-splitters, half and quarter wave plates, and phase-shifters. The quadratic contribution describes active devices such as squeezers and down-converters. The second term in Eq.~\eqref{eqn:monras_sld_form} has a linear dependence on the operators. These operations describe displacements in phase space. The coefficients $\smash{A^{(k)}}$ and $\smash{B^{(k)}}$ are completely determined from the first and second moments of the probe state and their derivatives. Finally, we include a constant term $-\frac{1}{2}\tr{A^{(k)}\bm{\Sigma}}$.

Monras~\cite{Monras2013_arxiv} found a closed form for these coefficients as a solution to the Stein-equation
\begin{align}
A^{(k)} = \sum_{j=0}^\infty \bm{F}^{\top j}\partial_k(\bm{\Sigma}^{-1})\bm{F}^j, \quad  B^{(k)} = 2\bm{\Sigma}^{-1}(\partial_k\bm{\lambda}),
\label{eqn:monras_sld_form_consts}
\end{align}
with $\smash{\bm{F} = (i\bm{\Sigma}\bm{\sigma}_y)^{-1}}$. Notice that the constant $\smash{A^{(k)}}$ is defined only for non-singular states, which excludes pure states. The \textsc{qfim} then reads
\begin{align}
[I_Q]_{jk} = \frac{1}{2} \frac{\bar{n}^2}{1 + \bar{n}^2}\Tr{(\partial_j\bm{\Sigma}) \bm{\Sigma}^{-1}(\partial_k\bm{\Sigma})\bm{\Sigma}^{-1}} + 2 (\partial_j \bm{\lambda})^\top\bm{\Sigma}^{-1}(\partial_k \bm{\lambda}),
\label{eqn:monras_qfim}
\end{align}
where $\bar{n}$ is the average photon number of the thermal states before symplectic transformations. The \textsc{qfi} in Eq.~\eqref{eqn:monras_qfim} is not valid for all mixed states, but only for `isotropic' or `isothermal' states that have the same symplectic eigenvalues for all modes, which includes the class of pure state models. An elegant form for the \textsc{qfim} for arbitrary mixed Gaussian states was derived by \v{S}afr\'anek~\cite{Safranek2018_JPA}, from which the origin of precision enhancements were assigned to three qualitatively different terms: changes to the orientation and squeezing of the Gaussian state, changes in purity, and changes in displacement.

Independently, Gao and Lee used phase space methods to derive an exact form of the \textsc{sld} and the \textsc{qfi}~\cite{Gao2014_EPJD}. Their use of a quadratic Ansatz for the \textsc{sld} lead to the result
\begin{align}
L_j = \frac{1}{2}\bm{\mathfrak{M}}^{-1}_{\gamma\kappa, \alpha\beta}\left(\partial_j\bm{\Sigma}^{\alpha\beta}\right)\left(\hat{\tilde{a}}^\gamma\hat{\tilde{a}}^\kappa - \bm{\Sigma}^{\gamma\kappa}\right) + \bm{\Sigma}^{-1}_{\mu\nu}\left(\partial_j\bm{\lambda}^\nu\right)\hat{\tilde{a}}^\mu,
\label{eqn:gao_lee_sld_form}
\end{align}
where $\smash{\bm{\mathfrak{M}} = \bm{\Sigma}\otimes\bm{\Sigma} + \bm{\Omega}\otimes\bm{\Omega}/4}$. The \textsc{qfi} for any $n$-mode bosonic Gaussian system was then determined through application of Wicks theorem for Gaussian states to yield
\begin{align}
\left[I_Q\right]_{jk} = \frac{1}{2}\bm{\mathfrak{M}}^{-1}_{\alpha\beta, \mu\nu}\partial_j\bm{\Sigma}^{\alpha\beta}\partial_k\bm{\Sigma}^{\mu\nu} + \bm{\Sigma}^{-1}_{\mu\nu} \partial_j\bm{\lambda}^{\mu}\partial_k\bm{\lambda}^{\nu}.
\label{eqn:gao_lee_qfi_form}
\end{align}
Calculating the \textsc{qfi} increases exponentially in computational time with the number of modes owing to the inversion of large dimensional matrices. This result is limited to mixed Gaussian states.  

Finally, Banchi, Braunstein and Pirandola derived analytical forms for the fidelity between two arbitrary Gaussian states~\cite{Banchi15}, which can be related to the \qfi\ via the quantum fidelity introduced in Eq.~\eqref{eq:gh94weuohisd}. A closed form of the \textsc{sld} has recently been derived using the quantum fidelity formalism~\cite{Oh2019_PRA}. The optimal measurement described by these \textsc{sld}s does not always correspond to Gaussian measurements~\cite{Oh2019_NPJ}.

\subsection{Hamiltonians with non-multiplicative factors}
\label{subsec:multiplicative_hamiltonians}

\noindent
So far we have considered the estimation precision of a given probe into the dynamics that imprints the parameter $\theta$. This is a channel estimation scheme, where an optimisation over the possible input probe states can improve the estimation precision~\cite{Fraisse2017_PRA}. Within this regime, we have reviewed parameter estimation schemes where the Hamiltonian is either a constant or has multiplicative dependence on the parameter of interest (e.g., a coupling strength). This regime is known as the phase-shift or phase-like Hamiltonian estimation, where the parameter to estimate multiplies a parameter-independent Hermitian generator $G$~\cite{Braunstein95, Holevo1978_RMP}, taking the unitary form $U = \exp(-i\theta G)$. This regime has been extensively studied, and these single parameter phase-shift Hamiltonians define the optimal probe state~\cite{Giovannetti2011_NP} as discussed in subsection~\ref{subsec:optimal_implementation}. It has been applied for Hamiltonian characterisation in the absence of noise ~\cite{Yurke1986_PRA, Sanders1995_PRL, Dorner2009_PRL, Skotiniotis2015_NJP}, and frequency measurements and atomic spectroscopy in the presence of decoherence~\cite{Huelga97,Escher2011}. 

More general Hamiltonians have recently started to attract attention~\cite{Brody2013_E}, since they permit the application of quantum metrology to a more general class of problems such as time-varying fields~\cite{Tsang2011_PRL, Magesan2013_PRA} and in gradient magnetometry~\cite{Urizar2013_PRA}. It is well known that a pure state parameterised by a multiplicative Hamiltonian of the form $\smash{{H}_j = \theta_j{G}_j}$ for a time $t$, the \textsc{qfi} is given by~\cite{Braunstein95,Pang2014_PRA}
\begin{align}
I_Q(\theta_j) = 4t^2(\Delta G_j)^2.
\label{eqn:qfi_pure_time_t}
\end{align}
The definition of the generator when the state is parameterised by a more general Hamiltonian for the form $H(\theta)$ becomes unclear. Assuming the Hamiltonian has $n$ unique eigenvalues $E_j$ with $j=1, \ldots,n$, degeneracy $d_j$ and corresponding eigenvectors $\ket{\smash{E_j^{(k)}}}$, $k \in \{1, \ldots, d_j\}$ satisfying $\smash{\braket{E_\alpha^{(\beta)}\vert E_\gamma^{(\delta)}} = \delta_{\alpha\gamma}\delta_{\beta\delta}}$, then the generator of translations in $\theta_j$ can be written as~\cite{Pang2014_PRA, Pang2016E_PRA,Sidhu2018_arxiv}
\begin{align}
\begin{split}
{G}_j(\bm{\theta}) & = \sum_{k = 1}^{n_g} \partial_j E_k{P}_k + 2\sum_{k\neq l}\sum_{m=1}^{d_k}\sum_{n=1}^{d_l} \exp\left[-i (E_k - E_l)/2\right] \\
& \qquad \times \sin\left[\frac{E_k - E_l}{2}\right] \Braket{E_l^{(n)}|\partial_j E_k^{(m)}}\ket{E_k^{(m)}}\bra{E_l^{(n)}},
\end{split}
\label{eqn:final_generator_general_form}
\end{align}
where $\smash{{P}_k = \sum_j\ket{\smash{E_k^{(j)}}}\bra{\smash{E_k^{(j)}}}}$ is the projector onto the $E_k$-eigenspace. The form of the generator in Eq.~\eqref{eqn:final_generator_general_form} implies that the \textsc{qfi} can be separated into two parts. The first part is due to the dependence of the eigenvalues on $\theta_j$, and the second due to the dependence of the eigenstates on $\theta_j$. An upper bound on the \textsc{qfi} was derived by Pang and Brun~\cite{Pang2014_PRA}
\begin{align}
\label{eqn:oscillatory_qfi_t_dependence}
I_{Q}^{\text{max}} \leq & ~2t^2 \sum_{j=1}^n d_j (\partial_j E_j)^2 \\ \nonumber
&+ \sum_{k\neq l}\sum_{m=1}^{d_k}\sum_{n=1}^{d_l} \Abs{\sin\left[\frac{(E_k - E_l)t}{2}\right]}^2 \Abs{\Braket{E_l^{(n)}\vert \partial_j E_k^{(m)}}}^2 .
\end{align}
Increasing the channel \textsc{qfi} can be achieved by enhancing the sensitivity of the generator through additional terms in the Hamiltonian~\cite{Fraisse2017_PRA}. Specifically, for time evolutions with parameter independent eigenvalues, Eq.~\eqref{eqn:final_generator_general_form} exhibits a periodic time dependence of the channel \textsc{qfi}~\cite{Pang2014_PRA}. Generally, this alone does not saturate the Heisenberg limit precision, but can with feedback controls~\cite{Yuan2015_PRL, Yuan2016_PRL}. In section~\ref{subsec:noisy_metrology}, we see how the precision achievable in channel estimation schemes can be improved with ancilla states.

The general result in Eq.~\eqref{eqn:final_generator_general_form} implies that for any unitary encoding, the optimal probe states can be assigned through the prescription described by Giovannetti \emph{et al.}~\cite{Giovannetti2011_NP}. However, this method requires that the spectrum for the Hamiltonian is known, and this is generally difficult. This complication can be avoided if the non-multiplicative estimation problem can be related to an equivalent, phase-like estimation problem through re-parameterisation of the parameters $\smash{\bm{\theta}}$, since the \textsc{qfi} matrix elements transforms as a metric tensor~\cite{Tsang2014_O}, as we have seen in section \ref{subsec:qfim_intro}. If a nonlinear function of the parameters can be transformed into a  phase-like scheme this approach may simplify the calculation of the \textsc{qfi}. We will return to functions of multiple parameters in section~\ref{subsec:estimating_functions}.


\section{Noisy quantum metrology}\label{subsec:noisy_metrology}
\begin{figure}[t!]
\centering
\includegraphics[width =0.55\columnwidth]{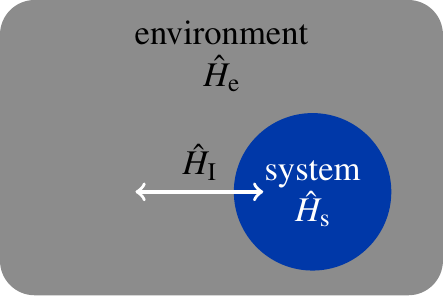}
\caption{System-bath interaction model where ${H}_\text{s}$, ${H}_\text{e}$, and ${H}_\text{I}$ are the system, environment, and interaction Hamiltonians.}
\label{fig:system_enviroment}
\end{figure}

\noindent	
We have seen that quantum resources can improve sensing capabilities over classical methods. However, very few systems can be prepared in complete isolation from its environment in practice. Therefore, applications of quantum metrology must include a description of the interaction of a system with its environment as illustrated in figure~\ref{fig:system_enviroment}. Generally, the performance of sensors are limited by decoherence in a familiar fashion to most applications of quantum technologies.

\subsection{Metrology in noisy quantum channels}
\noindent
Various methods to calculate parameter precision bounds in the presence of noise have been developed. A common approach is to use the Kraus decomposition of the quantum channel~\cite{Sarovar2006_JPA, Escher2011, Kolodynski2013_NJP}. The completely positive, trace preserving (\textsc{cptp}) channel $\smash{\rho(\theta) = \Lambda[\rho(0)](\theta)}$, satisfying $\smash{\Lambda[\rho(0)](\theta) \geq 0}$ and $\smash{\tr{\rho(\theta)} = \tr{\rho(0)}}$, can be expanded in terms of the Kraus operators~\cite{Choi1975_LAA, Kraus1983_book}
\begin{align}
 \rho(\theta) = \Lambda\left[\rho(0)\right] = \sum_{j=1}^q {K}_j(\theta) \rho(0) {K}^\dagger_j(\theta).
\label{eqn:kraus_theorem}
\end{align}
The set of $q$ Kraus operators $\mathscr{K}(\theta) = \{{K}_j(\theta)\}$ is referred to as a $q$-Kraus decomposition of $\Lambda$, satisfying $\smash{\sum_{j=1}^q {K}_j^\dagger(\theta){K}_j(\theta) = \mathbbm{1}}$. This representation is not unique, since for every $q$-Kraus decomposition of a channel, all the other $q$-Kraus decompositions can be constructed via a unitary transformation 
\begin{align}\nonumber
\sum_{k}u_{jk}(\theta){K}_j(\theta)\, ,
\end{align}
where $u_{ij}(\theta)$ is a unitary matrix. The set of all $q$-Kraus decompositions of a channel is called the $q$-Kraus ensemble and is noted $\smash{\mathscr{K}_q(\theta)}$. The smallest possible number $q$ of Kraus operators is known as the Kraus rank. It can be obtained as the number of non-vanishing eigenvalues of the Choi-matrix of the channel~\cite{Bengtsson2008}. The generality of this representation is that it can be used to model any system decoherence. Specifically, adding arbitrary environmental degrees of freedom to the probe system, the probe dynamics can be written as a unitary evolution for the combined probe-bath system. This process is not unique since it requires the environment to be defined; the probe evolution depends on the unitary evolution associated with the enlarged Hilbert space. A fixed choice of the unitary evolution and environmental degrees of freedom corresponds to choosing a Kraus operator $\smash{{K}_j(\theta)}$, which leads to an upper bound for the \textsc{qfi}~\cite{Escher2011}
\begin{align}
I_{Q,{\rm max}}\left[\rho(0), {K}_j(\theta)\right] \geq 4\left\{\braket{{H}_1(\theta)} - \braket{{H}_2(\theta)}^2\right\},
\label{eqn:open_system_qfi}
\end{align}
where 
\begin{align}
{H}_1(\theta) &= \sum_j \frac{\dif {K}_j^\dagger(\theta)}{\dif \theta}\frac{\dif {K}_j(\theta)}{\dif x}\, ,\\
{H}_2(\theta) &= i\sum_j \frac{\dif {K}_j^\dagger(\theta)}{\dif \theta} {K}_j(\theta)\, ,
\label{eqn:open_system_qfi_defos}
\end{align}
and the expectation is with respect to the initial probe state. For unitary processes, Eq.~\eqref{eqn:open_system_qfi} reduces to the common generator variance bound. The Kraus decomposition can be applied to ancilla assisted schemes, where the bound can be similarly determined from the reduced density matrix of the probe. Determining the \textsc{qfi} requires optimising Eq.~\eqref{eqn:open_system_qfi} over all Kraus operators, which is generally difficult. Alternative methods have been introduced to determine error bounds in noisy systems. This includes using a variational approach~\cite{Escher2012_PRL}, a generalised Bures angle approach to quantum channels~\cite{Yuan2017_NPJQI}, and through the geometry of quantum channels, which has been used to explore the effects depolarisation, dephasing, spontaneous emission and photon loss channels~\cite{DemkowiczDobrzanski12, Kolodynski2013_NJP}.

For non-unitary probe encoding, few analytic expressions for the \textsc{qfi} exist~\cite{Monras2007_PRL, Dorner2009_PRL, Dobrzanski2009_PRA} given that the calculation involves complex optimisation procedures which becomes increasingly difficult with increasing system size. In this case, it is more informative to report on asymptotic lower bounds to the estimation precision for specific noise models. For the estimation of some parameter $\theta$, the error in its estimate depends on the initial probe state. Noise generally degrades the optimal quadratic enhancement of precision estimates to one that is a constant improvement over classical methods~\cite{DemkowiczDobrzanski12}. In fact, it has been demonstrated that for parallel Markovian dephasing noise in the probe preparation, the precision scaling is reduced to the classical \textsc{sql} $\smash{\Delta\theta \sim 1/\sqrt{N}}$~\cite{Huelga97,Escher2011,DemkowiczDobrzanski12}, and transverse Markovian dephasing noise reduces the scaling bound to $\smash{\Delta\theta \sim 1/N^{5/6}}$ if an entangled probe is used~\cite{Chaves13}. For the entangled \textsc{ghz} probe, these scalings were generalised to Markovian noise in nonlinear quantum metrology with many-body open systems by Beau and del Campo~\cite{Beau2017_PRL}. Specifically, for a $k$-dimensional Hamiltonian and $p$-dimensional Lindblad operator, the variance of a Hamiltonian parameter scales as $\smash{N^{-[k-(p/2)]}}$, which surpasses the shot-noise limit for $2k > p+1$. The system-environment coupling parameter can be estimated with a precision that scales as $\smash{N^{-(p/2)}}$, while many-body decoherence enhances the precision to $\smash{N^{-k}}$ in the noise-amplitude estimation of a fluctuating $k$-body Hamiltonian. For non-Markovian noise, the precision bound can scale as $\smash{\Delta\theta \sim 1/N^{3/4}}$~\cite{Matsuzaki11,Chin12}. Although these reduced precision scalings lower the prospects of quantum metrology and sensing, some gain has been demonstrated with non-Markovian dephasing and probe readout schemes~\cite{Dooley2016_PRA}. We also review alternative methods to attempt to diminish the effects of noise later in this section.

Most of these established precision bounds are based on educated guesses or numerical methods on the integrated form of the probe dynamics. For non-unitary dynamics, where extra terms have to be added to fully describe the dynamics of the probe state, even the integration is not straightforward. The specific dynamics are described by the Lindblad operators $\smash{{\mathsf{L}}_j}$ through the Lindblad equation~\cite{Breuer2002_book}
\begin{align}
\partial_t\rho = -i[\rho, {H}(\theta)] + \sum_j {\mathsf{L}}_j\rho {\mathsf{L}}_j^\dagger - \frac{1}{2}\bigl\{\rho, {\mathsf{L}}_j^\dagger {\mathsf{L}}_j\bigr\} \, ,
\label{eqn:master_equation}
\end{align}
which is often referred to as the Master Equation. One method to obtain general precision bounds from Eq.~\eqref{eqn:master_equation} is through numerical methods, which does not match the insight provided by analytical approaches. Fortunately, this limitation has been addressed by Sekatski \emph{et al.}, who developed a framework to provide analytic results for general decoherence models described through the master equation~\cite{Sekatski2017_Q}. This was followed by a more qualitative approach that addresses the effects of arbitrary Markovian dynamics on the precision scaling of unitary parameter estimation problems. Specifically, if the generator of translations $G$ in $\theta$ can be written in terms of the operators~\cite{Dobrzanski2017_PRX, Zhou2018_NC}
\begin{align}
  \left\{\mathbbm{1}, {\mathsf{L}}_j^{\text{H}}, i{\mathsf{L}}_j^{\text{AH}}, ({\mathsf{L}}_j^\dagger {\mathsf{L}}_j)^{\text{H}},i({\mathsf{L}}_j^\dagger {\mathsf{L}}_j)^{\text{AH}}\right\},
\label{eqn:generator_set_property}
\end{align}
where the superscript `H' and `AH' denote the Hermitian and the anti-Hermitian part of an operator respectively, then for some interrogation time $T$, the precision scales as at most
\begin{align}
\Delta\theta &\geq \frac{\alpha}{\sqrt{T}},
\label{eqn:generator_element_scaling}
\end{align}
where $\alpha$ is a real constant. Equivalently, the precision scales no better than $1/N$ with $N$ quantifying the resources used. This method does not depend on the integrated form of the master equation, but only on the geometric properties on the Lindblad operators. Furthermore, if the generator is not in the span of the Lindblad operators then the Heisenberg scaling can be recovered with application of quantum error correction procedures. This was realised experimentally by Kessler \emph{et al.}~\cite{Kessler14}, D\"ur \emph{et al.}~\cite{Dur14}, and Sekatski \emph{et al.}~\cite{Sekatski2017_Q}. We will review this in further detail in subsection~\ref{subsubsec:fault_tolerant_metrology}.

One approach to quantify and understand the impact of decoherence on parameter estimates is to have access to the environmental degrees of freedom to make estimates of both Hamiltonian and bath-coupling parameters. Monras and Paris showed that for Gaussian probes, this approach makes no improvement~\cite{Monras2007_PRL} in that there is always an environment such that observing the combined probe-environment state does not provide additional information gain when compared to simply observing the probe itself~\cite{Escher2011}. For small losses, the \textsc{qfi} scales with the loss parameter itself, which demonstrates a qualitative improvement over the shot-noise limit.

The impact of loss on optical interferometry has been explored in detail in the field. We observed in subsection~\ref{subsec:entanglement_for_quantum_metrology} that the highly entangled \textsc{noon} state saturates the Heisenberg limit on precision for phase measurements. However, this class of states is extremely susceptible to losses, and are outperformed by purely classical states for moderate losses~\cite{Huelga97,Rubin2007_PRA,Gilbert2008_JOS,Huver08}. Although for lossy interferometry the Heisenberg limit is not attainable~\cite{Kolodynski2010_PRA, Knysh2011_PRA}, certain quantum states have been engineered to outperform both standard and \textsc{noon} states for phase measurements in the presence of losses~\cite{Kacprowicz2010_NP}. Note that this strategy differs from quantum error correction techniques used for quantum computing and protecting quantum memories, where the approach is to protect the information encoded in light~\cite{Roffe2018_QST}. This demonstrates that a quantum enhancement is possible even in the presence of decoherences. Given that a diverse range of different physical quantities can be determined through phase measurements, it is important to place general bounds for phase estimation with lossy optical interferometers. A systematic approach to address the optimal states with definite photon numbers for interferometry in the presence of losses was first considered by Dorner \emph{et al.}~\cite{Dorner2009_PRL}. This idea was later extended to find the optimal input state for two-mode interferometry using a numerical optimisation of the \textsc{qfi}~\cite{Dobrzanski2009_PRA}. Due to the convexity of the \textsc{qfi} that we observed in subsection~\ref{subsec:quantum_fi_geometry}, miniminising the estimation error is a convex optimisation problem. The corresponding precision beats classical methods and lies between the \textsc{sql} and the Heisenberg limit, depending on the loss rates. It can not be improved by considering probes with indefinite photon number and states with photons distributed between distinguishable time bins.

Besides interferometry, further applications of noisy metrology have been considered. For example, the effects of phase flip and amplitude damping decoherence channels with $N$-qubit \textsc{ghz} probes have been explored. As we observed in subsection~\ref{subsec:entanglement_for_quantum_metrology}, in the absence of noise, the Heisenberg limit can be achieved through rotations along the $Z$ direction. For a phase noise channel, the \textsc{qfi} decreases with increasing decoherence rate due to information leak to the environment, although it can exhibit revivals~\cite{Falaye2017_SR}. For an amplitude damping channel, the \textsc{qfi} can be enhanced by adjusting the temperature of the environment. In bosonic quantum metrology, a constant scaling improvement over the \textsc{sql} was observed by Spedalieri \emph{et al.}, who demonstrated that correlated thermal states outperform coherent states for the estimation of a loss parameter~\cite{Spedalieri2018_QST}. Also, the case of simultaneous estimation of multi-parameters in the presence of noise has been investigated~\cite{Yue2014_SR,Wang2018_OC}. Since preparing noiseless probe states and encoding channels is experimentally challenging, efforts to view noise as a utility to introduce correlations into the system have been explored~\cite{Braun2002_PRL, Benatti2003_PRL, Streltsov2011_PRL}. These correlations can lead to entanglement~\cite{Piani2011_PRL}, which may provide precision enhancements. Hence decoherences can be used to protect precisions against noise. However, too much noise is detrimental. Understanding the interplay between relaxing the noiseless criterion for ease of experimental realisation and the amount of useful noise in the system remains an open question. 	

\begin{figure}[t!]
\centering
\includegraphics[width =0.78\columnwidth]{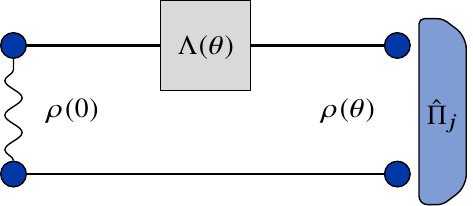}
\caption{Ancilla assisted quantum channel extension scheme, described by $\smash{\Lambda(\theta) \otimes \mathbbm{1}}$ and a \povm\ measurement $\Pi_j$.}
\label{pic:ancilla_assisted}
\end{figure}

\subsection{Ancilla-assisted schemes and channel estimation}\label{subsubsec:ancilla_assisted_approach}
\noindent
Assuming that noise cannot be eliminated from the system, a natural question to ask is whether it is possible to negate its effects. One strategy is to use ancillary systems, and allow the probe system to arbitrarily interact with the ancilla. There are two ways in which an ancillary state can be introduced to the state. In the first, the ancilla state is entangled with the probe but does not itself participate in the estimation~\cite{Dobrzanski2014_PRL}. This is known as channel extension and is illustrated in Fig.~\ref{pic:ancilla_assisted}. The objective is to consider how changes to the dynamics that imprints the parameters to the probe state improves or limits the estimation precision. Given that quantum channels are completely positive trace preserving maps, channel extension can be described through
\begin{align}
\Lambda(\theta) \rightarrow \Lambda(\theta) \otimes {A},
\label{eqn:channel_extension}
\end{align}
where ${A}$ is an arbitrary channel acting on an ancilla system. The channel quantum Fisher information $I_{Q,{\rm ch}}(\Lambda; \theta)$ for $\Lambda(\theta)$ is defined as
\begin{align}
I_{Q,{\rm ch}}(\Lambda; \theta) = \max_{\rho}  I_Q\left(\Lambda\left[\rho\right];\theta\right).
\label{eqn:channel_qfi}
\end{align}
Using the monotonicity of the \qfi, it is enough to consider extensions by the identity~\cite{Fujiwara2001_PRA}. This identity channel extension has been demonstrated to generate increased estimation precisions~\cite{Fujiwara2003_JPA,Fujiwara2004_PRA}. The underlying principle is to enlarge the Hilbert space of the probing system, such that noisy components can be readily separated from the signal. It has been shown that for phase estimation, the ancilla is useful for arbitrary values of the noise parameter~\cite{Huang2016_PRA,Huang2018_PRA}. This has recently been demonstrated experimentally~\cite{Wang2018_PRA,Sbroscia2018_PRA}.

The second method of extending the Hilbert space of the system through ancillary states is the Hamiltonian extension scheme and involves adding an operator to the Hamiltonian. This is a more fundamental approach, since the Hamiltonian describes the probe dynamics during the parameterisation process. It differs to channel extension since it can include an interaction term between the probe and the ancilla. This provides a general framework for the treatment of open quantum systems. However, it has been shown that this approach does not improve the sensitivity of phase shift-type measurements when considering the quantum Fisher information optimised over input states~\cite{Fraisse2017_PRA}. Several experiments have demonstrated the advantage of entanglement-assisted schemes in other applications including channel tomography~\cite{Altepeter2003_PRL} and channel capacity estimation~\cite{Hotta2005_PRA}.

The precision limit for more general parameter estimation schemes---including adaptive protocols---can be addressed by defining the encoding channel $\Lambda(\theta)$. Gaussian channel estimation has been explored by Takeoka and Wilde~\cite{Takeoka2016_arxiv} using the notion of `environment-parameterised' channels. This includes the set of thermal-loss~\cite{Papanastasiou2018_PRA}, Pauli~\cite{Flammia2019_arxiv}, erasure~\cite{Bennett1997_PRL}, and noisy amplifier channels. The \textsc{qfi} for the adaptive estimation of a noise parameter $\theta$ in a channel is determined through
\begin{align}
I_{Q, \text{ch}}\left(\Lambda;\theta\right) = \frac{8\left[1 - F\left(\rho_{\Lambda_\theta}, \rho_{\Lambda_{\theta + \dif\theta}}\right)\right]}{\dif\theta^2}.
\label{eqn:teleportation_covariant_chan_QFI}
\end{align}
Equivalent results to bound the performance of continuous-variable adaptive channel estimation schemes have been derived for these channels using the notion of `teleportation-covariant' quantum channels~\cite{Pirandola2017_NC}
\begin{align}
\Lambda\left[U \rho U^\dagger\right] = V\Lambda[\rho]V^\dagger,
\label{eqn:telport_covariant_chan}
\end{align}
where $V$ is not necessarily equal to $U$. Channel estimation bounds determined in this manner use the teleportation simulation method introduced by Bennet \emph{et al.}~\cite{Bennett1996_PRA}. In this technique, all adaptive operations on many uses of a channel can be recast as a non-adaptive protocol if it is possible to simulate every channel by teleportation. Hence, due to the property~\eqref{eqn:telport_covariant_chan}, the precision scaling can be written in terms of the Choi matrices of the encoding teleportation-covariant channel~\cite{Pirandola2017_PRL}. The estimation of noise parameters in these channels is limited to the \textsc{sql}. These channels equivalently describe the class of physically motivated quantum channels described earlier and bosonic Gaussian channels~\cite{WeedbrookRMP_2012}.

It is worth noting that studies of channel estimation and channel discrimination are inextricably linked. For the prototypical quantum channel $\Lambda(\theta)$, the former scheme corresponds to continuous $\theta$, where the task is to estimate the unknown parameterisation. Channel discrimination corresponds to discrete $\theta$, where it is more meaningful to distinguish the quantum channels. The symmetric discrimination between two arbitrary qudit channels was considered by Pirandola \emph{et al.} using a port-based teleportation (\textsc{pbt}) scheme for channel simulation~\cite{Pirandola2019_npjQI}. The \textsc{pbt} scheme is a quantum teleportation protocol introduced by Ishizaka and Hiroshima, to transmit an unknown state of a subsystem to a receiver~\cite{Ishizaka2008_PRL}. It differs from conventional teleportation protocol~\cite{Bennett1993_PRL} by removing the requirement for the receiver to perform conditional unitary operations. The \textsc{pbt} scheme for channel simulation has also be used to derive general limits to adaptive quantum metrology~\cite{Laurenza2019_QMQM}. Specifically, any adaptive protocol of channel estimation is bounded in terms of the channel's Choi matrix and saturates the \textsc{hl} in the number of probings. 

Via bounds on state discrimination, one can immediately determine informative bounds for a range of other tasks. This has been illustrated in quantum communication and key distribution~\cite{Liao2018_NJP}, estimation theory~\cite{Cope2017_QMQM}, quantum illumination tasks~\cite{Ray2019_PRA}, and quantum hypothesis testing~\cite{Takeoka2016_arxiv, Pirandola2019_npjQI}.


\subsection{Fault-tolerant metrology}\label{subsubsec:fault_tolerant_metrology}
\noindent
Tools from quantum error correction can also be used to enhance and potentially recover estimation precisions in the Heisenberg scaling regime for sensing under Markovian noise. These methods are often used in conjunction with, or in addition to, ancilla-assisted schemes~\cite{Kessler14,Arrad2014_PRL,Dur14,Zhou2018_NC} and are referred to as fault tolerant quantum metrology. Their purpose is to correct errors in the probe state and/or the measurement procedure against noise while permitting the signal to be encoded on the sensor. 

Quantum error correction (\textsc{qec}) has found widespread application in quantum information processing and has evolved into a selfcontained and well developed field of active research~\cite{Devitt2013_RPP,Terhal2015_RMP}. However, migrating these tools to implement fault tolerant quantum metrology is not straightforward. The characteristic difficulty is in designing \textsc{qec} codes that detect and correct the noise in the sensor, while leaving the encoded signal intact. This constraint is uncommon in quantum computing applications and consequently many existing \textsc{qec} codes are not suitable for quantum sensing applications~\cite{Layden2019_PRL}. It is possible to identify \textsc{qec} codes that can be used to protect quantum sensing protocols against noise, based on certain criteria. First, if the signal Hamiltonian and the error operators commute, it has been shown that the application of error correction can enhance the sensing precision~\cite{Arrad2014_PRL, Kessler14, Dur14}. Quantum sensing in this regime has been demonstrated using nitrogen-vacancy centers~\cite{Kessler14}, trapped ions systems~\cite{Reiter2017_NC}, and has been experimentally demonstrated for field sensing~\cite{Unden2016_PRL} without the use of feedback control methods. Second, provided the generator exists outside the span of the Lindblad operators and assuming access to noiseless ancillas, there exists a quantum error correction protocol that protects the sensor against noise~\cite{Sekatski2017_Q, Dobrzanski2017_PRX, Zhou2018_NC}. This has been extended to more realistic settings, e.g., where the signal and noise are in the same direction~\cite{Layden2018_NPJ}, and where noiseless ancilla states were shown to be unnecessary when the signal Hamiltonian and the error operators commute~\cite{Layden2019_PRL}. 

The overheads necessary to realise fault-tolerant metrology schemes makes them unfeasible with current technology. However, these methods are gaining increasing attention, and the next goal is to determine realisable fault tolerant quantum metrology protocols. 


\section{Distributed quantum sensing}\label{sec:quantum_networks}
\noindent
A natural extension to multi-parameter quantum estimation is to distribute the parameters over different sensors and use multiple probe systems to interrogate the whole system. This networked approach distributes sensing protocols over multiple resources. Distributed quantum sensing has recently gained increased attention since it provides a promising platform to address a wider class of applications and compare existing results. For example, this framework can describe distributed interferometric phase sensing~\cite{Ge2018_PRL}, estimate the properties of a multi-dimensional field where each sensor is an ensemble of atoms~\cite{Proctor2018_PRL}, and calibrating continuous-variable quantum key distribution networks~\cite{Zhuang2018_PRA}. Architectures based on a network of sensors have also been reported to realise the quantum internet~\cite{Kimble2008_N}, scalable quantum computing~\cite{Nickerson2013_NC}, and national~\cite{Sasaki2011_OE} and international~\cite{Liao2017_N} quantum cryptography. Proctor \emph{et al.}\ developed a general framework for the treatment of photonic and atomic sensor networks~\cite{Proctor2017_arxiv}. This approach is also important since it helps to understand the optimal strategies that saturate the \textsc{qcrb}.  

In section~\ref{subsec:entanglement_for_quantum_metrology} we reviewed work that demonstrated the necessity of entangled probe states to facilitate \textsc{hl} precision scalings. However, in section~\ref{subsec:nonentangling_strategies} we saw recent work that demonstrate non-entangling strategies to achieve the same scaling as entangled probes~\cite{Braun2018_RMP}. Beyond a few counter examples that have reported too much entanglement as detrimental~\cite{Baumgratz2016_PRL}, little research to date has addressed the question when and how much entanglement is necessary as a resource in order to saturate fundamental quantum precision bounds~\cite{Kok2017_PRA}. Distributed quantum sensing holds the promise to demonstrate the utility of entanglement and elucidate its contribution in providing precision enhancements. A separate question that shares a similar dilemma is the choice between simultaneous and sequential estimation strategies. These two strategies generate different precision scalings in the count resource~\cite{Zwierz2012_PRA, Friis2017_NJP, Yousefjani2017_PRA} and the time resource~\cite{Magesan2013_PRA, Yuan2015_PRL, Dooley2016_PRA}. However, no clear results have been found that determine when either strategy should be preferred. In this section, we review efforts in distributed quantum sensing to address these two open questions. 

\subsection{Framework for distributed sensing}\label{subsec:framework_dist_ses}
\begin{figure}[t!] 
\begin{center}
\includegraphics[width =0.8\columnwidth]{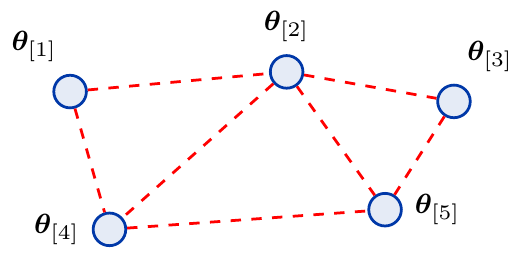}
\caption{Networked approach to quantum sensing for a decomposition of the parameters across 5 sensors with potentially entangled states (dashed lines). The $j^{\rm th}$ node is encoded by the parameter subset $\smash{\bm{\theta}_{[j]}}$.  The dashed lines represent (quantum) correlations between the nodes. This approach is well suited to field sensing, where each node is a spin system.}
\label{pic:networked_sensing_approach}
\end{center}
\end{figure}

\noindent
To formally define distributed sensing, consider a quantum system comprised of $n$ subsystems, labelled through an index $j = 1,2,\ldots,n$, with an associated Hilbert space $\smash{{\mathscr{H}}_j}$. The Hilbert space associated with two different subsystems are not required to be identical. Additionally, each space is not limited to a single field mode. The total Hilbert space is ${\mathscr{H}} = \otimes_{j=1}^n {\mathscr{H}}_j$. This product structure of the total system Hilbert space describes a spatially distributed sensor network as illustrated in~\ref{pic:networked_sensing_approach}. The physical system associated with each Hilbert space $\smash{{\mathscr{H}}_j}$ is referred to as a quantum sensor, and the collection of all sensors a quantum sensing network. 

Similar to single probe states, the parameters can be imprinted on the quantum network via unitary evolutions~\cite{Proctor2018_PRL} and more general dynamics~\cite{Sekatski2019_arxiv}. A unitary evolution of the total system is written as $\smash{{U}(\bm{\theta}) = \exp[-i\bm{H}\cdot\bm{\theta}]}$ where the vector of Hermitian operators $\smash{{\bm{H}} = ({H}_1,\ldots,{H}_D)}$ is the system Hamiltonian, and each node $j$ has $d_j$ parameters such that $\sum_j d_j = D$. The $j^{\rm th}$ Hamiltonian element ${H}_j = {h}_j \otimes_{k\neq j} \mathbbm{1}_k$ acts non-trivially on the $j^{\rm th}$ quantum sensor. From this, the product nature of the unitary is explicit:
\begin{align}
{U}(\bm{\theta}) = \bigotimes_{j=1}^n{U}_j(\bm{\theta}_{[j]})\, ,
\label{eqn:total_network_unitary}
\end{align}
where $[j]$ indicates the set of $d_j$ parameters at node $j$. Any system with a tensor product structure can be reconstructed in the language of quantum networks.

To understand how distributed sensing framework can be used to evaluate the role of entanglement, it is necessary to introduce local and global networked sensing strategies. For a given decomposition of a network into different sensors, an estimation procedure is local if~\cite{Knott2016_PRA} (1) the input probe state is separable with respect to the network of quantum sensors, and (2) the measurement and the construction of the estimator can be implemented with only local operations and classical communication (\textsc{locc}), along with local (classical) computations. A global estimation strategy is defined as a non-local estimation procedure, where either a probe that is entangled over the quantum sensors and/or a measurement requiring non-local quantum operations is used. Note that in local strategies, each parameter is estimated individually, while in global strategies parameters are estimated simultaneously. Local estimation strategies are more robust against local estimation failure than global strategies and often involve more realistic methods of state preparation~\cite{Etesse2015_PRL}, measurement, and control~\cite{Knott2016_PRA2}.

The use of correlations between different sensors in the network can provide precision enhancements beyond unentangled systems~\cite{Zhuang2018_PRA}. However, this is true only in specific scenarios. When each quantum sensor is used to estimate a single parameter, entanglement between different sensors has a detrimental impact on the estimation precision. Intuitively, when the parameters are encoded locally on each sensor, it does not make sense to use global states or measurements that exhibit entanglement between the sensors for improved precisions. Hence, under the single parameter networked sensing remit, separable state and measurements over the sensors are preferred~\cite{Proctor2018_PRL}. This reflects the results of simultaneous estimation in optical multi-parameter estimation problems~\cite{Humphreys2013_PRL,Yue2014_SR,Liu2016_JPA2}. It is also consistent with the results in~\cite{Baumgratz2016_PRL} where too much entanglement was reported as being detrimental, and the calibration of optical quantum gyroscopes, where the optimal level of entanglement depends on the parameters~\cite{Kok2017_PRA}. If instead each sensor is used to estimate multiple parameters, the effect of entanglement on the estimation precision depends on properties of each of the parameter generators~\cite{Proctor2018_PRL}. If all the generators corresponding to the parameters commute, then entanglement between sensors is still detrimental to the estimation precision. In contrast, if the generators do not all commute then entanglement between sensors may in some cases give a small constant reduction in the estimation uncertainty. However, this advantage diminishes when each sensor is coupled to an ancillary system. Practically, this is important for examples where the resources used must be limited to avoid damaging a sample whose properties are being probed~\cite{Wolfgramm2012_NP,Taylor2013_NP,Taylor2014_PRX}.

The original scheme for simultaneous quantum-enhanced phase estimation by Humphreys \emph{et al.}\ was realised using highly entangled \textsc{noon} states. However, the same precision enhancements exhibited by the global strategies can be obtained with mode-separable states and local measurements alone~\cite{Knott2016_PRA}. This is not surprising since in multimode optical systems with commuting phase generators, the crucial resource for enhanced metrology is a large particle number variance within each mode, which can be obtained without multimode entanglement.

\subsection{Estimating functions of parameters}\label{subsec:estimating_functions}
\noindent
So far we have considered the multi-parameter estimation of a vector of parameters: for a well-defined decomposition of a network of quantum sensors, different subsets $\smash{\bm\theta_{[j]}}$ of the parameters are locally encoded onto the space of the $j^{\rm th}$ sensor. An alternative approach to estimating the vector of parameters is to estimate a function of these elements instead. This is better suited for many practical applications and can be more informative than estimating the parameters. Immediate examples of this include vectorial magnetometry for imaging using a network of atomic sensors~\cite{Pham2011_NJP, Alfasi2016_AIP}, the realisation of a global quantum network of clocks~\cite{Komar2014_N}, and field interpolation~\cite{Qian2019_arxiv}. Field interpolation exemplifies the advantage of estimating a function of parameters since the the objective is to estimate the field at a spatial region not covered by a sensor within the network.

Determining precision bounds for the estimation of functions requires a strategy that minimises estimate variances of general functions of $\bm{\theta}$. One can construct a ${D}$-dimensional vector of parameter functions
\begin{align}
\bm{\vartheta} = \left(f_1(\bm{\theta}), \ldots, f_{D}(\bm{\theta})\right)
\label{eqn:function_of_parameters}
\end{align}
where $\smash{f_j: \mathbbm{R}^{D}\rightarrow\mathbbm{R}, j \in \{1,{D}\}}$ are continuously differentiable functions. In section~\ref{sec:multi_parameter_est}, we observed that the \textsc{qfi} elements transform as a metric tensor. Hence the single-shot \textsc{qcrb} for an estimator of $\smash{\bm{\vartheta}}$ is given by Eq.~\eqref{eqn:qcrb_for_functions}. The attainability of this bound for function estimation has a strict caveat. Extreme care should be taken when estimating a scalar function of multiple parameters $f(\theta_1,\ldots,\theta_D)$. This can not be treated as a single-parameter estimation $f$, since the mapping of many parameters to one implicitly assumes that the remaining $D-1$ parameters, whatever they are, are known perfectly. This assumption is generally not satisfied and the procedure therefore leads to unattainable bounds~\cite{Gross19}. This unattainability has been explored for a network of single qubits by Eldredge \emph{et al.}, where for the specific problem of nanoscale magnetic resonance imaging, \textsc{ghz} and spin-squeezed states are found to be optimal~\cite{Eldredge2018_PRA}.

Work in this direction has to date been limited to estimating the class of linear functions, where the vector $\smash{\bm{\vartheta}}$ is constructed from linear transformations of $\smash{\bm{\theta}}$ through $\smash{\bm{\vartheta} = {J \bm\theta}}$. This class of functions is useful for modeling optical and atomic sensing applications, and the estimation of parameter averages. The last application has been considered for a network of precision clocks by Komar \emph{et al.}~\cite{Komar2014_N}. To facilitate convenient comparisons between these works, Proctor \emph{et al.}\ demand each row of the transformation matrix $\smash{{J}}$ be a normalised vector~\cite{Proctor2017_arxiv}. In the case of a single parameter encoded on each sensor, entanglement can help to perform better estimates of linear functions. The precise effect of entanglement depends on the explicit chose of linear function. Denoting the first row of $\smash{{J}}$ with the vector $\smash{\bm{v}}$ such that $\smash{\vartheta_1 = \bm{v}\cdot\bm\theta}$, then Proctor \emph{et al.}\ find that an entangled probe across the sensors always provides a precision enhancement compared with a separable state, unless the function $\smash{\bm{v} = \|\bm{v}\|\bm{e}_k}$~\cite{Proctor2018_PRL}. Intuitively, if $\smash{\bm{v}}$ contains more than one non-zero element, then $\vartheta_1$ describes a global process across multiple sensors and naturally, a globally correlated state will be most sensitive to changes in $\vartheta$. Conversely, if only one element is non-zero, the process is local and entanglement is not required and can indeed be detrimental for the estimation performance. 

To conclude, the use of simultaneous estimation schemes and entangled resources are not always necessary to achieve quantum-enhanced sensing. In particular, the use of entanglement can be detrimental and any precision enhancements depend on whether the parameters of interest are local or global properties of a set of systems.  As future work, it would be interesting to observe how these results can be extended to the class of non-linear functions. The analogy between the choice of appropriate weight/importance matrices in the class of non-distributed sensing to the freedom in choice of arbitrary functions in distributed sensing can also be explored.


\section{Conclusions and outlook}\label{sec:conc_outlook}
\noindent
Without exaggeration, our ability to devise novel approaches for high precision measurements has been one of the greatest drivers of fundamental science and technologies. A plethora of historical examples in optics, communication, computation, imaging, and metrology bear substantial evidence to support this statement. However, we are reaching various limits in metrology, including the shot noise limit and the diffraction limit. The latter directly impacts our ability to keep up with Moore's law in the manufacturing of electronics. Quantum metrology provides a natural extension to classical methods to continue driving performance improvements using quantum systems. 

Quantum metrology has matured into a broad field with many active areas of theoretical and experimental research. In this review, we provide a summary of the main techniques in sensing and metrology. We use geometric arguments stemming from information theoretic concepts to motivate the key quantities in classical and quantum estimation theory. This is a powerful approach that provides a method to visualise the estimation process, and helps compare different sensing protocols. It is important to understand what resources are necessary to improve the precision of measurements. Precision bounds derived using geometric arguments can also provide conceptual connections between quantum estimation and general quantum information tasks. We review methods to generate general bounds for qubit and multi-particle systems using query complexity arguments for the estimation procedure. 

Central to information geometry is the information matrix, which assigns a quantitative measure on the performance of parameter estimates. We show that the quantum Fisher information (\textsc{qfim}) is part of a wider family of different information matrices that can be found via the Schwarz inequality. These information matrices can be used to derive bounds for elements of the covariance matrix. For multiple parameters, we review the symmetric logarithmic derivative (\textsc{sld}) and the right logarithmic derivative \textsc{qfim}, the Kubo-Mori and Wigner-Yanase skew information matrices. We review how they can be connected via their Riemannian metrics. We provide a comparison of these bounds and discuss their attainability. Specifically, we report that the commonly used \textsc{qcrb} for multiple parameters is often not simultaneously saturable, which adds to the difficulty that optimal measurements often depend on the true, unknown values of the parameters. The Holevo Cram{\'e}r-Rao bound is then found to be the most informative alternative bound, although it is generally difficult to determine for an arbitrary probe (note that this is an area of current research interest~\cite{Albarelli19}). The specific case of Gaussian state estimation and non phase-like parameters is considered. Numerical methods to solve the \textsc{qcrb} have also been discussed. Although we focus mostly on unbiased Fisher estimation schemes, we comment on the effect that biased estimators have on the estimation precision. As a first in the literature, we also show how to construct the Bayesian analogue of the \textsc{qfim}. We show that this procedure follows the same method as the Fisher estimation scheme using the prior-weighted density operator and the posterior mean operator to define a corresponding symmetric logarithmic posterior mean operator.

The prevailing thought of entanglement as a necessary resource for enhanced quantum sensing has been challenged, and  several works attribute the precision enhancement instead to quantum correlations. Further, the optimal amount of entanglement in the probe state depends on the parameters and the estimation scheme. This demonstrates that entanglement provides limited use in certain cases. This could be seen as welcoming news, given the difficultly to produce robust states with high entanglement. We review efforts that explore non-entangling strategies to saturate the \textsc{qcrb}. 

We conclude by summarising key results in current research efforts in noisy metrology schemes. A longstanding impediment of noisy quantum metrology is the reduction of quantum enhancements to constant factor improvements over classical precision bounds. This has often reduced expectations of practical applications of quantum sensing. The use of fault-tolerant schemes to reinstate quantum scaling has renewed efforts to explore the most general bounds in noisy quantum metrology with error correction. This strategy introduces its own unique difficulties: first, well-known methods in quantum error correction for quantum computing cannot be straightforwardly implemented for use in quantum metrology since they would also ``correct'' the signal we want to measure. Second, the physical overheads required make most uses currently impractical. We review research that addresses these issues. Another current area of intense research is distributed quantum sensing. This field demonstrates how methods in quantum estimation theory are often interwoven with related efforts in quantum communication. We summarise our current understanding of the most efficient methods to distribute resources over different nodes in a network of sensors. 

The progress of practical applications presented in this review highlights quantum sensing as a frontrunner to the emergence of quantum technologies. However, developing realisable protocols remains a non-trivial problem with many open questions.

\begin{acknowledgments}
\noindent The authors would like to thank Jonathan A. Gross, Rafal Demkowicz-Dobrzanski, and Jes{\'u}s Rubio Jimenez for useful discussions, and Cosmo Lupo for valuable comments on the manuscript. JSS and PK acknowledge EPSRC for funding via the Quantum Communications Hub (EP/M013472/1).
\end{acknowledgments}

\section*{References}
\bibliographystyle{apsrev4-1}
%

\end{document}